\documentclass[
  aps,prd,
  reprint,
  superscriptaddress,
  amsmath,amssymb,
  preprintnumbers,
  nofootinbib,
]{revtex4-2}

\usepackage{subfig}
\usepackage{graphicx}
\usepackage{dcolumn}
\usepackage{bm}
\usepackage{comment}
\usepackage{xcolor}
\usepackage{array}
\usepackage{xurl}
\usepackage[pdftex,bookmarks,hidelinks]{hyperref}
\providecommand{\cmsTable}[1]{\resizebox{\linewidth}{!}{#1}}
\usepackage{xspace}
\usepackage{enumitem}
\usepackage{booktabs}
\usepackage{listings}
\usepackage{caption}
\usepackage{ragged2e}
\lstdefinelanguage{XML}{
  morestring=[b]",
  morecomment=[s]{<!--}{-->},
  morekeywords={xmlns,version,type,id},
  keywordstyle=\color{blue},
  stringstyle=\color{cyan},
  commentstyle=\color{blue},
  identifierstyle=\color{red}
}
\begin{document}
\newcommand{\numu}{\ensuremath{\nu_\mu}\xspace}
\newcommand{\nue}{\ensuremath{\nu_e}\xspace}
\newcommand{\numubar}{\ensuremath{\bar{\nu}_\mu}\xspace}
\newcommand{\nuebar}{\ensuremath{\bar{\nu}_e}\xspace}
\newcommand{\sig}{\ensuremath{1\mu \text{N}p}\xspace}
\newcommand{\dmsq}{\ensuremath{\Delta m^2_{41}}\xspace}
\newcommand{\thmumu}{\ensuremath{\theta_{\mu\mu}}\xspace}
\newcommand{\thee}{\ensuremath{\theta_{ee}}\xspace}
\newcommand{\thmue}{\ensuremath{\theta_{\mu e}}\xspace}
\newcommand{\sinsqtwothmumu}{\mathrm{sin}\ensuremath{^22\theta_{\mu\mu}}\xspace}
\newcommand{\sinsqtwothee}{\mathrm{sin}\ensuremath{^22\theta_{ee}}\xspace}
\newcommand{\sinsqtwothmue}{\mathrm{sin}\ensuremath{^22\theta_{\mu e}}\xspace}
\newcommand{\chisq}{\ensuremath{\chi^2}\xspace}
\newcommand{\GeVcc}{\mathrm{GeV c}\ensuremath{^2}}

\preprint{FERMILAB-PUB-26-0194-PPD}

\title{\textbf{First search for sterile neutrino oscillation leading to \texorpdfstring{$\nu_\mu$}{nu\_mu} disappearance in the Booster Neutrino Beam at ICARUS} 
}

\affiliation{Brookhaven National Laboratory, Upton, NY 11973, USA}
\affiliation{CBPF, Centro Brasileiro de Pesquisas Fisicas, Rio de Janeiro, Brazil}
\affiliation{CERN, European Organization for Nuclear Research 1211 Gen\`eve 23, Switzerland, CERN}
\affiliation{University of Chicago, Chicago, IL 60637, USA}
\affiliation{Centro de Investigacion y de Estudios Avanzados del IPN (Cinvestav), Mexico City}
\affiliation{Colorado State University, Fort Collins, CO 80523, USA}
\affiliation{Fermi National Accelerator Laboratory, Batavia, IL 60510, USA}
\affiliation{University of Houston, Houston, TX 77204, USA}
\affiliation{Indian Institute of Science, Bengaluru, India}
\affiliation{INFN Sezione di Bologna and University, Bologna, Italy}
\affiliation{INFN Sezione di Catania and University, Catania, Italy}
\affiliation{INFN Sezione di Genova and University, Genova, Italy}
\affiliation{INFN GSSI, L’Aquila, Italy}
\affiliation{INFN LNGS, Assergi,  Italy}
\affiliation{INFN LNS, Catania, Italy}
\affiliation{INFN Sezione di Milano, Milano, Italy}
\affiliation{INFN Sezione di Milano Bicocca and University, Milano, Italy}
\affiliation{INFN Sezione di Napoli, Napoli, Italy}
\affiliation{INFN Sezione di Padova and University, Padova, Italy}
\affiliation{INFN Sezione di Pavia and University, Pavia, Italy}
\affiliation{INFN Sezione di Pisa, Pisa, Italy}
\affiliation{University of Pittsburgh, Pittsburgh, PA 15260, USA}
\affiliation{University of Rochester, Rochester, NY 14627, USA}
\affiliation{SLAC National Accelerator Laboratory, Menlo Park, CA 94025, USA}
\affiliation{Southern Methodist University, Dallas, TX 75275, USA}
\affiliation{University of Texas at Arlington, Arlington, TX 76019, USA}
\affiliation{Tufts University, Medford, MA 02155, USA}
\affiliation{Virginia Tech, Blacksburg, VA 24060, USA}
\affiliation{York University, Toronto, Canada}

\author{F. Abd Alrahman}\altaffiliation{}\affiliation{University of Houston, Houston, TX 77204, USA}
\author{P. Abratenko}\altaffiliation{}\affiliation{Tufts University, Medford, MA 02155, USA}
\author{N. Abrego-Martinez}\altaffiliation{}\affiliation{Centro de Investigacion y de Estudios Avanzados del IPN (Cinvestav), Mexico City}
\author{A. Aduszkiewicz}\altaffiliation{}\affiliation{University of Houston, Houston, TX 77204, USA}
\author{F. Akbar}\altaffiliation{}\affiliation{University of Rochester, Rochester, NY 14627, USA}
\author{L. Aliaga Soplin}\altaffiliation{}\affiliation{University of Texas at Arlington, Arlington, TX 76019, USA}
\author{M. Artero Pons}\altaffiliation{}\affiliation{INFN Sezione di Padova and University, Padova, Italy}
\author{J. Asaadi}\altaffiliation{}\affiliation{University of Texas at Arlington, Arlington, TX 76019, USA}
\author{W. F. Badgett}\altaffiliation{}\affiliation{Fermi National Accelerator Laboratory, Batavia, IL 60510, USA}
\author{B. Baibussinov}\altaffiliation{}\affiliation{INFN Sezione di Padova and University, Padova, Italy}
% ...
\collaboration{ICARUS Collaboration}

\date{\today}

\begin{abstract}
We present a search for muon neutrino disappearance in the Booster Neutrino Beam (BNB) at Fermilab using the ICARUS detector. Neutrino interactions identified as muon neutrinos interacting with argon nuclei via the charged current interaction and having only a muon and at least one proton in the final state (1$\mu$Np) have been selected from data collected in 2022-2023 (ICARUS Run 2) and compared with a simulation-based expectation. In the context of a fit to a two-neutrino approximation of the sterile 3+1 model, including the impact of systematic uncertainty from the flux, neutrino interaction, and detector models, we find no statistically significant muon neutrino disappearance at the ICARUS baseline of 600 meters from the BNB target. Corresponding 90\% C.L. exclusion contours in $\Delta m^2_{41}$ - sin$^22\theta_{\mu\mu}$ space are presented. This is the first oscillation analysis produced by ICARUS exposed to the BNB. We note that the analysis is systematics limited due to large unconstrained uncertainties from the flux and interaction models. In future joint analyses, data from ICARUS and the SBND detector, exposed to the BNB at 110 meters from target, will be combined to provide significant constraint of these uncertainties, enabling a robust, world-leading two-detector analysis.
\end{abstract}

\maketitle


\section{\label{sec:intro}Introduction}
While three-flavor neutrino oscillation measurements are incompatible with short-baseline oscillation, a number of anomalous results~\cite{LSND:2001aii,MiniBooNE:2013uba,Kaether:2010ag,SAGE:2009eeu,Barinov:2022wfh, Serebrov:2020kmd, IceCubeCollaboration:2024nle} suggest the possibility of neutrino oscillation at a scale of L/E$\sim$1~km/GeV. Definitive investigation of this possibility is one of the primary motivations of the Short-Baseline Neutrino (SBN) program~\cite{SBNProposal}, consisting of three liquid argon time projection chamber \cite{Rubbia:LarTPC} (LArTPC) detectors in the Booster Neutrino Beam at Fermilab. Detectors observing similar flux in the same beamline, having the same target nucleus, and having similar detector designs allows for significant constraint of systematic uncertainty when data from multiple detectors is analyzed together. 

The ICARUS LArTPC, located 600 meters downstream of the BNB target, serves as the far detector of the SBN  program 
and has been taking physics data since 2022. Here, we present a search for muon neutrino disappearance using only the ICARUS detector~\cite{ICARUS:T600,ICARUS:InitialFNALOperation}, which is an important first step towards a more sensitive two-detector SBN analysis. Our results are consistent with other disappearance analyses~\cite{IceCubeCollaboration:2024dxk,MINOS:2017cae,NOvA:2024imi,MiniBooNE:2012meu,Dydak:1983zq,Stockdale:1984cg} and demonstrate the quality of ICARUS data and the performance of reconstruction and analysis tools developed by the collaboration. As expected in a single-detector oscillation search, the sensitivity is limited by large \textit{a priori} uncertainties in modeling the neutrino flux and neutrino interactions on argon; the addition of near detector data will provide \textit{in situ} measurements designed to constrain uncertainty in future SBN analyses. 

Data is evaluated in the context of the 3+1 sterile neutrino model, described in Section~\ref{sec:sterile}. For this analysis we consider neutrinos produced in the Booster Neutrino Beam (BNB) at Fermilab and observed in the ICARUS detector, which are described in Section~\ref{sec:beamanddet}. 
The details of the experimental operating conditions are described in Section~\ref{sec:operations}. Simulated data are generated using LArSoft~\cite{larsoft} with SBN-specific detector descriptions; the simulation is described in Section~\ref{sec:simreco}.

We present analyses based on two different reconstruction frameworks: Pandora~\cite{PandoraFirstPaper} and SPINE~\cite{Spine}, which are described in Section~\ref{sec:simreco} and in Appendices~\ref{app:pandora} and \ref{app:spine}, respectively. These reconstruction and selection parts of the analysis are based on separate code with very different underlying algorithms, but many aspects of the analysis, such as simulation, signal processing, calibration, cosmic rejection, systematic uncertainty, and statistical methods are shared. Observed differences in the performance of the two reconstruction approaches have been used to help improve them and/or to help evaluate systematic uncertainty in the analysis. For these reasons, the results should not be treated as fully independent, but rather considered together as a single result.

Charged-current (CC) interactions of muon neutrinos on argon nuclei with a final state consisting of one muon and at least one proton ($1\mu$Np), in which all final-state particles are reconstructed as contained within the detector fiducial volume, are considered as signal events. This final state is selected due to its favorable reconstruction characteristics (contained muon and proton tracks can be robustly identified, and their momenta can be estimated from range) and its strong rejection of cosmic background (the presence of a vertex naturally eliminates many cosmic muon tracks).
Details of the event selection are described in Section~\ref{sec:selection}. The impact of a comprehensive suite of systematic uncertainties, including contributions from flux, interaction, and detector modeling, is evaluated. The systematic variations included in the analysis and studies to confirm appropriate coverage of uncertainties are described in Section~\ref{sec:syst}. A new fitting framework, PROfit, has been developed for SBN and is described in Section~\ref{sec:fit} and Appendix~\ref{app:profit}.
Results of the analyses are shown in Section~\ref{sec:results}.

The analysis presented here has been performed according to a blind procedure, with the reconstruction tools and selection criteria developed using only 10\% of on-beam data and larger samples of non-neutrino data. Data-Monte Carlo (MC) comparisons of neutrino candidates using the full dataset were performed only after the analysis was frozen and reviewed by the collaboration. The fitting framework was developed using only simulation, with no fits performed on data until after the analysis was frozen. The unblinding fit was performed using a staged approach, as described in Appendix~\ref{app:profit}.

\section{\label{sec:sterile}Short-Baseline Oscillation and Sterile Neutrinos}
Neutrino oscillation driven by a mass-squared splitting $\Delta m^2_{41} \sim$1~$\mathrm{eV}^{2}$, in a 3+1 sterile neutrino extension to the Pontecorvo–Maki–Nakagawa–Sakata (PMNS) formalism, has been proposed as an explanation for anomalous electron-neutrino-like appearance seen by LSND~\cite{LSND:2001aii} and MiniBooNE~\cite{MiniBooNE:2013uba}. Gallium-based detectors have also reported anomalous electron-neutrino disappearance which could arise from a similar mass-squared splitting~\cite{Kaether:2010ag,SAGE:2009eeu,Barinov:2022wfh}. 
The mass-squared splittings of the PMNS three-neutrino model are known to be far smaller than $1~\mathrm{eV}^{2}$, and thus oscillations at this scale would require an additional neutrino mass state in the model. As measurements from the Large Electron-Positron (LEP) Collider at CERN have ruled out more than three light neutrino states coupling to the Z boson~\cite{ALEPH:2005ab}, the fourth flavor in the minimally extended 3+1 model is assumed to not couple to the weak force and is called ``sterile." The PMNS mixing matrix is extended to a $4\times4$ unitary matrix with six real valued mixing angles $\theta_{ij}$ with $1 \leq i < j \leq 4$~\cite{Kopp:2013vaa}.

Potential appearance signals suggested by the anomalous results are in tension with other measurements from accelerator neutrinos~\cite{Giunti:2019aiy, Hardin:2022muu}; most recently, results from MicroBooNE~\cite{MicroBooNE:2025nll} and KATRIN~\cite{KATRIN:2025lph} exclude at 95\% confidence level the single light sterile neutrino interpretation of the LSND and MiniBooNE anomalies and most of the gallium allowed regions, respectively.

A number of other experiments have placed limits in $\dmsq$ and $\sinsqtwothmumu$ space using
$\numu$ disappearance, as is done in the analysis reported here. 
At short baselines, the survival probability for muon neutrinos ($\numu$), of energy $E_\nu$, traveling a distance $L$, is well approximated as
\begin{equation}\label{eq:numu_survival}
P\left(\nu_\mu\rightarrow\nu_\mu\right) \approx  1 - \sinsqtwothmumu \mathrm{sin}^2 \left(1.267  \frac{\dmsq L}{E}\right),
\end{equation}
where the PMNS parameters are expressed in an effective mixing angle
\begin{equation}\label{eq:theta_numu}
\sin^2\left(2\theta_{\mu\mu}\right) \equiv 4\cos^2\theta_{14}\sin^2\theta_{24}\left(1 - \cos^2\theta_{14}\sin^2\theta_{24}\right).
\end{equation}
Most experiments have set limits on $\numu$ disappearance, while IceCube reports a closed contour at 95\% C.L.~\cite{IceCubeCollaboration:2024nle}. The existing searches for $\numu$ disappearance are summarized alongside the ICARUS result presented here in Section~\ref{sec:results}.

\section{\label{sec:beamanddet}Beam and Detector}
\begin{figure}[!htb]
    \centering
    \includegraphics[width=\linewidth]{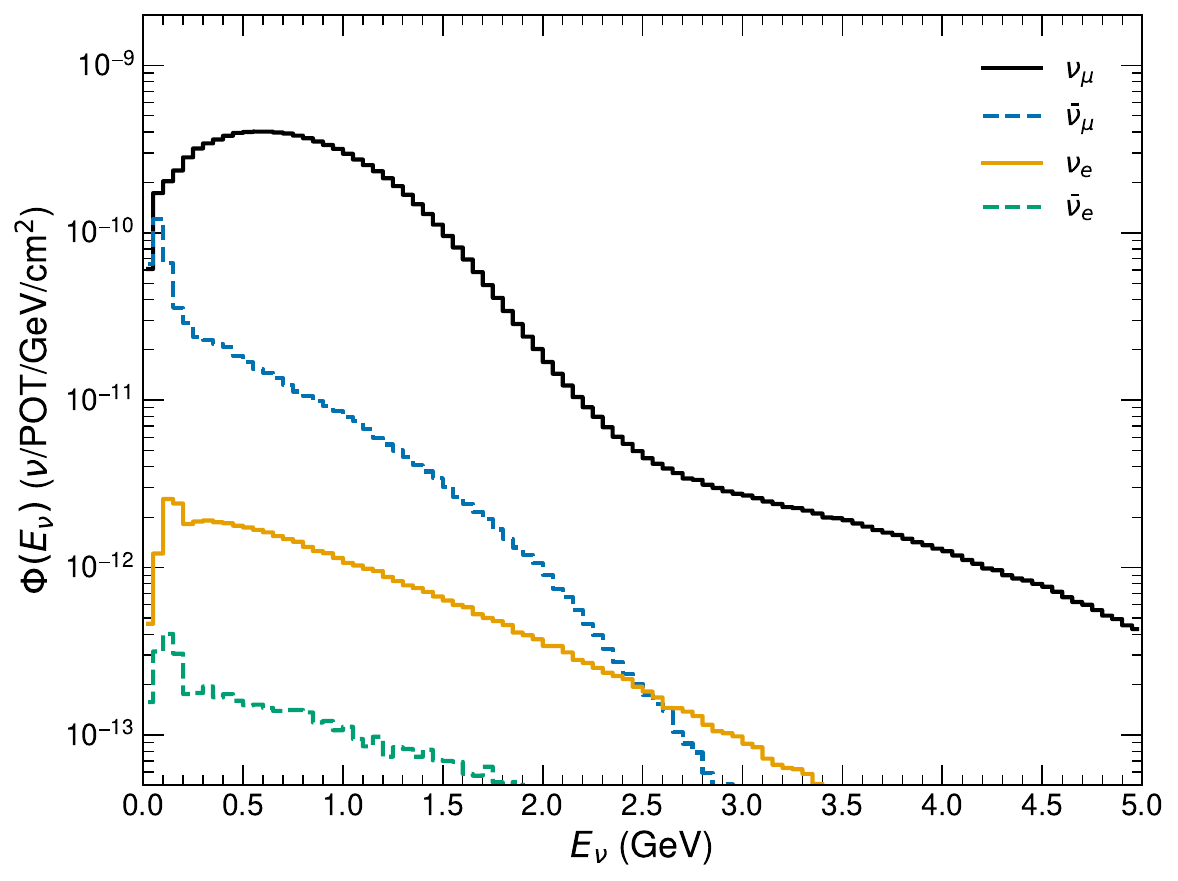}
    \caption{\justifying Predicted BNB flux by neutrino species at ICARUS. The flux simulation was developed by MiniBooNE~\cite{MiniBooNE:2008hfu} and MicroBooNE~\cite{uboonefluxpubnote} and is based on \textsc{Geant4}~\cite{Agostinelli2003Geant4}, with corrections applied to the pion production cross section using a Sanford--Wang parameterization fitted to HARP~\cite{Schmitz:2008zz} and E910~\cite{E910:2007puw} data. Parent hadron decays to neutrinos are sampled over the front face of the ICARUS detector, where the flux is predicted to vary by less than 4\%.}
    \label{fig:bnbflux}
\end{figure}

The Booster Neutrino Beam (BNB) is generated from protons accelerated to a kinetic energy of 8 GeV impinging a beryllium target. Each 1.6 $\mu$s spill of the Booster is nominally composed of $\sim$5$\times10^{12}$ protons split into 81 bunches. These bunches are 2 ns wide and approximately 19 ns apart. The collisions from these protons on the BNB target produce a beam of hadrons, primarily pions with a small kaon fraction. The charged hadrons are focused by a magnetic horn into a 50~m-long decay tunnel where the hadrons decay in flight, producing neutrinos. ICARUS is located on-axis of the BNB and 600~m from the BNB target.  The expected flux by particle type at ICARUS is shown in Fig.~\ref{fig:bnbflux};
the flux at ICARUS is composed primarily of $\nu_\mu$ ($\sim$93\%) with small backgrounds of $\bar\nu_\mu$ ($\sim$6\%) and $\nu_e/\bar\nu_e$ ($<1$\%). The flux has an average kinetic energy of approximately 0.8 GeV.
In addition, ICARUS is exposed to the Main Injector neutrino beam (NuMI) at a 5.75° far-off-axis position, 800~m downstream of the NuMI target.

The primary ICARUS detector is a LArTPC
containing a total of 760~tons of ultra-pure liquid argon. 
ICARUS consists of two identical adjacent cryogenic modules (``east" and ``west" cryostats), with each module having two TPCs separated by a common cathode. The maximum drift length is 1.5~m, corresponding to $\sim1$~ms at the nominal drift field of 500~V/cm. The anode structure has three readout wire planes with 3~mm pitch, spaced 3~mm apart, with wires oriented at 0° (Induction 1), +60° (Induction 2), and –60° (Collection) with respect to horizontal. With appropriate voltage biasing, the first two planes (Induction 1 and Induction 2) provide nondestructive charge measurements, while the ionization charge is fully collected on the final (Collection) plane. Each TPC identifies neutrino interactions through the detection of ionization charge deposited in tracks and electromagnetic showers by charged particles produced in neutrino-argon interactions. The ionization charge is used to reconstruct charged particle trajectories with good calorimetry and precise ($\sim$mm) spatial resolution. More details are available in~\cite{ICARUS:T600,ICARUS:InitialFNALOperation}. 
Scintillation light is detected by 360 8" photomultiplier tubes (PMTs) installed behind the TPC wire planes and used for event selection and triggering~\cite{ICARUS:PMT2020}.
A cosmic ray tagging
(CRT) system~\cite{ICARUS:CRT} surrounds the cryostat with two layers of wavelength-shifting fiber embedded plastic scintillators and concrete overburden of $\sim$3~meters is installed above the detector to mitigate background from cosmic rays. 
The ICARUS trigger system~\cite{ICARUS:TriggerPaper} issues a global trigger that opens data-acquisition windows sized to capture the full charge drift and all associated scintillation light. The primary BNB ``on-beam" trigger relies on coincident PMT activity inside the beam gate timed to the corresponding $1.6~\mu s$ long proton-extraction spills.
Analogous ``off-beam" triggers collect cosmic-ray events using gates shifted 33~ms after each beam gate. Complementary ``minimum-bias" triggers are independent of scintillation activity, so provide unbiased data samples at reduced rates to support calibration and performance studies.

\section{\label{sec:operations}Data}

The analysis presented here uses data collected with the BNB of Fermilab from winter 2022 to spring 2023 (ICARUS Run 2), corresponding to an exposure of $\sim$2.05$\times 10^{20}$ proton on target (POT). 
Data with nominal detector and beam conditions, and thus suitable for analysis, are identified using recorded run history and detector conditions, data quality metrics, and beam quality metrics.

\begin{figure}[!htb]
    \centering

    \includegraphics[width=1.0\linewidth]{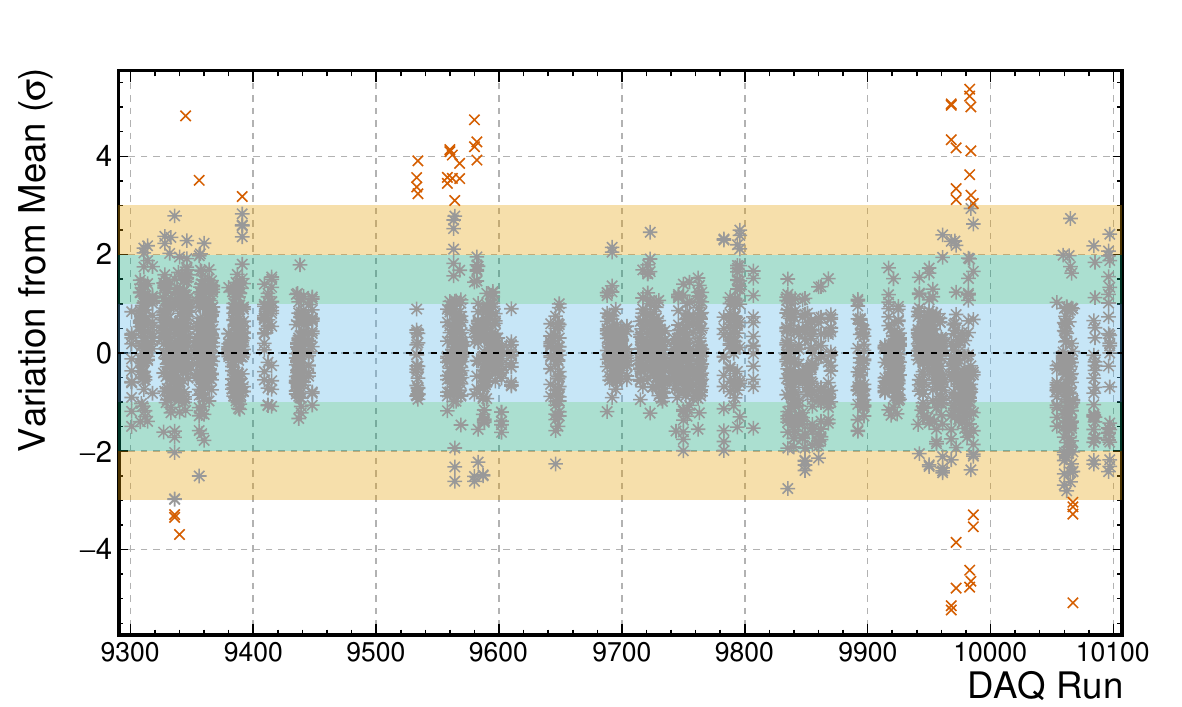}
    \caption{\justifying The deviation of the data quality metrics from their respective means for each run that passed pre-filtering. The dashed line represents no deviation from the global mean of that quantity. The blue, green, and yellow bands represent the regions 1, 2, and 3 standard deviations from the mean, respectively. Any metric marked in red falls outside out the 3$\sigma$ region and the corresponding runs are flagged as poor data quality.}
    \label{fig:detector_dq_summary}
\end{figure}

Good runs are defined as those in which
the DAQ was properly working, all detector components (TPCs, PMTs, and top and side CRT) were present in the collected data, the run had a minimum duration of 1~hour, and no other detector or DAQ anomalies were recorded that could spoil the quality of the collected data. 
Data quality metrics are evaluated for each run compared to the global mean for the dataset, removing runs which have a metric more than 3$\sigma$ away from the global mean. The adopted metrics are chosen to reflect low-level activity in the detector, such as the average number of hits per track and the average total photoelectrons (PE) per scintillation light burst, as well as higher-level reconstructed quantities such as the number of distinct interactions per event. 
Additionally, checks are placed on the wire bias voltage and cathode voltages to ensure they were within nominal ranges.
The deviation from the mean for all quality metrics for each run can be seen in Fig.~\ref{fig:detector_dq_summary}.

Good spills within good runs are identified using beam quality metrics adopted from MicroBooNE that are based on information recorded by beamline monitoring devices in the BNB. 
The
requirements are non-zero beam intensity, nominal horn current, and a ``good" value for a metric that quantifies the centering of the spill on the beam target.

After data and beam quality cuts, the total POT used for the analysis is $\sim$1.596$ \times 10^{20}$ POT.
The reduction of available exposure due to the data and beam quality is  shown in Table~\ref{tab:DataBeamQualityStatistics}.

 \begin{table}[!htbp]
 \centering
 \begin{tabular}{lcccc}
 \hline
        & \multicolumn{2}{c}{POT } & \multicolumn{2}{c}{ Spills}\\
      &  tot (10$^{20}$)& frac.  & tot (10$^{7}$) & frac. \\
      \hline
      Total analyzed & 1.957 & 1.000 & 5.232 & 1.000\\
      Good Runs & 1.752 & 0.895 & 4.630 &0.885\\
      Beam Quality & 1.783 & 0.911 & 4.491 & 0.858\\
      Beam Quality + Good Runs& 1.596 & 0.816 & 4.014 & 0.767\\ 
      \hline
 \end{tabular}
 \caption{Effect of the data and beam quality conditions on the POT and spills available.}
\label{tab:DataBeamQualityStatistics}
 \end{table} 
 \setlength{\tabcolsep}{1em}

\begin{figure}[!htb]
    \centering
    \includegraphics[width=0.9\linewidth]{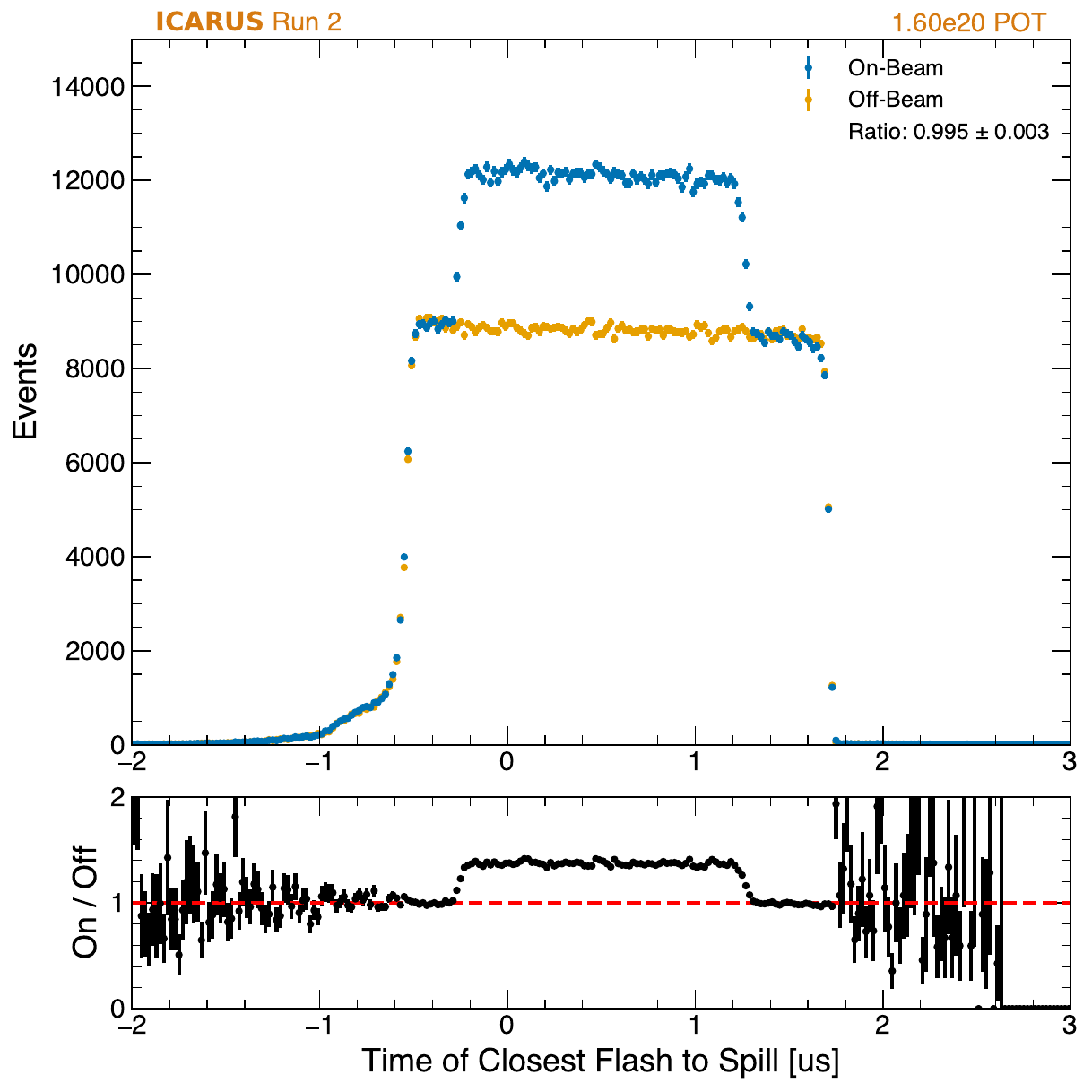}
    \includegraphics[width=0.9\linewidth]{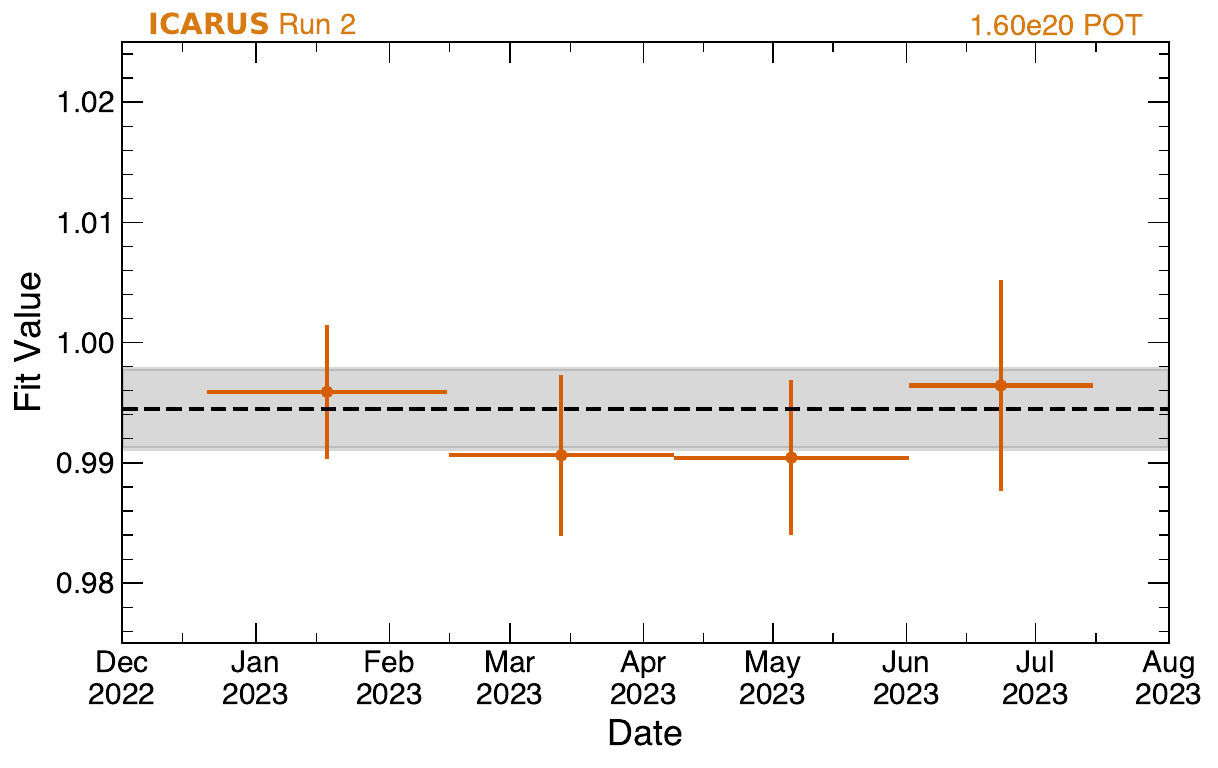}
    \caption{ \justifying (Top) Triggering flash time distribution for 
    on-beam (blue) and off-beam (orange) Run 2 majority triggers (top), normalized to the same exposure, and the corresponding on-beam/off-beam ratio (bottom). A clear beam induced excess is observed in the [-.3,1.3]$~\mu s$ beam spill window. The on-beam/off-beam ratio in the region $[1.3,1.6]~\mu$s is calculated and found to be consistent with 1.0 as expected. 
    (Bottom) On-/off-beam ratio in the region $[1.3,1.6]~\mu$s of the top panel for four two-month periods, with statistical uncertainties. The dotted line and red band indicate the fitted ratio and 1$\sigma$ error for the full period.
   }
    \label{fig:OnOffBeamNormalization}
\end{figure}

 The POT corresponding to the analyzed datasets is extracted using spill and POT information recorded for each triggered event and validated by comparing triggering–flash time distributions in on- and off-beam data normalized to equal exposure, as shown in the top panel of Fig.~\ref{fig:OnOffBeamNormalization}. A clear excess from BNB interactions is observed in on-beam triggers during the  1.6$~\mu s$ neutrino beam spill, while identical cosmic-ray rates are seen outside the spill window.  
The normalized on- and off-beam cosmic contributions agree within statistical uncertainties throughout  Run 2 data taking, as seen in the bottom panel of Fig.~\ref{fig:OnOffBeamNormalization}, differing only by $(0.5\pm 0.3)\% $ when averaged over the whole Run 2.

\section{\label{sec:simreco}Simulation and Reconstruction}
\subsection{Simulation}
\label{sec:Simulation}
A Monte Carlo (MC) simulation of neutrino interactions coming from the BNB, corresponding to 1.74$\times10^{21}$ POT, i.e., 10 times larger than the analysed data, is produced. MC events include a BNB neutrino interaction in the liquid argon in the 1.6~$\mu$s of the BNB spill and cosmic ray interactions crossing the detector in a $\pm 1$~ms time window around the BNB spill. 
The neutrino interactions are simulated using the GENIE event generator~\cite{Andreopoulos:2009rq,geniereleases}; the \verb|AR23_20i_00_000|
comprehensive model configuration,
which is developed within the context of DUNE~\cite{DUNE:2020lwj}, is chosen as a reference model. Additional details of the interaction model and its uncertainties are discussed in Section~\ref{sect:InteractionSystematics}.
Cosmic rays are modeled by CORSIKA~\cite{corsika}.
MC events are processed, reconstructed and selected using the same \emph{icaruscode} software used for data, based on the \emph{LArSoft} framework~\cite{larsoft}, which is  used in all LArTPC experiments.

The central value (CV) simulation  includes a modeling of the response of all subdetectors (TPC, PMTs, CRT) and the corresponding generation of signals and noise. Many known  effects impacting the detector response have been modeled and calibrated with a data-driven approach, as described in more detail in~\cite{ICARUS:TPCCalibration}. The effect of unmodeled or mis-modeled effects are included as systematic uncertainties, as described in Section~\ref{sect:DetectorSystematics}.
The main detector aspects addressed in the simulation are: 
\begin{itemize}
\item{TPC gain and signal shape:} The  wire signal gain, i.e., the signal amplitude corresponding to a known number of drift electrons, is computed using cosmic muons as described in~\cite{ICARUS:TPCCalibration}. Variations of around $1\%$ among the average gain of different TPCs are observed and are included in the simulation. 

\item{TPC noise:}
The spectra and RMS of both coherent (correlated across a DAQ board) and intrinsic noise components are simulated independently for each board, based on the corresponding noise observed in data. 
\item{Electron lifetime:}
The average values observed in data during Run 2 are 4(8)~ms for the East(West) cryostat, while the simulation assumes a conservative value of 3~ms, identical for both cryostats.
\item{Recombination model:}
Several models have been proposed in literature to parameterize the probability of drift electron recombination as a function of both electric field and dE/dx. The CV assumes the Modified Box model~\cite{ArgoneutRecombination}.
\item{Space charge:}
The accumulation of ions from liquid argon ionization, mainly induced by cosmic rays, generates a distortion of the TPC electric field and consequently of the reconstructed coordinates, extending up to a few millimeters for a detector on surface~\cite{SpaceChargePavia}. The simulation includes this effect.
\item{Scintillation light modeling:}
A quantum efficiency (QE) of 7.3\% is used in the simulation of light collected by the PMTs; this value is selected to match the average light signal amplitude between data and MC.
\end{itemize}

\subsection{Raw Signal Processing and Calibration}
\label{sec:RawSignalProcessing}
The processing of the recorded TPC wires signals follows the logic described in more detail in~\cite{ICARUS:InitialFNALOperation}, based on the ``deconvolution" of the electrostatic field and electronic response from the wire signal waveform, ideally recovering the original current of drift electrons in the TPC.
Deconvolution is applied wire by wire, after removing coherent noise across the 64 channels of each front-end board, using the expected field and electronics response functions~\cite{ICARUS:TPCCalibration}.

The field response, describing the drift of electrons in the vicinity of the TPC wires and the generation of signals via electrical induction or charge collection, has been tuned in a data-driven way, minimizing the difference between predicted and observed signals, as illustrated in~\cite{ICARUS:TPCCalibration}. 
The time intervals containing physical signals (``hits") are searched for in each wire waveform, based on a threshold algorithm. Hits are then fitted with a Gaussian, whose area is proportional to the number of drift electrons generating the signal.

The calibration of TPC wire signals converts the area of signal waveforms into an accurate measurement of the energy deposited on each wire by a physical track, through the ionization of liquid argon and electron drift~\cite{ICARUS:TPCCalibration}. This requires a quantification of detector effects impacting signal generation, such as electron recombination, attenuation of the drift electron current,
diffusion, and possible nonuniformities of the TPC calorimetric response throughout the TPC volume, as described in~\cite{ICARUS:EllipsoidalRecombinationModel} and~\cite{ICARUS:TPCCalibration}. Calibrated hits provide the local energy deposition for a track segment, 
typically a few millimeters in length, as measured on each wire, along with the corresponding 2D localization from the wire coordinate and drift time. Overall, the efficiency for detecting a wire signal and correctly associating it with the originating track is greater than 90\% in the induction planes and exceeds 98\% in the collection plane.

Each PMT waveform is processed with a 
threshold-based algorithm to identify fired PMTs and reconstruct the detected light.
All signals above threshold within 100~ns are then clustered into a ``flash", which nominally corresponds to the light produced by a single interaction, whether from cosmic rays or neutrinos, with few ns time resolution.

CRT hits—space–time points associated with tracks crossing the CRT—are formed by time-ordering CRT signals above the 
thresholds in both adjacent layers and grouping them by region, with several ns and cm resolution. 
CRT timing is cross-calibrated with the PMT system and is validated by studying the CRT-PMT time of flight measured with incoming cosmic muons~\cite{ICARUS:CRT}.  

\subsection{Track, shower and event reconstruction}
\label{sec:TrackEventReco}

The reconstruction chain uses the calibrated TPC hits, PMT-based hits and flashes, and CRT hits from the raw-signal processing stage. Low thresholds for defining hits are chosen to maintain high hit efficiency but introduce noise-induced false signals; these are mitigated
by forming 3D spacepoints from combinations of 2D hits 
with intersecting wires and consistent timing.
Starting from these objects, the pattern recognition stage identifies the tracks and showers present in each event, hierarchically grouped into separate interactions and associated to a primary interaction vertex which should ideally correspond to possible neutrino interactions or individual cosmic rays in the detector. In this reconstruction step, the physical properties associated to each track/shower and to the reconstructed interaction are also determined, which serve as inputs to the event selection described in the following section. 
The Pandora and SPINE reconstruction algorithms start from the same input, 
but adopt different strategies to resolve 3D objects and identify and reconstruct the interactions in the detector. Pandora first reconstructs separate 2D clusters in the different wire planes, which are then combined into 3D objects as described in appendix \ref{app:pandora}, while SPINE considers all possible combinations of 2D hits as  described in appendix \ref{app:spine}. 

Pandora  is a pattern recognition software package widely used in LAr-based detectors, especially LArTPCs (e.g.,~\cite{PandoraMB2017, PandoraProtoDune2022}). 
It provides a fully-automated, end-to-end reconstruction of events, generating a hierarchical set of tracks and electromagnetic showers to represent final-state particles and any subsequent interactions or decays. 
Pandora adopts a multi-algorithm approach, in which complex pattern recognition tasks are decomposed into specialized subtasks. 
Each algorithm operates independently within the processing chain and addresses a specific aspect in a given topology - for example, interaction vertex identification or shower 3D reconstruction. The collective output of this ensemble of algorithms progressively builds up a global representation of the underlying events, providing a robust and comprehensive reconstruction. All the 3D objects deemed to be associated with a single interaction in the detector - either a neutrino interaction or a cosmic ray - are grouped together in a so called \textit{slice}. A schematic of the reconstruction flow is shown in Fig.~\ref{fig:Pandora_scheme}. More details about its application to ICARUS are present in Appendix~\ref{app:pandora}.

The SPINE (Scalable Particle Imaging with Neural Embeddings)~\cite{Spine} reconstruction is an end-to-end, Machine Learning (ML) based reconstruction chain. It consists of a hierarchical set of neural networks that can be trained as a cohesive whole. Two types of neural networks are employed: CNNs (Convolutional Neural Networks) and GNNs (Graph Neural Networks). The CNNs are used to extract point-level information from voxelized 3D images of the detector activity, while the GNNs are used to aggregate this point-level information and extract higher-level features at the particle- and interaction-level. 
The neural networks in the SPINE reconstruction chain are trained using supervised learning techniques on a large set of simulated events independent from those used in the analysis. This training set uses generic, physics-agnostic event generators to produce neutrino-like and cosmic ray-like interactions in order to cover a wide range of possible event topologies and avoid biases towards specific interaction models. The simulated events are processed through the full ICARUS simulation to ensure that the training data accurately reflects the detector response and noise characteristics. More details of the SPINE framework are provided in Appendix~\ref{app:spine}.

For both reconstruction paradigms, the energy of each particle is estimated through its range, where the distance traveled by the particle its related to its momentum or kinetic energy via the continuous slowing down approximation (CSDA)~\cite{PDG2025}. 
The energy of the neutrino is reconstructed by adding the total energy of the muon, the kinetic energy of the proton(s),
and an empirical average removal energy of 30.9~MeV~\cite{Bodek2019} for each proton above 50~MeV. 

\section{\label{sec:selection}Event Selection}
\subsection{Selection Criteria}
\begin{figure}[!htb]
    \centering
    \includegraphics[width=\linewidth]{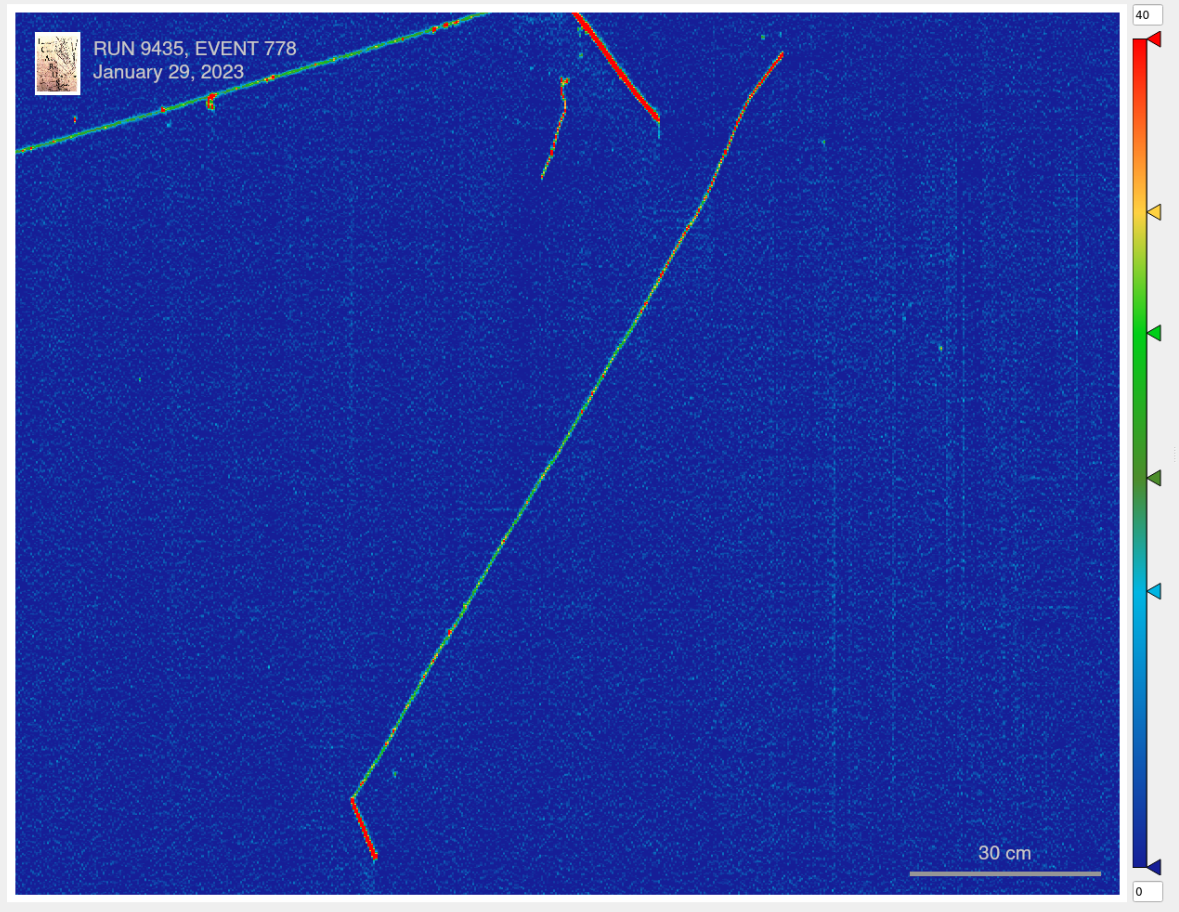}
     \includegraphics[width=\linewidth]{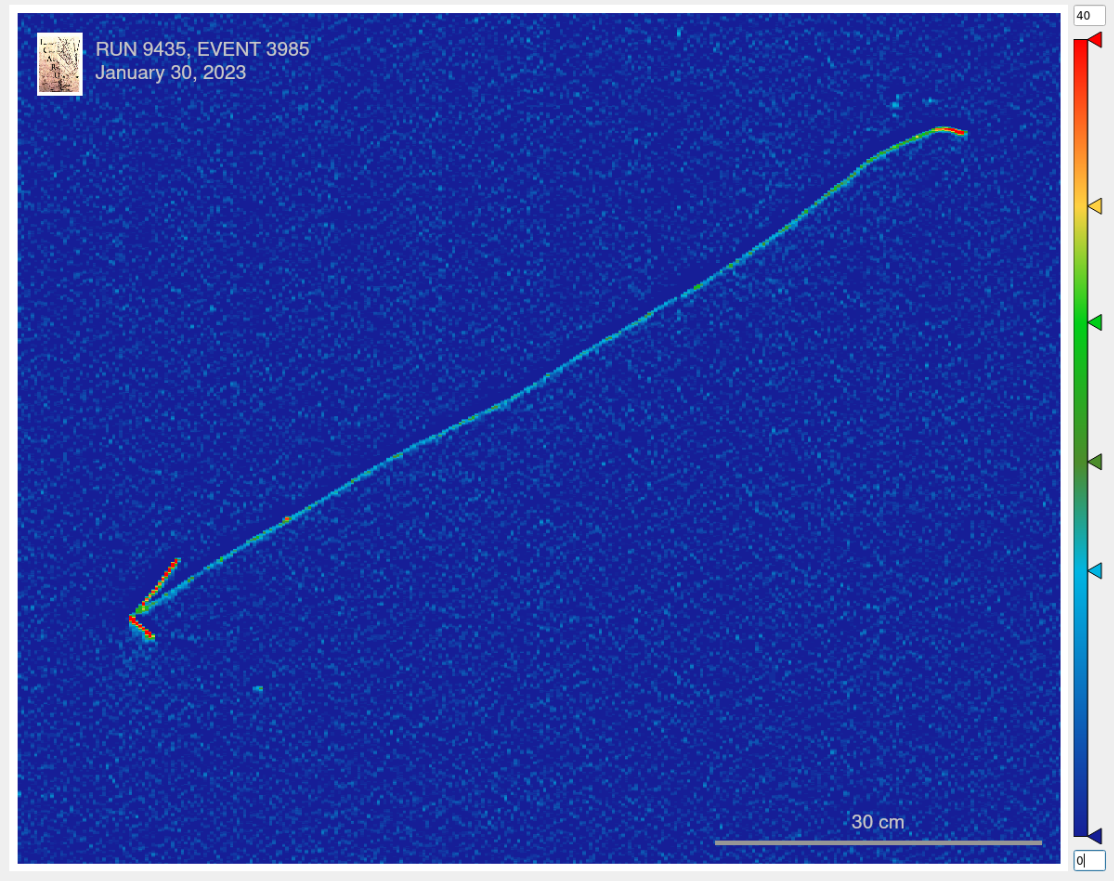}
    \caption{\justifying Examples of 1$\mu$1p (top) and 1$\mu$2p (bottom) candidate events observed in Run 2 data. In the upper part of the top figure overlapping cosmic activity is also visible. The color scale indicates the energy deposition of each particle. 
    }
    \label{fig:1muNp_evd}
\end{figure}

A 1$\mu$Np final-state topology is chosen to simplify reconstruction and minimize cosmic contributions. The target topology has a charged-current vertex in the fiducial volume with exactly one contained muon longer than 50~cm in length, at least one proton with deposited energy $>50$~MeV, and no additional charged particles or photons. All reconstructed charged particles must be contained within the active volume, 
to allow for range-based measurements of particle momenta.
The muon length requirement is intended to reduce the contamination from NC interactions with a pion in the final state. The proton threshold ensures that the track is long enough to be reconstructed with good efficiency and sufficiently energetic to be well separated from the muon in the final state.
A veto threshold is applied to extra particles to ensure that the particles are sufficiently energetic to be visible in the final state if they are indeed present.
Example event displays illustrating the target final state are shown in Fig.~\ref{fig:1muNp_evd}. 

Event selection criteria are applied to the reconstructed events with the aim of selecting 1$\mu$Np candidates while rejecting other neutrino interactions as well as cosmic rays associated events. Some selection criteria are common between the Pandora and SPINE-based analyses while others differ due to the intrinsic nature of both algorithms. The common criteria applied are: 
\begin{itemize}
    \item Fiducial volume cut: requires the reconstructed event vertex to be more than 25~cm from the edge of the active volume in the horizontal ($x$) and vertical ($y$) coordinates, and more than 30 (50)~cm from the upstream (downstream) edge of the active volume in the $z$-coordinate (along the beam axis). To remove portions of the volume in which the drift field is distorted by a hanging cable, an additional cut is applied,
    removing $\sim$0.3\% of the liquid argon volume.
    \item Containment: all particles contained within the reconstructed interactions are required to be more than 5~cm from the edge of the active volume in all directions. 
    \item Single Muon in the Final State: presence of a single primary muon with a length of at least 50~cm. 
    \item Proton(s) in the Final State: presence of  at least one primary proton depositing at least 50~MeV of energy. 
    \item Pion/Shower Veto: the final state may not contain any charged pions or showers (electrons/photons). These particles are required to be primary particles and to have deposited at least 25~MeV of energy to be counted as final-state particles. 
\end{itemize}

In Pandora, tracks and showers are identified by 
a Boosted Decision Tree (BDT)~\cite{ArtificialIntelligenceBDT} classifier. 
Particle identification variables for tracks are calculated by comparing the measured energy deposition to a MC-based template of dE/dx vs.~residual range for a given particle type. 
While perfect data/MC agreement is not achieved in the PID classifier distributions, the applied cuts are loose and implemented in a region of the distribution that is well-modeled. 

The selection based on the Pandora reconstruction includes 
several additional cuts for rejection of cosmic ray background. 
The interaction must not be tagged as ``clear cosmic" by Pandora, as described in Appendix~\ref{app:pandora}.
Events are rejected in which 
PMT and CRT signals are in time with each other and with the beam spill, indicating a particle entering or exiting the cryostat. Compatibility between the  event reconstructed in the TPC and the triggering flash in the PMT system is checked by comparing the barycenter along the $z$-direction of the light associated with the triggering flash and the TPC charge barycenter of the neutrino candidate, also in the $z$-axis. The event is selected if the separation between the two barycenters is less than 100~cm. 

In SPINE, the selected interaction is required to be matched with an optical flash detected by the PMTs that is consistent with the time of the neutrino beam spill. 
This requirement provides significant background rejection of cosmic-induced interactions that are not in-time with the beam, but is not expected to provide significant background rejection for cosmic-induced interactions that are in-time with the beam and that trigger the detector readout. The latter might instead be targeted using the CRT system; however, given that contained events are selected in this analysis, there is very little background from in-time cosmics and so a CRT veto is not applied. 

 \subsection{Efficiency, purity and background estimation}
\begin{figure}[!htb]
    \centering
    \includegraphics[width=0.95\linewidth]{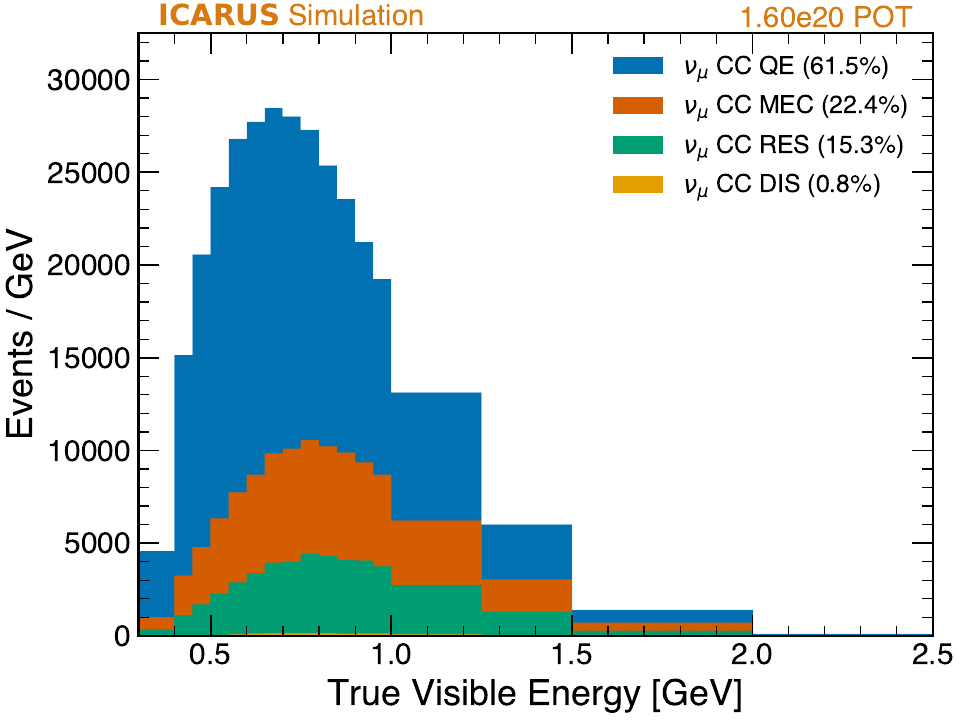}
    \caption{\justifying True \sig visible energy spectrum selected by truth criteria. The true interaction type is indicated by color. The rate is scaled to the exposure of Run 2, 1.6 $\times 10^{20}$ POT.
    }
    \label{fig:truespec}
\end{figure}

The selection efficiency, purity, and background levels are estimated primarily using MC simulation. Figure~\ref{fig:truespec} shows the defined \sig signal spectrum, with contributions from each true interaction type. Many of the true \sig events are from quasielastic interactions, 
but there is a significant contribution from meson exchange current 
and resonant pion production.

 Selection efficiency is computed using MC simulation as the ratio between the number of selected events and true number of events in the target sample. Similarly, the purity is defined as selected events truly in the target sample divided by all selected candidates. As a result, some events classified as ``background" are true $\numu$-CC interactions that have a different true final-state topology from the target sample. This choice does not impact the results of the analysis as the efficiency and purity are used only for informational purposes and not in extraction of physics results. 
 The Pandora selection achieves a 48\% efficiency and 82\% purity and the SPINE selection achieves 77\% efficiency and 92\% purity, as evaluated with MC simulation. In both selections, the primary background is mis-identified \numu charged-current interactions, with less than 3.8\%  and 1.1\% of the selected sample coming from NC interactions for the  Pandora and SPINE selections, respectively.

 The CRT-PMT cosmic rejection algorithm relies only on timing information and only isolated flashes are selected, so the main inefficiency for signal comes from random CRT-PMT coincidences within the 200 ns window. In order to evaluate this contribution, a data-driven analysis is performed, showing an inefficiency of $\leq 1.0\%$. This result is cross-checked with a visual study of identified events without the application of the CRT-PMT-based cut, where compatible results are obtained.

 The contribution of in-time cosmics is measured directly from data, by applying the event selection to the off-beam dataset. The in-time contribution is found to be $\sim$0.5\% for Pandora and $<0.1$\% for SPINE. The contribution from out-of-time comics in coincidence with neutrino events is estimated by MC and corresponds to only $\sim$0.5\% of the total selected events in Pandora and $<0.1$\% for SPINE, comparable to that from in-time cosmics. Thus the overall cosmic background accounts for only $\sim$1\% and $<$0.1\% of the total selected events for Pandora and SPINE respectively. The cumulative efficiency and purity of the selections are shown in Table~\ref{tab:Eff_table} and \ref{tab:Purity_table}, respectively, for each analysis.

\begin{table}[!htbp]
\centering
\begin{tabular}{lcc}
\toprule
Selection Cut -   Efficiency             & Pandora {[}\%{]} & SPINE {[}\%{]} \\ \toprule 
No Cut                   & 100                 & 100             \\
Reconstructed Vertex in FV       & 96.8                & 99.4             \\
Containment          & 92.4                & 98.9             \\
Cosmic Rejection         & 88.3                & 96.0            \\
Muon ID                  & 76.9                & 87.5            \\
Proton ID                & 55.2               & 80.6            \\
No Pions    & 50.7                & 78.2            \\
No Showers  & 48.5                & 77.0            \\ 
\bottomrule
\end{tabular}
\caption{\justifying Summary of the cumulative efficiency values for Pandora and SPINE \sig selection on a cut-by-cut basis.}
\label{tab:Eff_table}
\end{table}

\begin{table}[!htbp]
    \centering
        \begin{tabular}{lcc}
        \toprule
        Selection Cut -   Purity & Pandora {[}\%{]} & SPINE {[}\%{]} \\
        \toprule
        No Cut & $<0.1$ & $<0.1$  \\
        Reconstructed Vertex in FV & 0.8  & 0.8 \\
        Containment & 1.2 & 6.5  \\
        Cosmic Rejection & 24.9 & 26.1  \\
        Muon ID & 45.2 & 52.4  \\
        Proton ID & 65.1 & 68.1  \\
        No Pions & 71.4 & 80.1 \\
        No Showers & 81.5 & 91.6 \\
        \bottomrule
        \end{tabular}
        \caption{\justifying Summary of the cumulative purity values for Pandora and SPINE \sig selection on a cut-by-cut basis.}
        \label{tab:Purity_table}
    \end{table}
    
\begin{figure}[!htbp]
    \centering
    \includegraphics[width=0.9\linewidth]{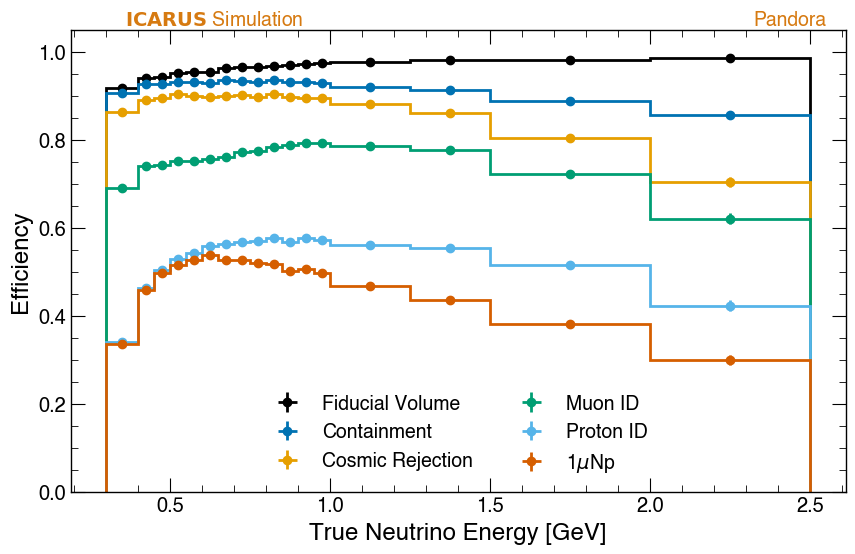}
    \includegraphics[width=0.9\linewidth]{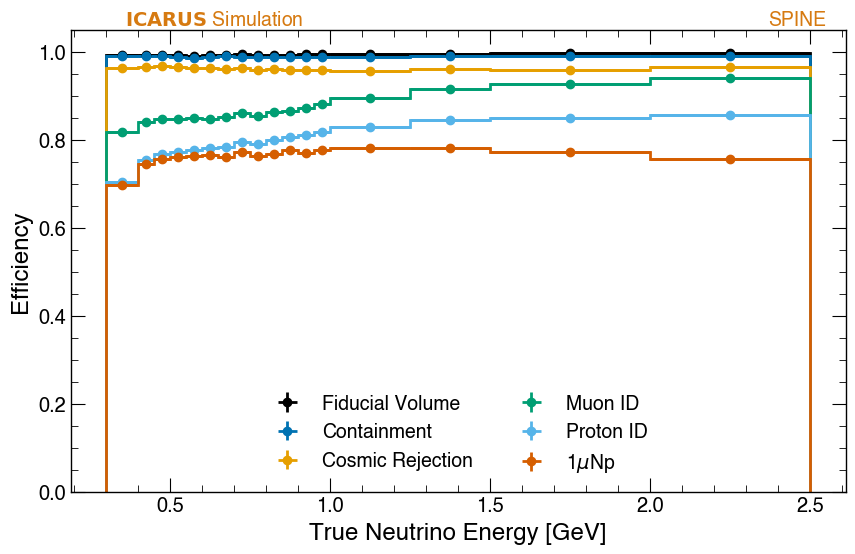}
    \caption{\justifying Cumulative efficiency of selection criteria for Pandora (top) and SPINE (bottom) as a function of true neutrino energy applied in the order appearing in the plot, with respect to true \sig signal, as defined in the text. Statistical uncertainties are included but in most cases are too small to be visible in the plots.}
    \label{fig:signaleff}
\end{figure}

\begin{figure}[!htb]
    \centering
    \includegraphics[width=0.9\linewidth]{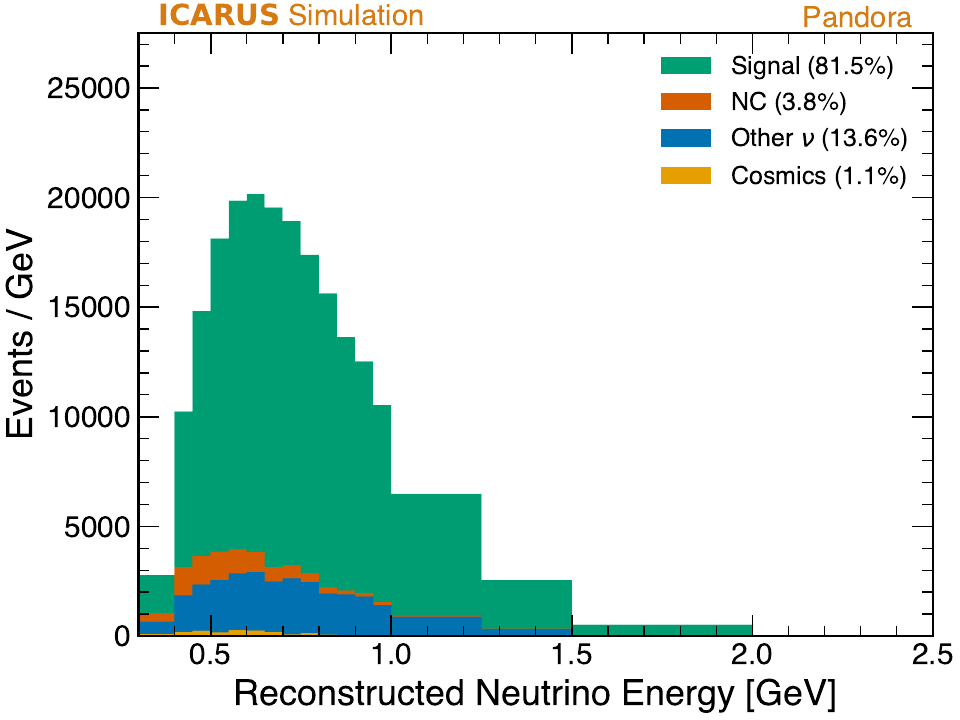}
    \includegraphics[width=0.85\linewidth]{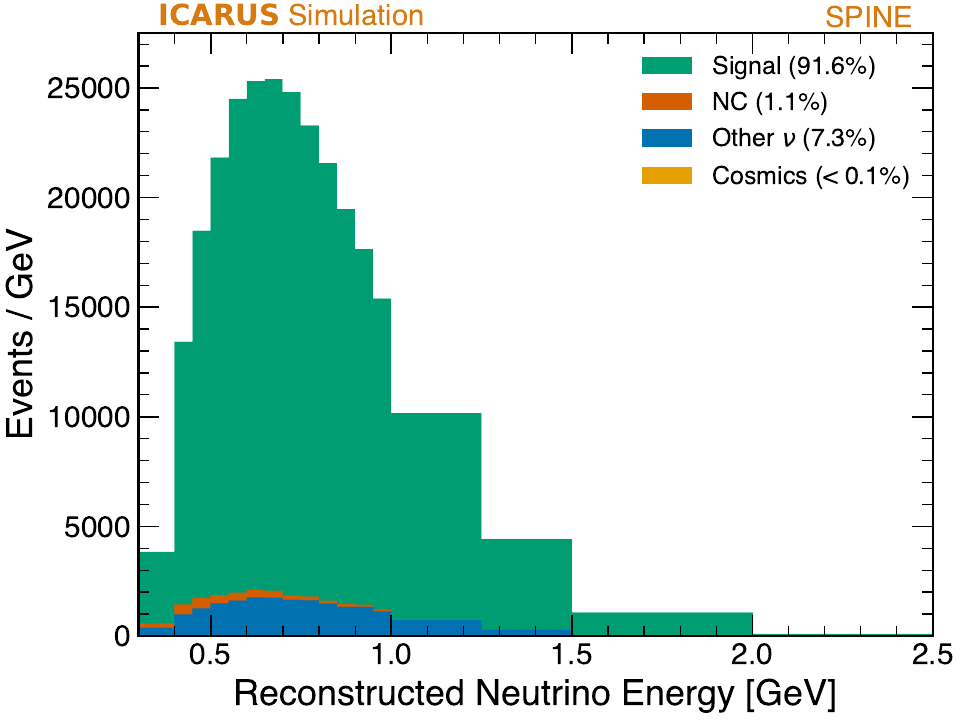}
    \caption{\justifying Selected spectrum from simulation sample highlighting the different final-state topologies for Pandora (top) and SPINE (bottom).}
    \label{fig:selected_mc}
\end{figure}

 Figure~\ref{fig:signaleff} shows the efficiency for each of the selection criteria with respect to the above true signal definition, applied cumulatively, as a function of true neutrino energy, for Pandora and SPINE. For Pandora, the largest source of inefficiency is the proton identification and is dominated by mis-reconstruction of few cm long proton tracks. The muon identification inefficiency is primarily from mis-reconstruction of the muon track, while the pion rejection inefficiency has a number of different sources. Note that the lower efficiency regions at low and high neutrino energy correspond to regions with low signal statistics; as such they are statistically limited even in the MC samples, as seen in the top plot of Fig.~\ref{fig:selected_mc}.
For SPINE, the selection efficiency falls off at lower muon/proton kinetic energies as expected, manifesting as a drop-off in efficiency at lower total neutrino energy, as seen in the bottom plot of Fig.~\ref{fig:selected_mc}. The main background is a collection of charged-current muon neutrino interactions that either contain charged pions in the final state or are actually non-contained events reconstructed as being contained.

\begin{figure}[!htb]
    \centering
    \includegraphics[width=0.9\linewidth]{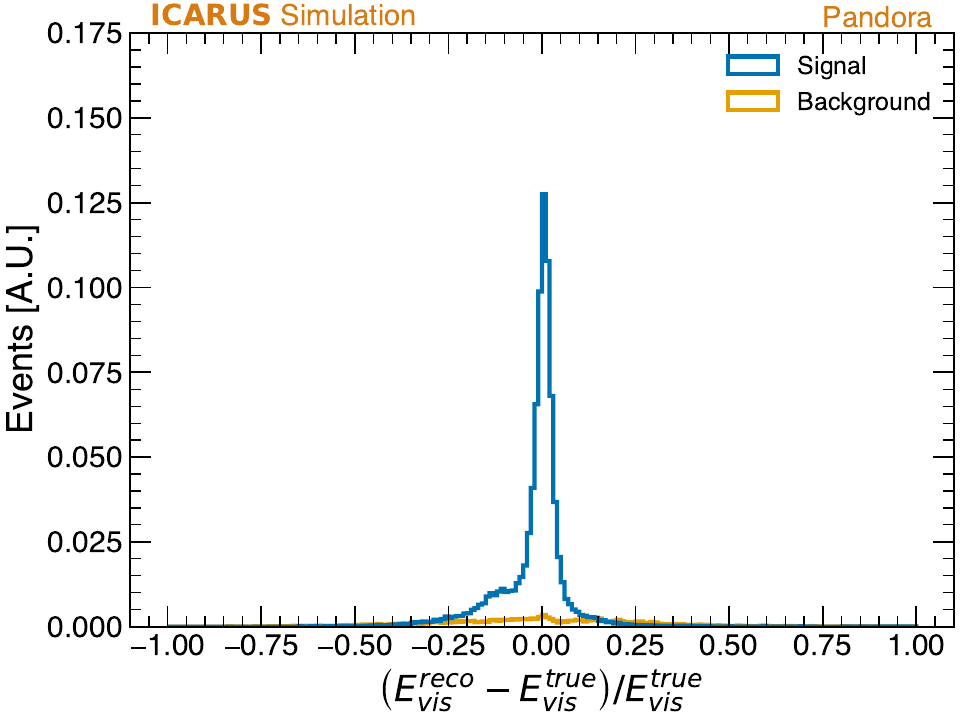}
    \includegraphics[width=0.9\linewidth]{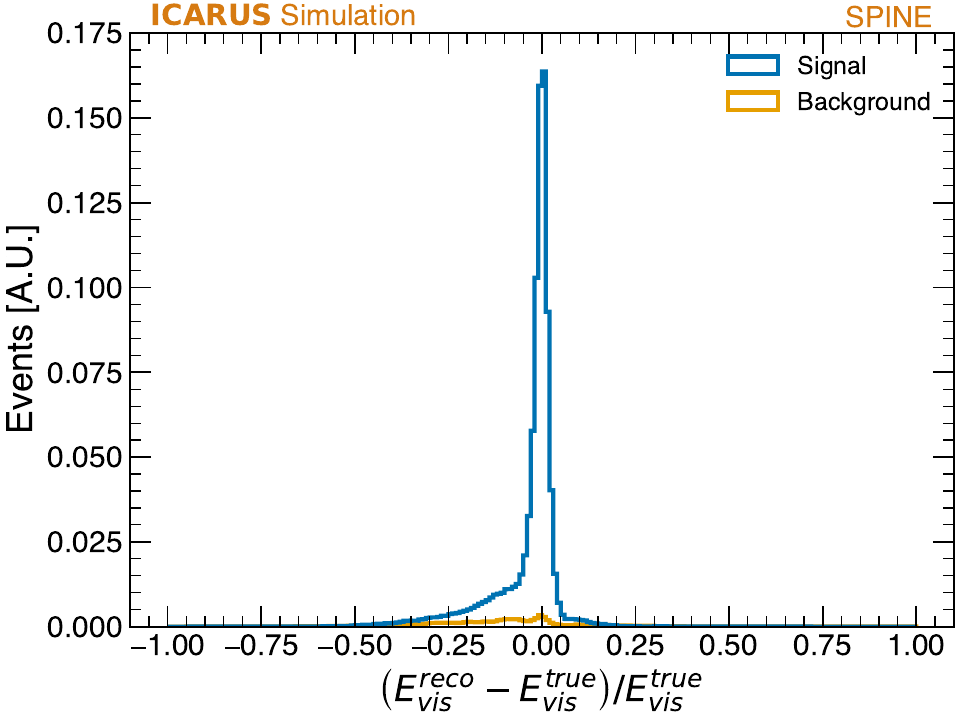}
    \caption{\justifying Energy resolution for the selected \sig sample, separately for true signal and neutrino background contributions. Visible true energy is chosen for both Pandora (top) and SPINE (bottom).}
    \label{fig:E_resolution}
\end{figure}

As shown in Fig.~\ref{fig:E_resolution}, the neutrino energy is well reconstructed for signal events and, as expected, the energy resolution is quite poor for neutrino events with a misreconstructed final-state topology. The ``low-side" tail of the resolution for signal events is primarily from misreconstruction of particle lengths and thus momenta. In Pandora this is primarily due to splitting of tracks, in which a single true track is reconstructed as two separate tracks, leading to underestimate of the muon track length. Protons that are reconstructed as a separate particle after a scattering event are also expected to contribute in both SPINE and Pandora.

\subsection{Selected Events}
\begin{figure}[!htb]
    \centering
    \includegraphics[width=\columnwidth{}]{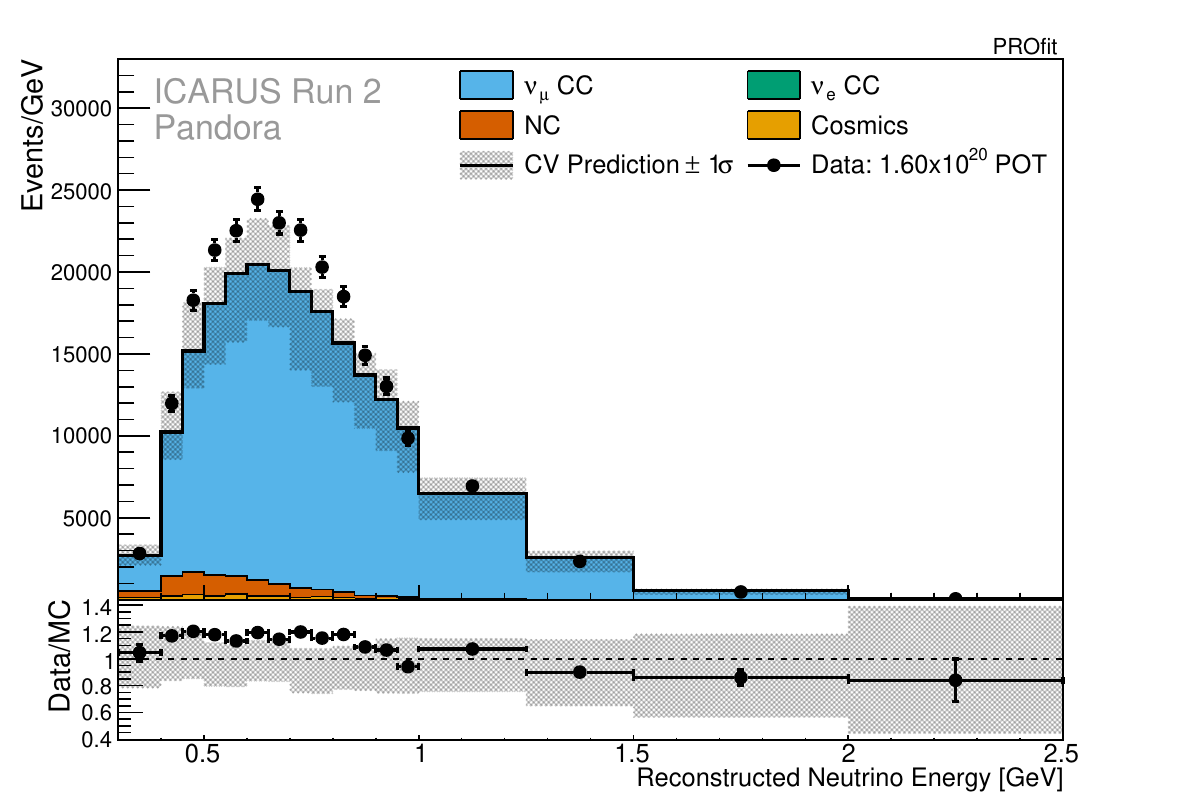}
    \includegraphics[width=\columnwidth{}]{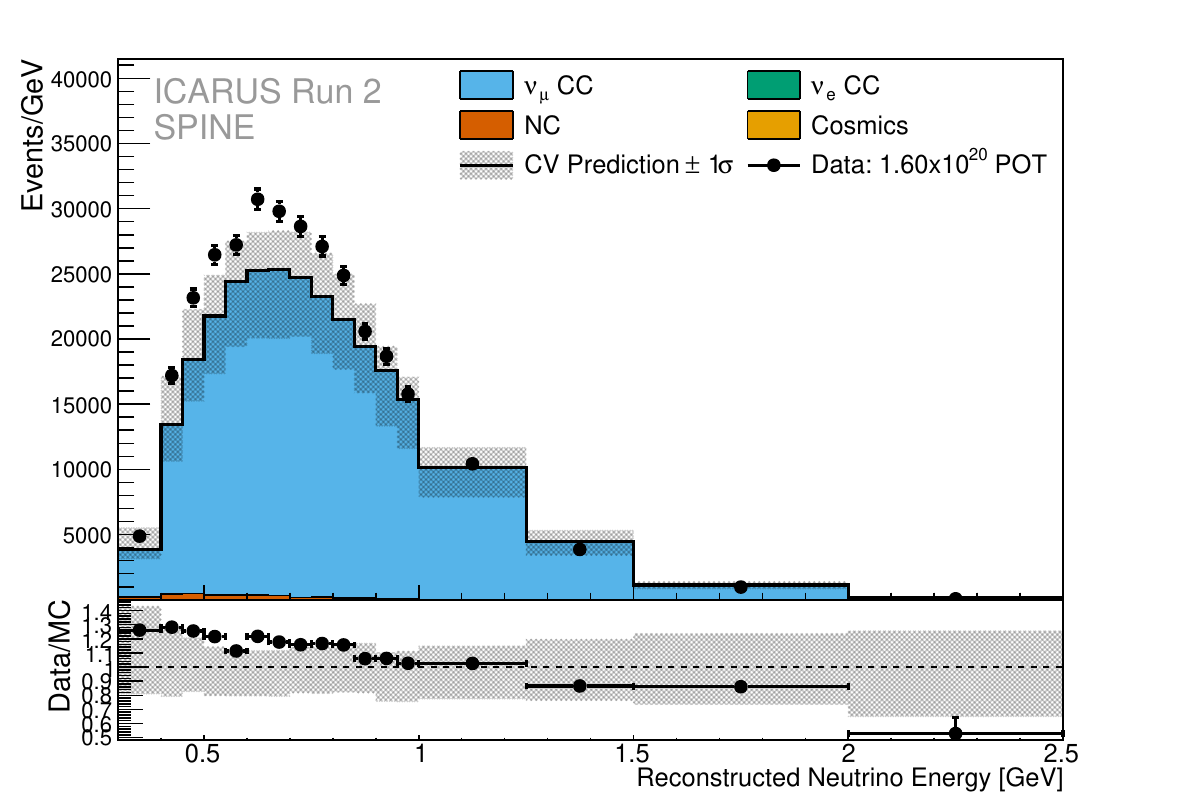}  
    \caption{\justifying Reconstructed neutrino energy distributions and data/MC ratios for selected events for Pandora (top) and SPINE (bottom).}
    \label{fig:selected_datamc_ereco}
\end{figure}

\begin{figure}[!htb]
    \centering
    \includegraphics[width=\columnwidth{}]{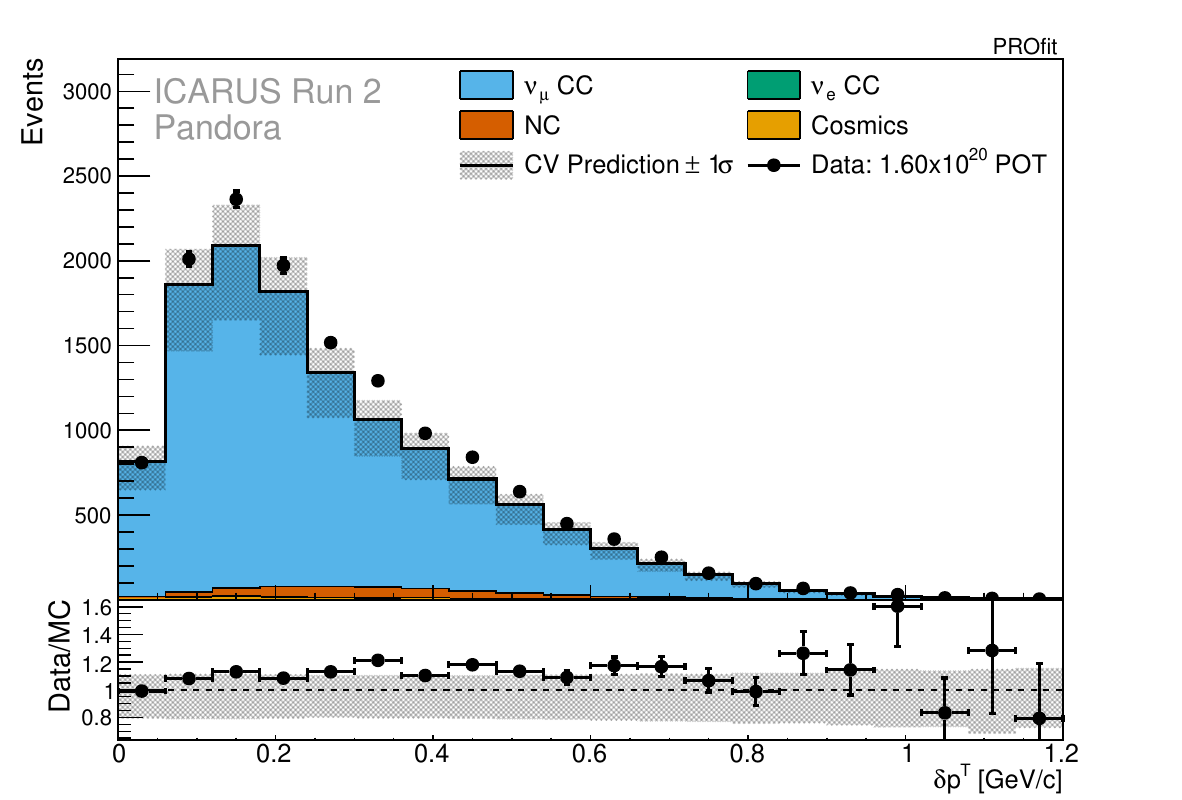}
    \includegraphics[width=\columnwidth{}]{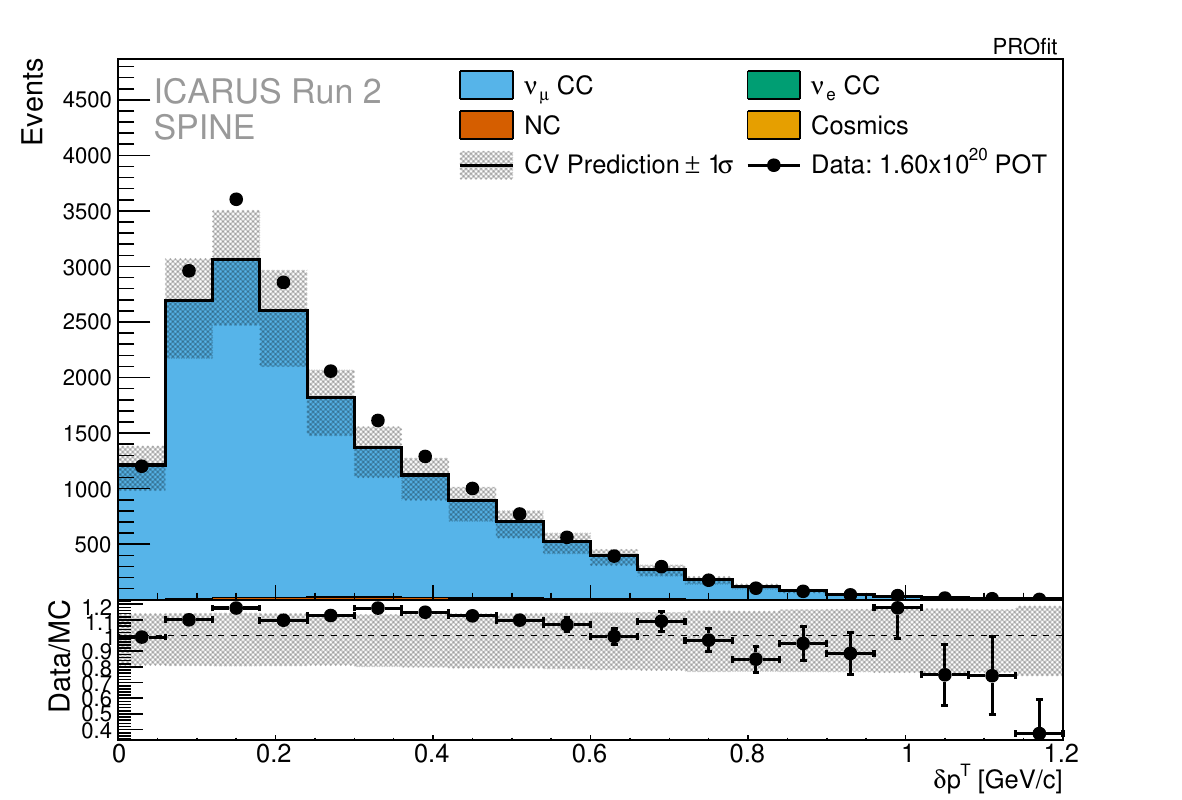}  
    \caption{\justifying Reconstructed $\delta p_T$ distributions and data/MC ratios for selected events for Pandora (top) and SPINE (bottom).}
    \label{fig:selected_datamc_dpt}
\end{figure}

\begin{figure}[!htb]
    \centering
    \includegraphics[width=\columnwidth{}]{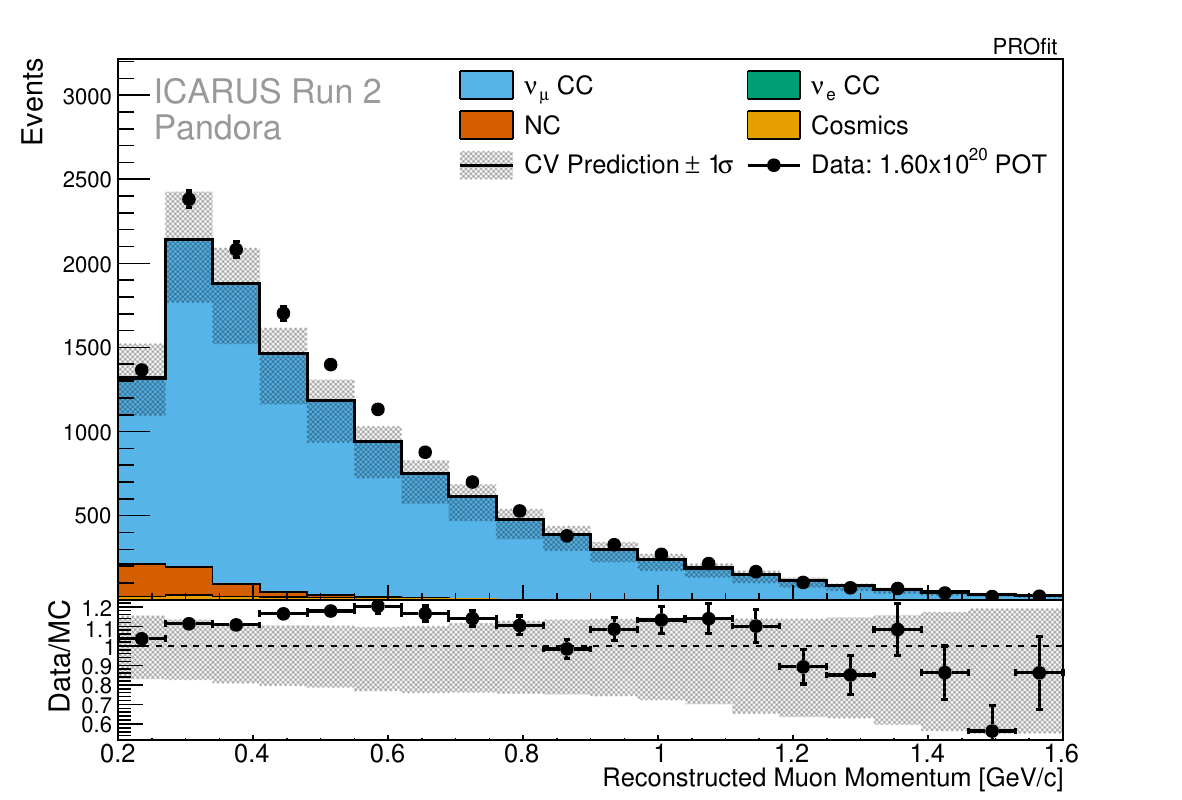}
    \includegraphics[width=\columnwidth{}]{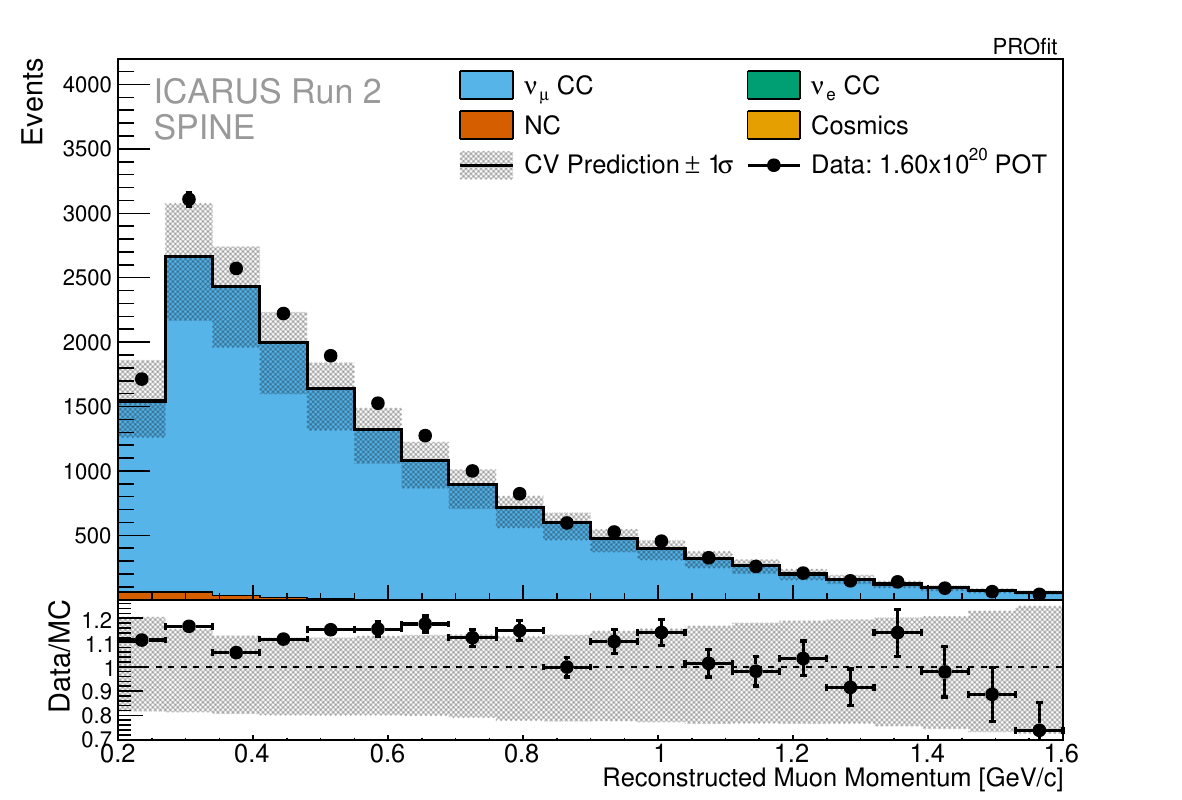}  
    \caption{\justifying Reconstructed muon momentum distributions and data/MC ratios for selected events for Pandora (top) and SPINE (bottom). }
    \label{fig:selected_datamc_pmu}
\end{figure}

\begin{figure}[!htb]
    \centering
    \includegraphics[width=\columnwidth{}]{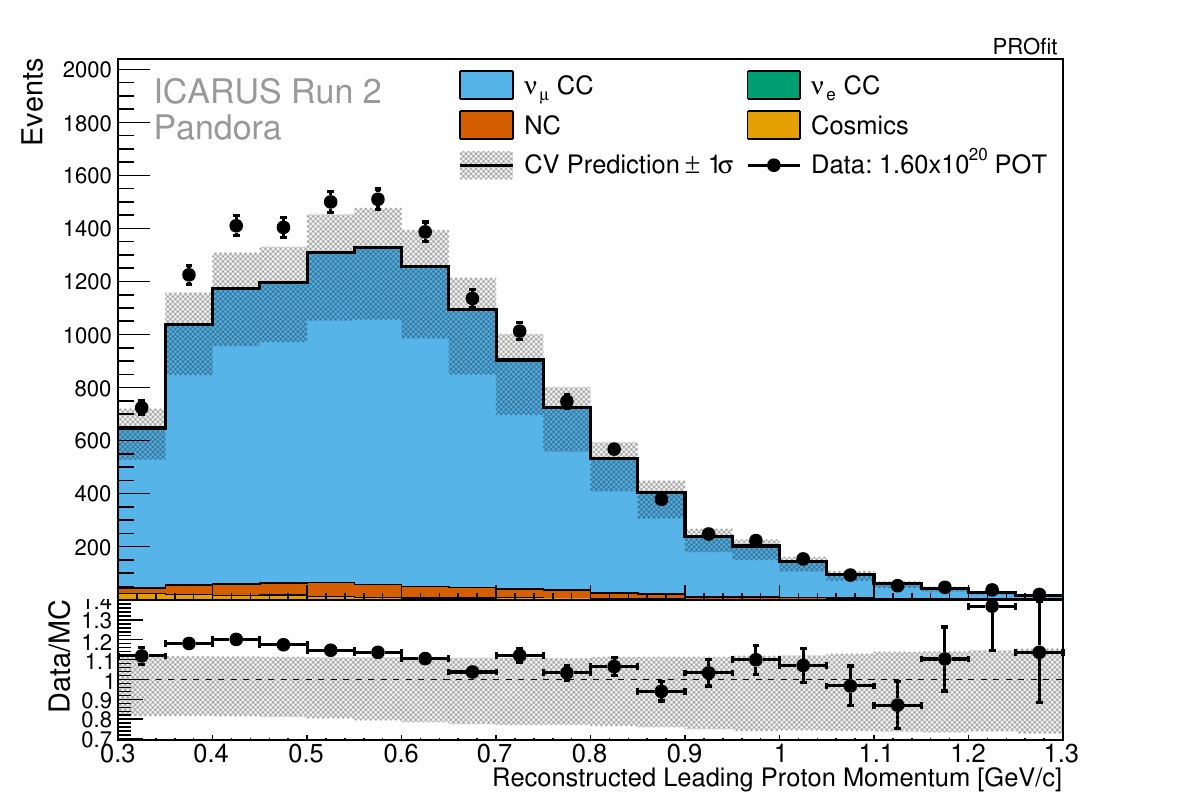}
    \includegraphics[width=\columnwidth{}]{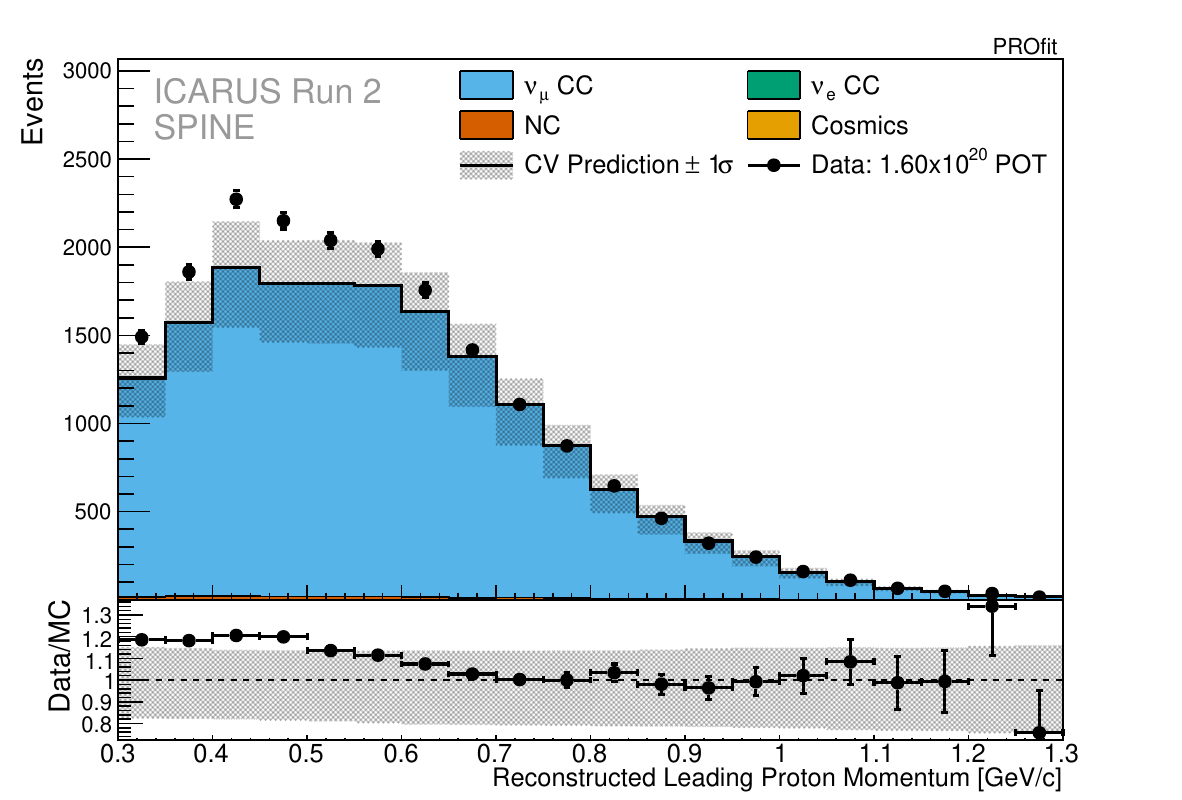}  
    \caption{\justifying Reconstructed leading proton momentum distributions and data/MC ratios for selected events for Pandora (top) and SPINE (bottom).}
    \label{fig:selected_datamc_pp}
\end{figure}

A total of 13,914 and 19,124 events are selected from Run 2 for the Pandora and SPINE analyses, respectively.
The reconstruction of physical quantities
shows good performance for signal events. 
Figure~\ref{fig:selected_datamc_ereco} shows the reconstructed neutrino energy distribution for selected events in data and MC, for the Pandora and SPINE analyses. Figures~\ref{fig:selected_datamc_dpt}, \ref{fig:selected_datamc_pmu}, and \ref{fig:selected_datamc_pp} show data-MC comparisons of the reconstructed transverse momentum difference between the muon and leading proton ($\delta p_{T}$~\cite{NuclEffIndepFromEnergy}), selected muon momentum, and leading proton momentum for both analyses. The error bars on the data points represent the statistical uncertainty and the shaded band represents the $1\sigma$ range of the central value MC prediction, based on the systematics uncertainties described in Section~\ref{sec:syst}. Small contributions to the prediction from misidentified $\nue$-CC, neutral current, and cosmic events are identified. As described in the previous section, the rate of selected cosmic events in-time with the beam spill is measured using data collected in off-beam triggers. The other background rates are predicted using MC truth information. In general, the data points are above the central value (CV) MC prediction; this tendency of the CV MC to underpredict the number of events is expected based on comparisons of the interaction model used in this analysis to other neutrino datasets~\cite{Filali:2024vpy,MicroBooNE:2025nll}. Most of this known data-MC discrepancy is covered by the systematic uncertainties included in the analyses and the shapes of the distributions are generally in good agreement with simulation.

\subsection{Validation and Pandora-SPINE Comparison}
A visual study campaign performed on data provides a rough evaluation of the selection purity and efficiency of both reconstruction paradigms and allows a comparison with the simulation-based efficiency and purity calculations described above.    
Selected events for a small sample of data ($\sim200$ events) are visually classified as signal or background, with purity defined as the ratio between the number of visually identified \sig candidates and the total number of selected events. The efficiency is defined as the fraction of events selected by each algorithm within a unbiased reference sample of visually selected events in the target sample. To evaluate the performance of the visual study, simulated events are also visually analyzed, and the visual classification is compared with the MC truth information. This study shows that the visual method slightly overestimates the purity, while providing a good prediction of the efficiency.

Using the visual scan method, efficiency and purity are found to be 51\% and 84\% for the Pandora-based selection, and 70\% and 94\% for the SPINE event selection. These values are in agreement with the efficiencies determined from MC, considering the small overestimation in purity and the reduced statistics available in the visual analysis. In this small subset of events, each selection accepts around half of the events identified by the other; the overlapping events are dominated by genuine \sig events. Pandora is observed to recover some good 1$\mu$Np candidates with short muon length not found by SPINE. However, Pandora is observed to have reduced identification efficiency relative to SPINE for events with low energy protons, and for tracks that are along the drift direction or parallel to the readout plane.

Additionally, data-data comparisons are made to check for any unexpected systematic effect as a function of detector position, operating conditions, and time within the run. Ratios of data with a vertex in the east vs. west cryostat and under various operating conditions such as whether or not the NuMI beam was operating are inspected. For cases where differences are expected, double ratios of SPINE to Pandora data are compared, for example, events with upstream vs. downstream vertices or with vertices near or far from the cathode. Event rate per POT as a function of run number is also inspected. No significant deviations from expectation are observed in any of these ratios.

\section{\label{sec:syst}Systematic Uncertainty}
Potential neutrino oscillations between the neutrino production point and the ICARUS detector are studied by comparing the observed spectrum at ICARUS to a prediction based primarily on MC simulation. Uncertainties in modeling the neutrino beam, neutrino interactions, and the detector lead directly to uncertainties in the predicted spectrum and thus the measured values of the oscillation parameters and so are included as systematic uncertainties in the result presented here. While the variations considered are the same for both the Pandora- and SPINE-based analyses, the impact on the predicted spectrum is expected to be different between two analyses as a result of differences in the reconstruction and selection efficiencies which lead to differing fractions of underlying event types in the selected samples. Details of the systematic uncertainties considered for the flux model, interaction model, and detector model are described in the following subsections.

\subsection{Flux Systematics}
\label{sect:FluxUncertainty}
\begin{figure}[!htb]
    \centering
    \includegraphics[width=\columnwidth{}]{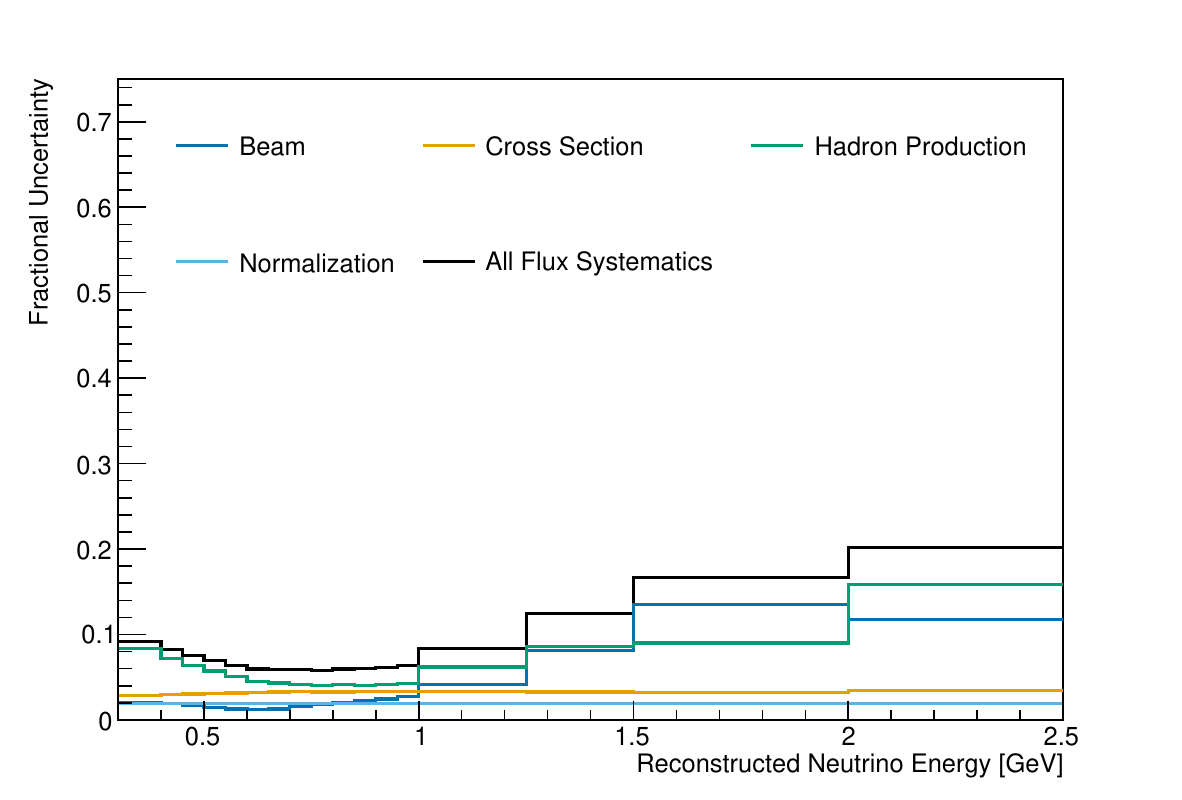}
    \includegraphics[width=\columnwidth{}]{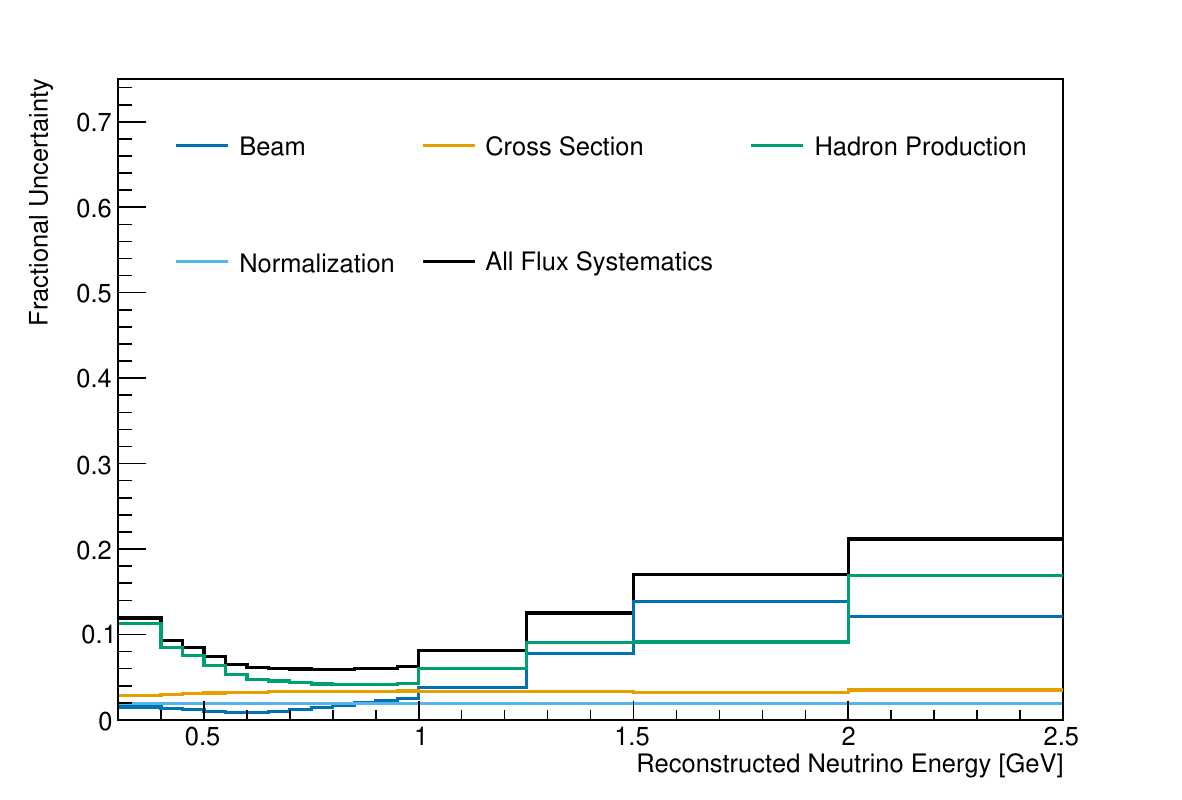}  
    \caption{\justifying Fractional size of the flux systematic uncertainty categories as a function of reconstructed neutrino energy for Pandora (top) and SPINE (bottom). The beam category includes the two horn current parameters and the cross-section category includes all the secondary interaction cross sections.}
    \label{fig:systfrac_flux}
\end{figure}

\begin{figure}[!htb]
    \centering
    \includegraphics[width=\columnwidth{}]{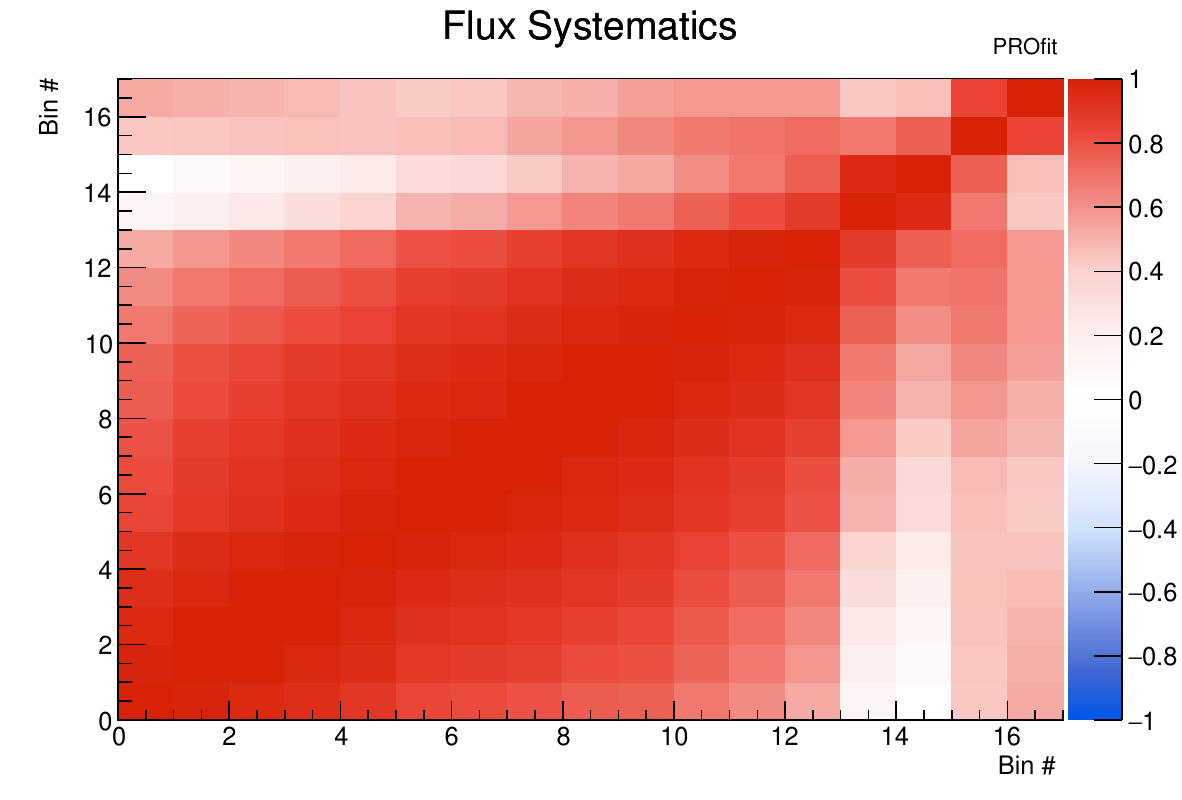}
    \includegraphics[width=\columnwidth{}]{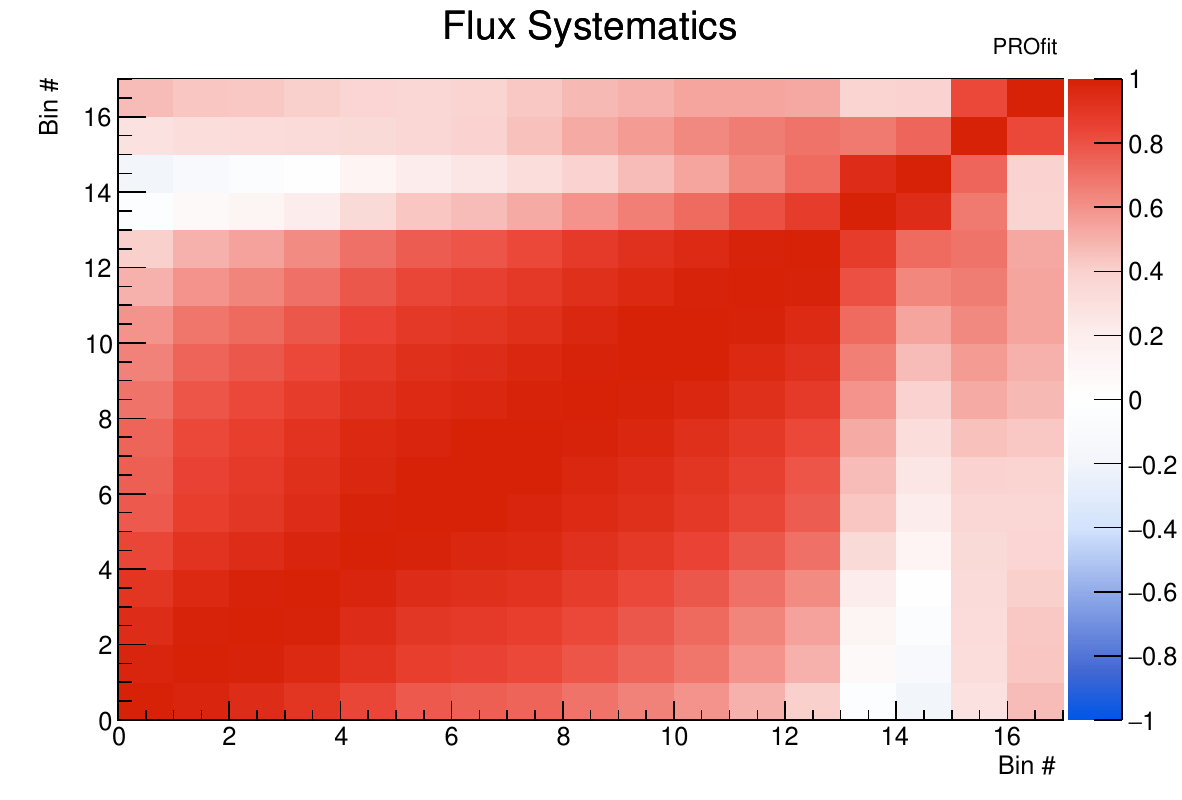}  
    \caption{\justifying Flux correlation matrix for Pandora (top) and SPINE (bottom). Axes are labeled with bin numbers using the fit binning in reconstructed neutrino energy.}
    \label{fig:syst_corr_flux}
\end{figure}

The neutrino flux in the Booster Neutrino beam has been studied extensively by previous experiments, including MiniBooNE and MicroBooNE. The ICARUS flux prediction and related uncertainties are based on those from MiniBooNE~\cite{MiniBooNE:2008hfu}, adopting some minor modifications from MicroBooNE~\cite{uboonefluxpubnote}. Uncertainties in the flux prediction arise from uncertainties in proton delivery, particle production, hadronic interactions, the horn magnetic field, and the beamline geometry. The dominant source of uncertainty for the $\nu_\mu$ flux is $\pi^+$ production, while both $\pi^+$ and $K^+$ production are important for the $\nu_e$ component. The systematics are applied within the fitting framework (see Section~\ref{sec:fit} and Appendix~\ref{app:profit}) as a covariance matrix generated with MC ``universes" in which the individual sources of flux uncertainty are varied randomly according to their prefit uncertainties. 
The parameters corresponding to these universes are varied independently in the fit. Table \ref{tab:fluxvariations} provides a description of each variation included in the flux uncertainty. Additionally, there is a 2\% uncertainty in the flux POT normalization, which is not included in the flux covariance matrix but rather treated as a separate flat normalization uncertainty in the fit.
\begin{table}[b]
\caption{\justifying \label{tab:fluxvariations}
Uncertainties on the flux. The three general categories are beam focusing uncertainties, uncertainties on hadronic secondary interactions in the target (where $\sigma^N$ are for secondary nucleon interactions and $\sigma^{\pi}$ are for secondary pion interactions), and uncertainties from hadron production. The Be and Al secondary interaction cross sections are varied together such that there are 6 total secondary interaction parameters in the fit.
}
\begin{ruledtabular}
\begin{tabular}{lc}
     Parameter & Prefit Unc.  \\ \hline
     Horn Current Depth & $<$18\% \\
     Horn Current & $\pm$6\% \\ \hline
     $\sigma^N_{inelastic}$ (Be, Al) & $\pm$5\%, $\pm$10\% \\
     $\sigma^N_{QE}$ (Be, Al) & $\pm$20\%, $\pm$45\% \\
     $\sigma^N_{Total}$ (Be, Al) & $\pm$15\%, $\pm$25\% \\
     $\sigma^{\pi}_{inelastic}$ (Be, Al) & $\pm$10\%, $\pm$20\% \\
     $\sigma^{\pi}_{QE}$ (Be, Al) & $\pm$11.2\%, $\pm$25.9\% \\
     $\sigma^{\pi}_{Total}$ (Be, Al) & $\pm$11.9\%, $\pm$28.7\% \\ \hline
     $\pi^+$ production (\numu, \numubar) & $\pm$11.7\%, $\pm$1.0\% \\
     $\pi^-$ production (\numu, \numubar) & 0\%, 11.6\% \\
     $K^+$ production (\numu, \numubar) & 0.2\%, 0.1\% \\
     $K^-$ production (\numu, \numubar) & 0\% 0.4\% \\
     $K^0$ production (\numu, \numubar) & 0\%, 0.3\% \\ \hline
     POT normalization& 2\% \\ 
\end{tabular}
\end{ruledtabular}
\end{table}
The fractional contributions to the uncertainty from each category of flux systematic are shown in Fig.~\ref{fig:systfrac_flux}. The flux correlation matrix is shown in Fig.~\ref{fig:syst_corr_flux}. In both the analyses,  lower energy events are highly correlated with
each other, but not with higher ($>1$~GeV) reconstructed
neutrino energy bins.

\subsection{Neutrino Interaction Systematics}
\label{sect:InteractionSystematics}
Neutrino interactions are simulated in the analysis presented here using GENIE AR23~\cite{Andreopoulos:2009rq, geniereleases}, with the central value MC using the nominal weights for all parameters. The model uses local Fermi gas (LFG) to describe the nuclear ground state (including a correlated high-momentum nucleon tail)~\cite{Hen:2016kwk,Weinstein:2010rt} with the Valencia~\cite{Nieves:2011pp} 1p1h (QE) and SuSAv2~\cite{Gonzalez-Jimenez:2014eqa, Megias:2016fjk,RuizSimo:2016rtu,RuizSimo:2016ikw} 2p2h meson exchange current (MEC) interaction models. The 1p1h model includes the use of a $z$-expansion form factor fit to deuterium data~\cite{Meyer:2016oeg}. 
The Berger-Sehgal~\cite{Berger:2007rq,Graczyk:2007bc,BergerSehgal:2009} model of neutrino-induced coherent pion production is used for single pion production (SPP) channels. The AGKY~\cite{Yang:2009zx} model covers multi-pion productions at low  hadronic invariant mass ($W<2.3$~GeV/c$^2$), including non-resonant backgrounds to the single-pion channels. The deep inelastic scattering (DIS) model uses the Bodek-Yang model~\cite{Bodek:2002ps} at higher $W$, with a linear AGKY to DIS transition at intermediate (2.3-3.0~GeV/c$^2$) $W$-values. The hA2018~\cite{Dytman:2011zz,Dytman:2021ohr} model is used for final-state interactions (FSI) and includes emission of deexcitation photons for argon nuclei. Some of the model parameters are tuned to free nucleon data \cite{GENIE:2021wox}.

Many systematic uncertainties and prior constraints on their variation are taken directly from GENIE as outlined in the GENIE Physics and User's Manual~\cite{geniemanual}. Additional uncertainties, not available in GENIE, are implemented using the NuSystematics package~\cite{nusystematics}. 
Reweights are calculated for each systematic, for each event, at a set of preset values: $\pm1,2,3\sigma$, and splines are used to interpolate systematic weight values from these node points. Systematics are implemented in the fitter either by using random throws of parameter values to construct a covariance matrix in bins of reconstructed neutrino energy or by associating each spline with an additional nuisance parameter with its own prior uncertainty, which are then constrained in the fit. Systematics which are expected to be significant, such that we are interested in understanding which values the fit prefers, and those which are expected to be non-Gaussian, are treated using the latter method. The remaining, less significant, systematics are included in a covariance matrix. Table \ref{tab:xsecvariations} provides a description of all the variations used in the analysis, including their prefit uncertainties and whether they are treated using splines or covariance in the fit. 

\begin{table*}
    \caption{List of interaction model dials used in the analysis. For each parameter, the central value weight, the nominal $\pm1\sigma$ uncertainties, and the implementation method within the fitter are noted. The $z$-expansion ``B" parameters represent a new basis after the principal component analysis (PCA) performed on the original A1~A4 parameters, which are highly correlated}
    \centering
\cmsTable{
    \begin{tabular}{llcccc}
\hline
Dial name & Short description & Central value & $+1\sigma$ & $-1\sigma$  & Implementation\\
\hline
\hline
ZExpB1CCQE & $B1$ Z-expansion parameter for axial-vector form factor on CCQE & \multicolumn{3}{l}{See caption} & spline \\
ZExpB2CCQE & $B2$ Z-expansion parameter for axial-vector form factor on CCQE & \multicolumn{3}{l}{See caption} & spline \\
ZExpB3CCQE & $B3$ Z-expansion parameter for axial-vector form factor on CCQE & \multicolumn{3}{l}{See caption} & spline \\
ZExpB4CCQE & $B4$ Z-expansion paramer for axial-vector form factor on CCQE & \multicolumn{3}{l}{See caption} & spline \\
RPA\_CCQE & RPA suppression is turned on (off) for dial=0 (1). & -- & -- & -- & spline \\
CoulombCCQE & Strength of EM potential for Coulomb corrections on CCQE & 1 & 20\% & 20\% & spline\\
\hline
NormCCMEC & Normalization of CC-MEC & 1 & 50\% & 50\% & spline  \\
NormNCMEC & Normalization of NC-MEC & 1 & 50\%  & 50\%  & covariance\\
XSecShape\_CCMEC & dial=1 for the reweight to Valencia 2p2h model & -- & -- & -- & spline \\
DecayAngMEC & dial=1 gives an alternative distribution proportional to $\cos{\theta}^{2}$ & -- & -- & -- & spline\\
FracPN\_CCMEC & CC MEC proton-neutron fraction & & 20\% & 20\% & spline \\
\hline
MaCCRES & Axial-vector mass of the dipole form factor on CCRes & 1.088962 & 20\% & 20\% & spline\\
MvCCRES & Vector mass of the dipole form factor on CCRes & 0.840 & 10\% & 10\% & spline \\
MaNCRES & Axial-vector mass of the dipole form factor on NCRes & 1.088962 & 20\% & 20\% & covariance\\
MvNCRES & Vector mass of the dipole form factor on NCRes & 0.840 & 10\% & 10\% & covariance \\ 
\hline
NonRESBGvpCC1pi & Scale for non-resonance background: $\nu$-p CC + 1$\pi$ & 1 & 50\% & 50\% & covariance\\
NonRESBGvpCC2pi & Scale for non-resonance background: $\nu$-p CC + 2$\pi$ & 1 & 50\% & 50\% & covariance\\
NonRESBGvpNC1pi & Scale for non-resonance background: $\nu$-p NC + 1$\pi$ & 1 & 50\% & 50\% & covariance\\
NonRESBGvpNC2pi & Scale for non-resonance background: $\nu$-p NC + 2$\pi$ & 1 & 50\% & 50\% & covariance\\
NonRESBGvnCC1pi & Scale for non-resonance background: $\nu$-n CC + 1$\pi$ & 1 & 50\% & 50\% & covariance\\
NonRESBGvnCC2pi & Scale for non-resonance background: $\nu$-n CC + 2$\pi$ & 1 & 50\% & 50\% & covariance\\
NonRESBGvnNC1pi & Scale for non-resonance background: $\nu$-n NC + 1$\pi$ & 1 & 50\% & 50\% & covariance\\
NonRESBGvnNC2pi & Scale for non-resonance background: $\nu$-n NC + 2$\pi$ & 1 & 50\% & 50\% & covariance\\
NonRESBGvbarpCC1pi & Scale for non-resonance background: $\bar{\nu}$-p CC + 1$\pi$ & 1 & 50\% & 50\% & covariance\\
NonRESBGvbarpCC2pi & Scale for non-resonance background: $\bar{\nu}$-p CC + 2$\pi$ & 1 & 50\% & 50\% & covariance\\
NonRESBGvbarpNC1pi & Scale for non-resonance background: $\bar{\nu}$-p NC + 1$\pi$ & 1 & 50\% & 50\% & covariance\\
NonRESBGvbarpNC2pi & Scale for non-resonance background: $\bar{\nu}$-p NC + 2$\pi$ & 1 & 50\% & 50\% & covariance\\
NonRESBGvbarnCC1pi & Scale for non-resonance background: $\bar{\nu}$-n CC + 1$\pi$ & 1 & 50\% & 50\% & covariance\\
NonRESBGvbarnCC2pi & Scale for non-resonance background: $\bar{\nu}$-n CC + 2$\pi$ & 1 & 50\% & 50\% & covariance\\
NonRESBGvbarnNC1pi & Scale for non-resonance background: $\bar{\nu}$-n NC + 1$\pi$ & 1 & 50\% & 50\% & covariance\\
NonRESBGvbarnNC2pi & Scale for non-resonance background: $\bar{\nu}$-n NC + 2$\pi$ & 1 & 50\% & 50\% & covariance\\
\hline
AhtBY & $A_{HT}$ higher twist parameter in BY model scaling $\xi_{W}$ & 0.538 & 25\% & 25\% & covariance\\
BhtBY & $B_{HT}$ higher twist parameter in BY model scaling $\xi_{W}$ & 0.305 & 25\% & 25\% & covariance\\
CV1uBY & $C_{\nu1u}$ $u$ valence GRV98 PDF correction parameter in BY model & 0.291 & 30\% & 30\% & covariance\\
CV2uBY & $C_{\nu2u}$ $u$ valence GRV98 PDF correction parameter in BY model & 0.189 & 40\% & 40\% & covariance\\
\hline
NormCCCOH & Normalization of CC-COH & 1 & $100\%$  & $100\%$ & covariance\\
NormNCCOH & Normalization of MC-COH & 1 & $100\%$ & $100\%$ & covariance\\
\hline
MFP\_pi & Scale for mean free path in the FSI of $\pi$ & 1 & 20\% & 20\% & covariance\\
FrCEx\_pi & Scale for fraction of charge-exchange fate in the FSI of $\pi$ & 1 & 50\% & 50\% & spline\\
FrInel\_pi & Scale for fraction of inelastic scattered fate in the FSI of $\pi$ & 1 & 40\% & 40\% & covariance \\
FrAbs\_pi & Scale for fraction of absorption fate in the FSI of $\pi$ & 1 & 30\% & 30\% & covariance\\
FrPiProd\_pi & Scale for fraction of pion production fate in the FSI of $\pi$ & 1 & 20\% & 20\% & covariance\\
MFP\_N & Scale for mean free path in the FSI of nucleon & 1 & 20\% & 20\% & covariance\\
FrCEx\_N & Scale for fraction of charge-exchange fate in the FSI of nucleon & 1 & 50\% & 50\% & spline\\
FrInel\_N & Scale for fraction of inelastic scattered fate in the FSI of nucleon & 1 & 40\% & 40\% & covariance\\
FrAbs\_N & Scale for fraction of absorption fate in the FSI of nucleon & 1 & 20\% & 20\% & covariance\\
FrPiProd\_N & Scale for fraction of pion production fate in the FSI of nucleon & 1 & 20\% & 20\% & covariance\\
\hline
MaNCEL & Axial-vector mass of the dipole form factor on NCEL & 0.994989 & 25\% & 25\% & covariance\\
EtaNCEL & Strange axial-vector mass of the dipole form factor on NCEL & 0.12 & 30\% & 30\% & covariance\\
\hline
\hline
    \end{tabular}
}
\label{tab:xsecvariations}
\end{table*}

The systematic uncertainties with the largest observed impact on the analysis are the random-phase-approximation (RPA) suppression of the quasi-elastic (QE) channel, the $z$-expansion parameters describing the QE axial form factor, the shape and normalization of charged-current meson exchanged current (CCMEC), the axial and vector masses describing resonance production, and FSI. These uncertainties impact the predicted spectra as follows: 
\begin{itemize}
    \item RPA QE adjusts the strength of the low-$Q^{2}$ suppression of the QE channel implemented by the RPA correction to the base model. This is the largest impact interaction systematic for reconstructed neutrino energy, $E_{\nu, \text{ reco}}<0.5$ GeV. 
    \item The $z$-expansion parameters control the QE form factor and are the second largest systematic for $E_{\nu \text{ reco}}<1$ GeV.
    \item CC MEC normalization adjusts the normalization of the 2p2h/MEC prediction. While the total contribution of the 2p2h events are relatively small compared to QE, the uncertainty on the process is large and this systematic absorbs the bulk of it. This is the largest impact interaction systematic $E_{\nu \text{ reco}}>0.5$ GeV and has been observed to be effective at covering unrelated variations that adjust the normalization at low to medium reconstructed energy.
    \item While the CC MEC normalization absorbs most of the uncertainty on the 2p2h process, there is also uncertainty on the shape of the energy-momentum ($q_0 - q_3$) distribution and the difference between the default SuSAv2 model and the alternate Valencia model is taken as an uncertainty. The result is additional uncertainty in the number of events in the  $0.5<E_{\nu \text{ reco}}<1$ GeV region.
    \item CCRES $M_A$ and $M_v$ are the axial and vector masses of the associated form factors in the resonance model used. These are the second largest interaction systematics for $E_{\nu \text{ reco}}>1$ GeV.
    \item FSI systematics change the kinematics of the hadronic system and can impact energy smearing and move protons and pion above/below the tracking threshold. FSI systematics have roughly the same impact as the CC MEC normalization systematic on the selected event sample, but impact the true-to-reconstructed energy smearing in important ways.
\end{itemize}

\begin{figure}[!htb]
    \centering
    \includegraphics[width=\columnwidth{}]{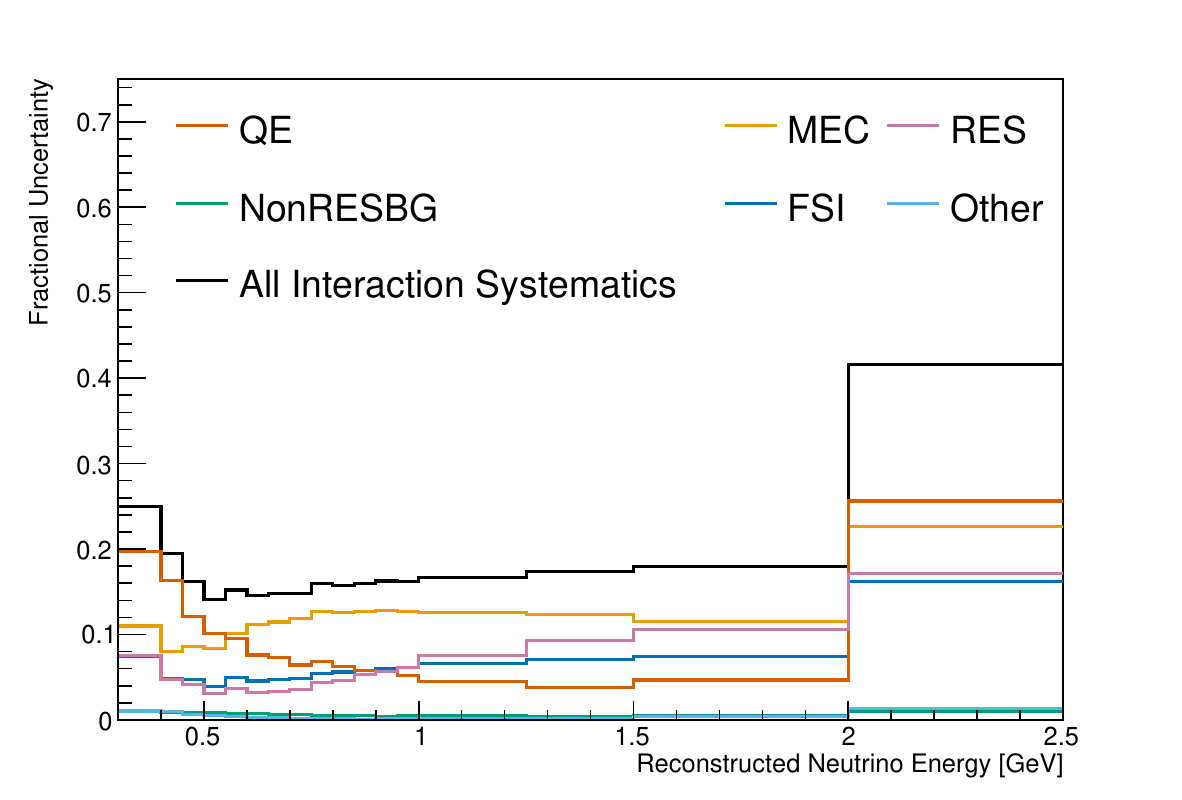}
    \includegraphics[width=\columnwidth{}]{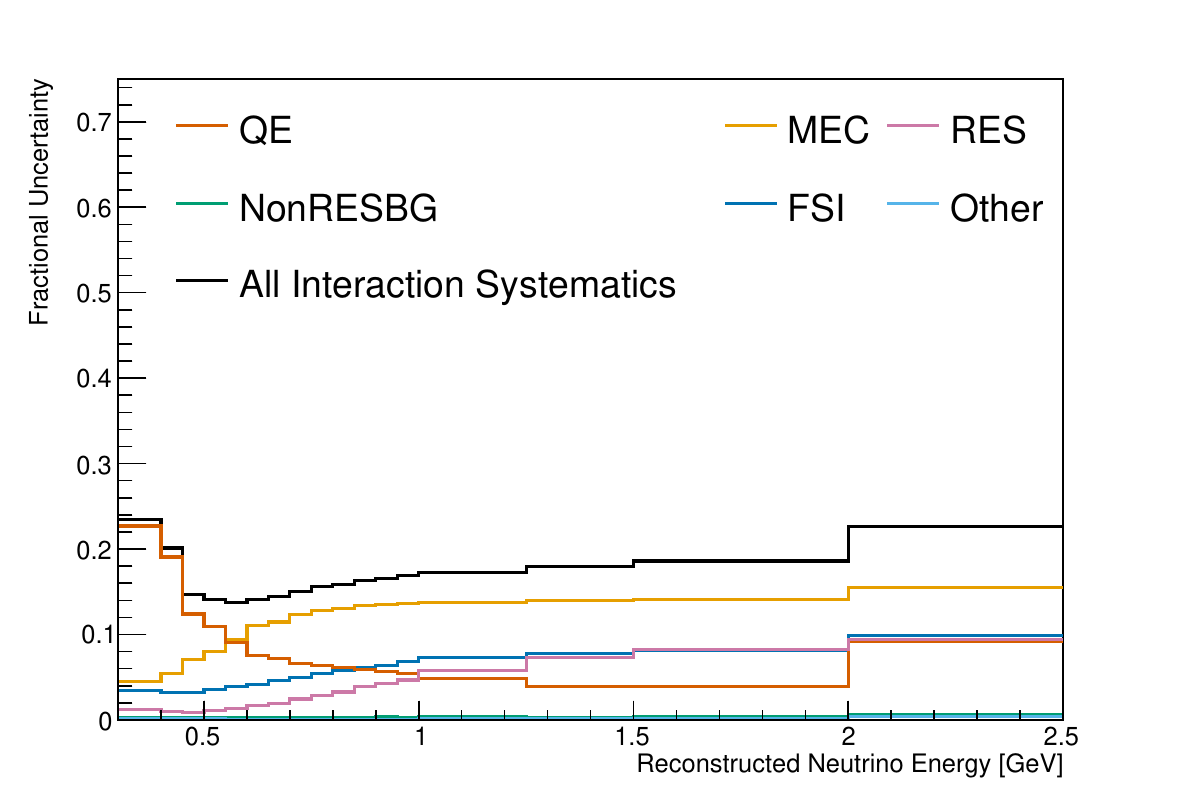}  
    \caption{\justifying Fractional size of interaction model systematic uncertainty categories as a function of reconstructed neutrino energy for Pandora (top) and SPINE (bottom).}
    \label{fig:systfrac_xsec}
\end{figure}

\begin{figure}[!htb]
    \centering
    \includegraphics[width=\columnwidth{}]{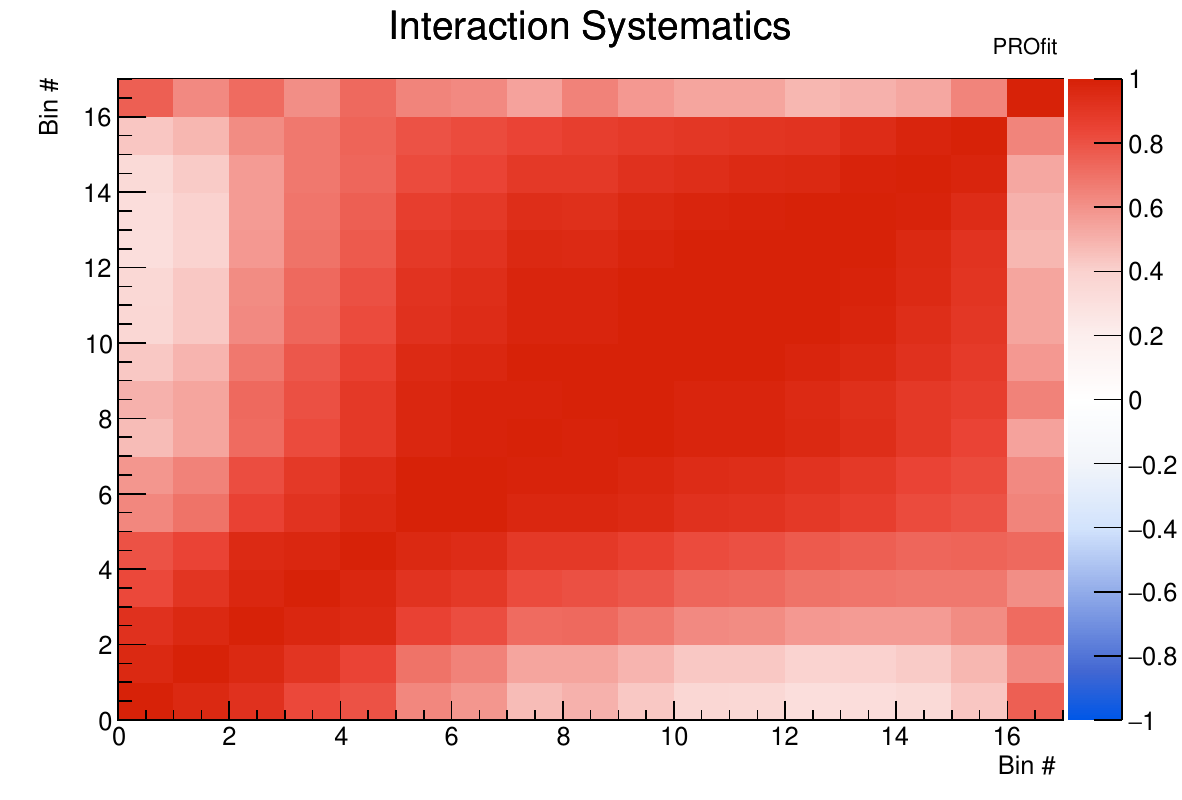}
    \includegraphics[width=\columnwidth{}]{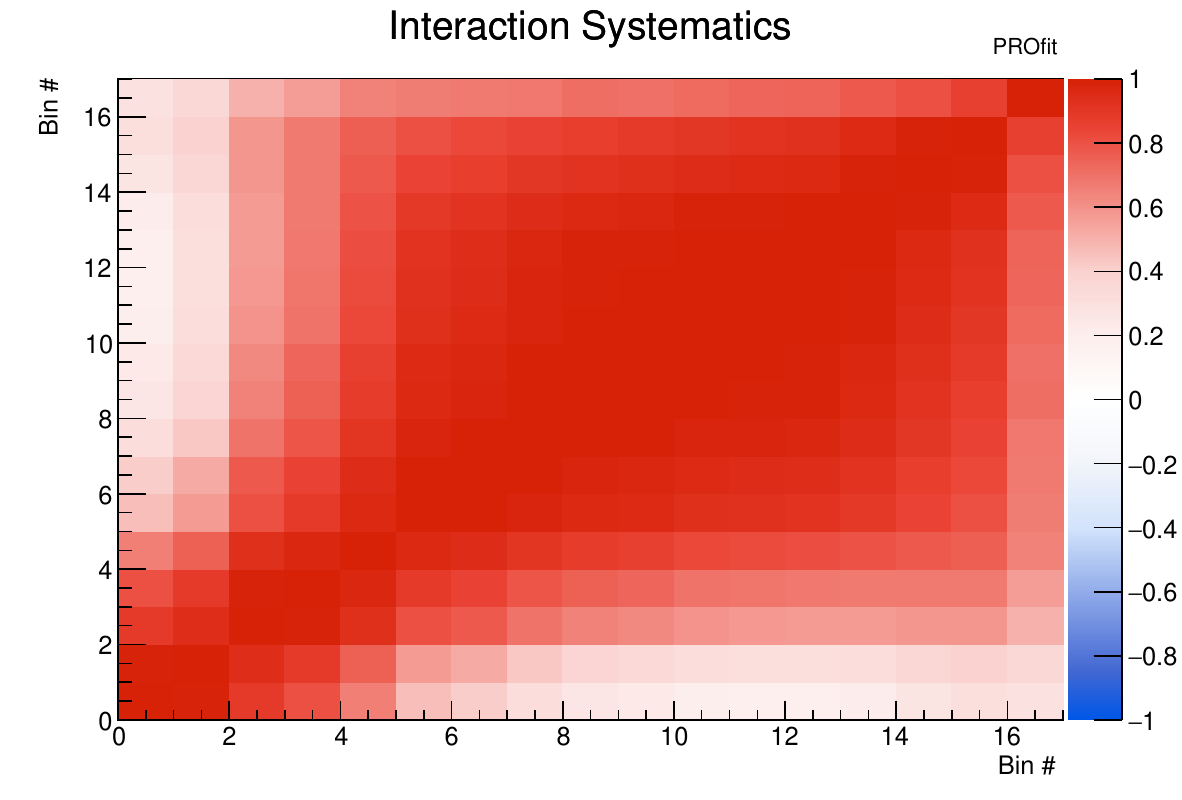}  
    \caption{\justifying Interaction model correlation matrix for Pandora (top) and SPINE (bottom). Axes are labeled with bin numbers using the fit binning in reconstructed neutrino energy.}
    \label{fig:syst_corr_xsec}
\end{figure}

The fractional contributions to the uncertainty from each category of interaction model systematic are shown in Fig.~\ref{fig:systfrac_xsec}. The largest contribution to the QE category is the RPA suppression parameter and the largest contribution to the MEC category is the CC-MEC normalization. The interaction model correlation matrix is shown in Fig.~\ref{fig:syst_corr_xsec}. In both analyses, there is a clear separation
in correlations, with reconstructed neutrino energy bins
$<0.8$~GeV being quite correlated with each other but
not with bins $>0.8$~GeV and vice versa.

\subsection{Detector Model Systematics}
\label{sect:DetectorSystematics}
Uncertainties arising from imperfections in the MC simulation of detector response are evaluated using alternative MC samples in which the detector model parameters are varied. The simulation samples used to evaluate detector modeling uncertainties include neutrino interactions uniformly distributed during the beam spill gate and cosmic rays overlapping out of time, as described in Section~\ref{sec:Simulation}. Each sample contains 200,000 neutrino events occurring within the ICARUS fiducial volume. The central value (CV) simulation represents the nominal detector model, while the variation samples reflect specific modifications to individual components of the detector model relative to this nominal configuration. To minimize statistical fluctuations between the CV and variation samples, an identical set of generator-level events is used for both the CV and the variation samples. Samples are analyzed using the same reconstruction and selection tools as in the nominal analysis and ratios of the resulting selected spectra are formed between the variation and CV samples. These binned ratios are used to produce spline functions used by the fitting framework, with the assumption that the weight will scale according to a Gaussian distribution (ie: the 2$\sigma$ ratio is twice the 1$\sigma$ ratio). Each spline is then associated with an additional nuisance parameter with its own prior uncertainty, which is constrained in the fit, as described above.

A number of potential detector variations were studied and ten of these were determined to be sufficiently impactful to include in the final analysis. The variations having the largest impact on the predicted spectra are:
\begin{itemize}
\item Intrinsic noise: the simulation does not include channel-to-channel noise variations which are measured to be of order 4\%. The level of intrinsic noise in the TPC readout is varied by $\pm$10\%, both to account for the impact of this unmodeled effect and to provide additional ad-hoc freedom in the detector model.
\item Shaping time: the shaping time in the TPC readout electronics is varied by $\pm$15\% to account for the impact of imperfections in simulation of the TPC field response. An additional variation of shaping time in the second induction plane readout electronics is included to cover a mis-modeling of this parameter.
\item Spatial non-uniformity: A detailed simulation of non-uniform detector response, believed to arise from modest incomplete transparency of the wire planes in the ICARUS detector, is implemented in this variation. In future analyses this simulation will be included in the CV MC; as this was not feasible for the analysis presented here, the impact is included as a systematic variation. 
\item Recombination: There is an angular dependence to the recombination model that is best described using an ellipsoidal description of the $\beta$ parameter in the Modified Box Model that depends on the angle between the track direction and the drift field~\cite{ICARUS:2024woh}. In this variation, the ellipsoidal modified box model replaces the original Modified Box Model used in the CV MC.
\item Free Electron Lifetime: The CV MC is generated with a nominal electron lifetime of 3 ms, while the measured electron lifetimes in ICARUS during Run 2 are 4~ms and 8~ms for the east and west cryostats, respectively. In this variation, the electron lifetime is set to 8~ms.
\end{itemize}

\begin{figure}[!htb]
    \centering
    \includegraphics[width=\columnwidth{}]{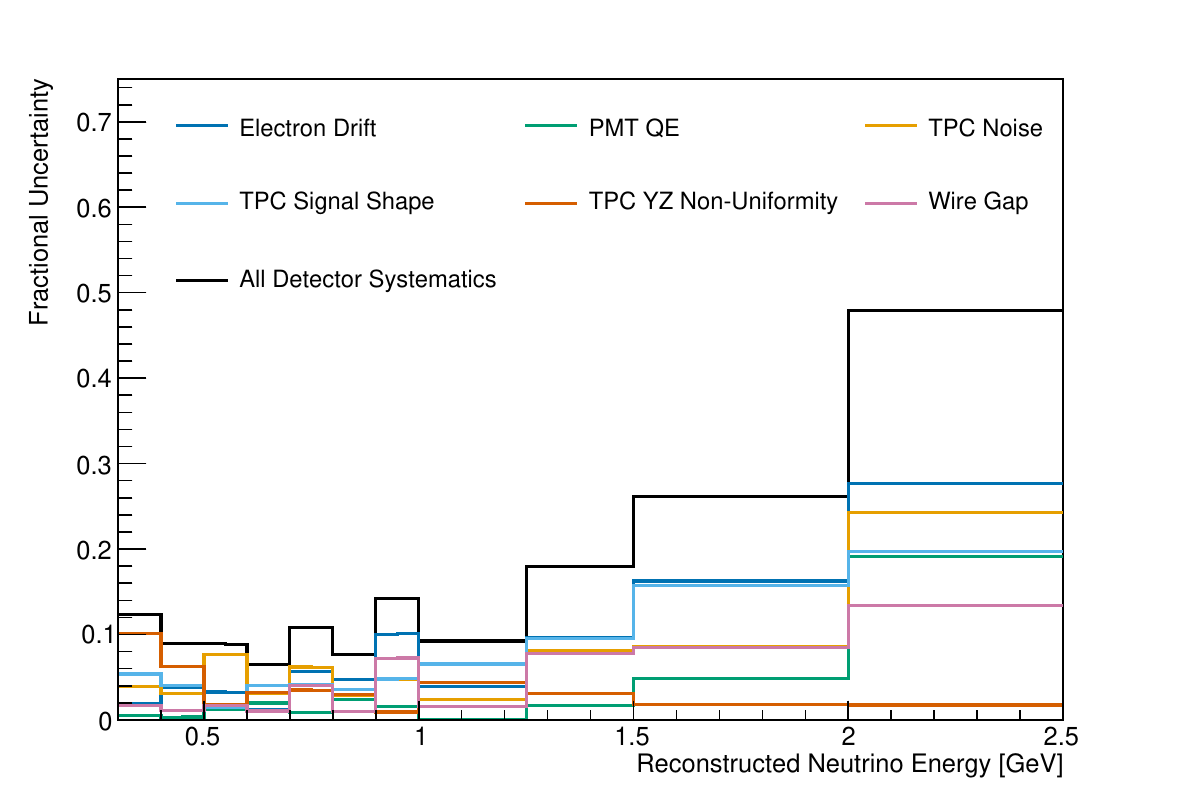}
    \includegraphics[width=\columnwidth{}]{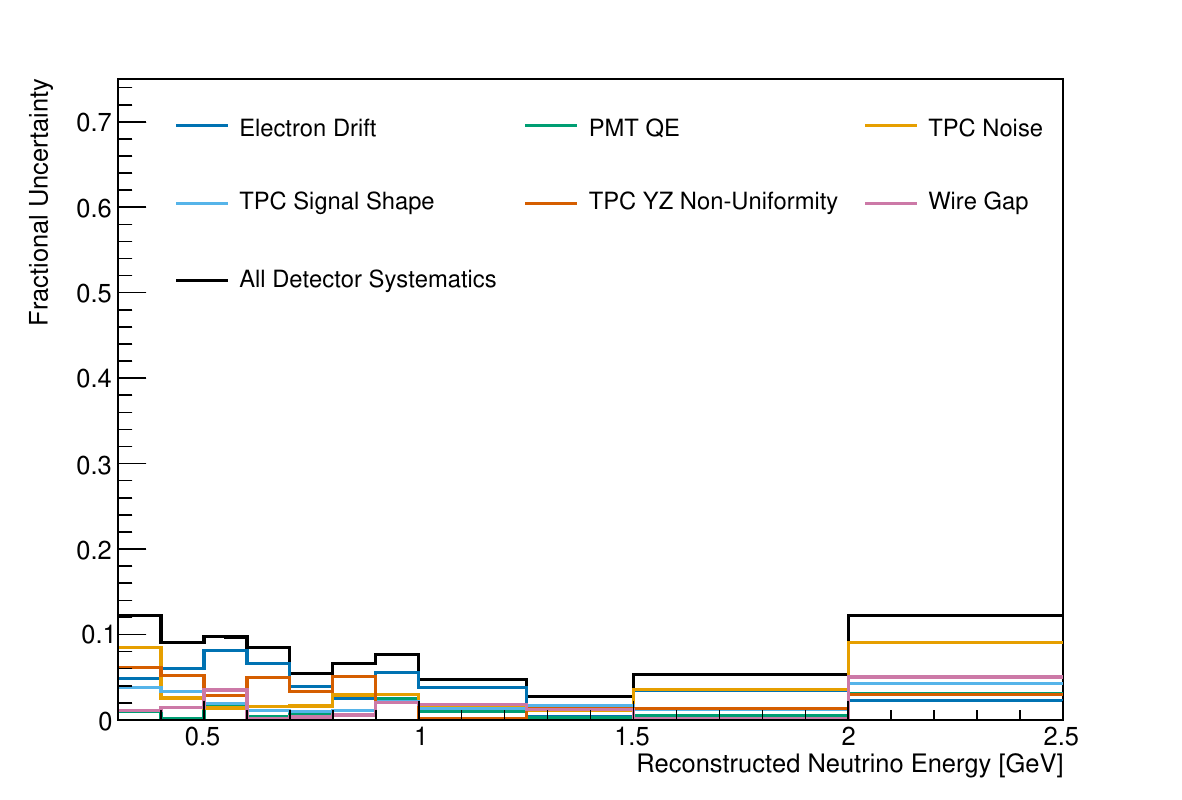}  
    \caption{\justifying Fractional size of detector model systematic uncertainty categories as a function of reconstructed neutrino energy for Pandora (top) and SPINE (bottom). The electron drift category includes electron lifetime and recombination.}
    \label{fig:systfrac_det}
\end{figure}

\begin{figure}[!htb]
    \centering
    \includegraphics[width=\columnwidth{}]{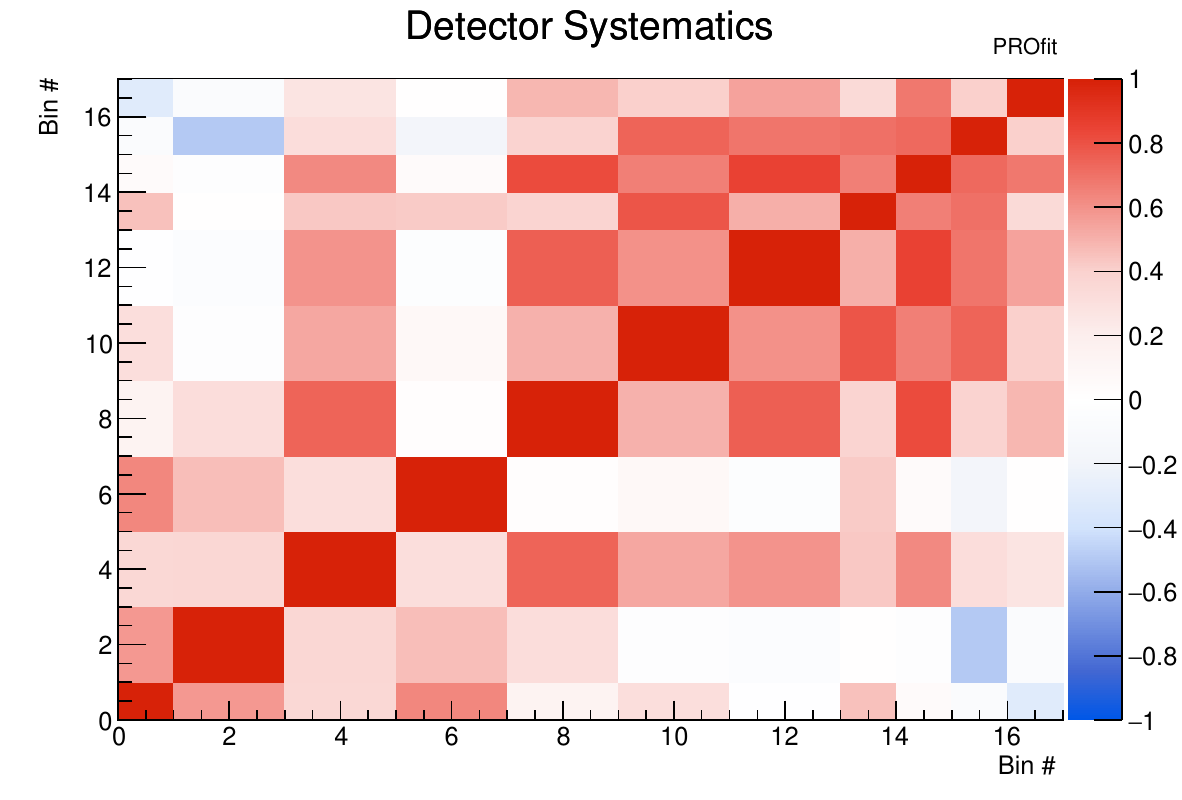}
    \includegraphics[width=\columnwidth{}]{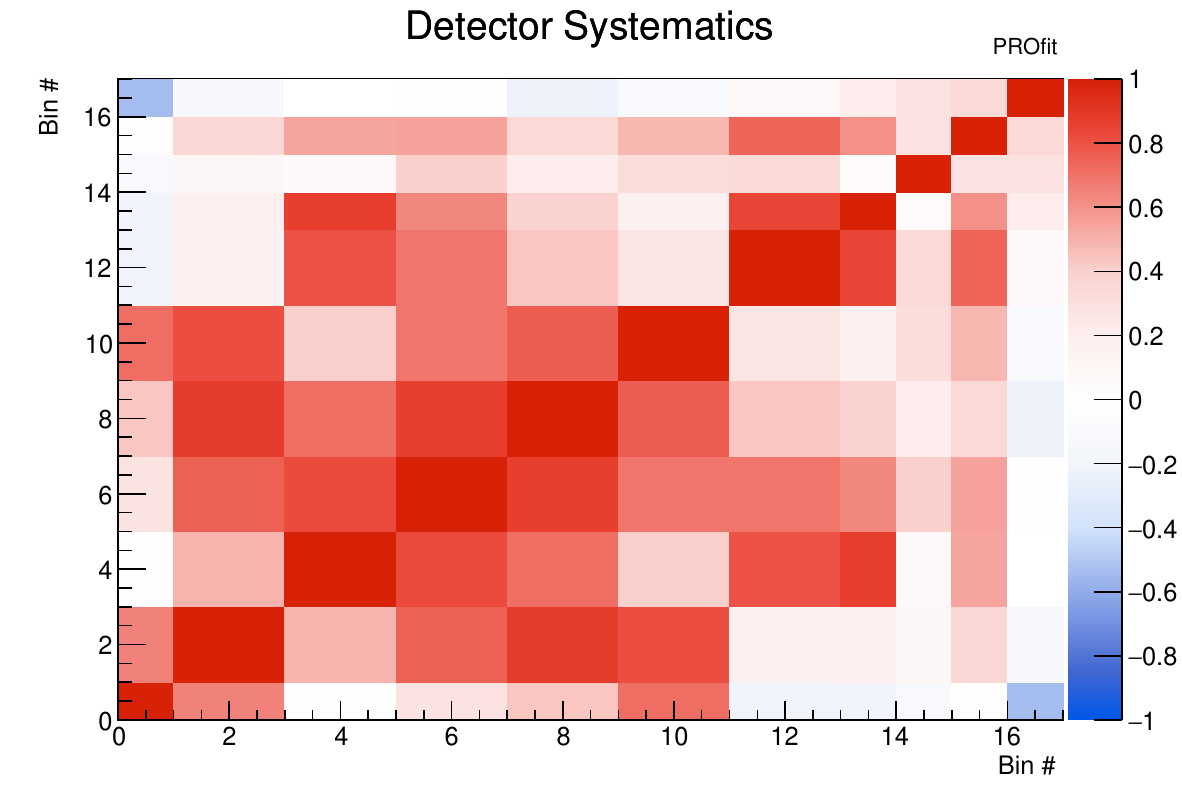}  
    \caption{\justifying Detector model correlation matrix for Pandora (top) and SPINE (bottom). Axes are labeled with bin numbers using the fit binning in reconstructed neutrino energy.}
    \label{fig:syst_corr_det}
\end{figure}

Other effects considered include variations in TPC coherent noise, variations in scintillation light yield, the impact of an unmodeled gap in the induction 1 wire plane, the impact of unmodeled cathode non-planarity, uncertainty in trigger efficiency, and variations of fiducial volume and containment cuts.
The fractional contributions to the uncertainty from each category of detector systematic are shown in Fig.~\ref{fig:systfrac_det}. The detector systematic correlation matrix is shown in Fig.~\ref{fig:syst_corr_det}.

\subsection{Systematics Validation}
To verify that the systematic variations included in the analysis are sufficient to describe the true uncertainty in the predicted spectra, data-MC comparisons are performed on detector performance quantities that are impacted by the detector model and fitter-level studies are performed on ``mock data" generated under alternative model assumptions. In the case of low-level detector quantities, which are common between the SPINE and Pandora analyses, the range of variations included as systematic uncertainties is compared to discrepancies between data and the central value MC. For the mock data studies, an evaluation is made as to whether systematic variations are sufficient to provide reasonable fits, such that none of the systematic parameters are pulled in unphysical or dramatic ways and no false oscillation signals are observed.

\begin{figure}[!htb]
    \centering
    \includegraphics[width=0.8\columnwidth{}]{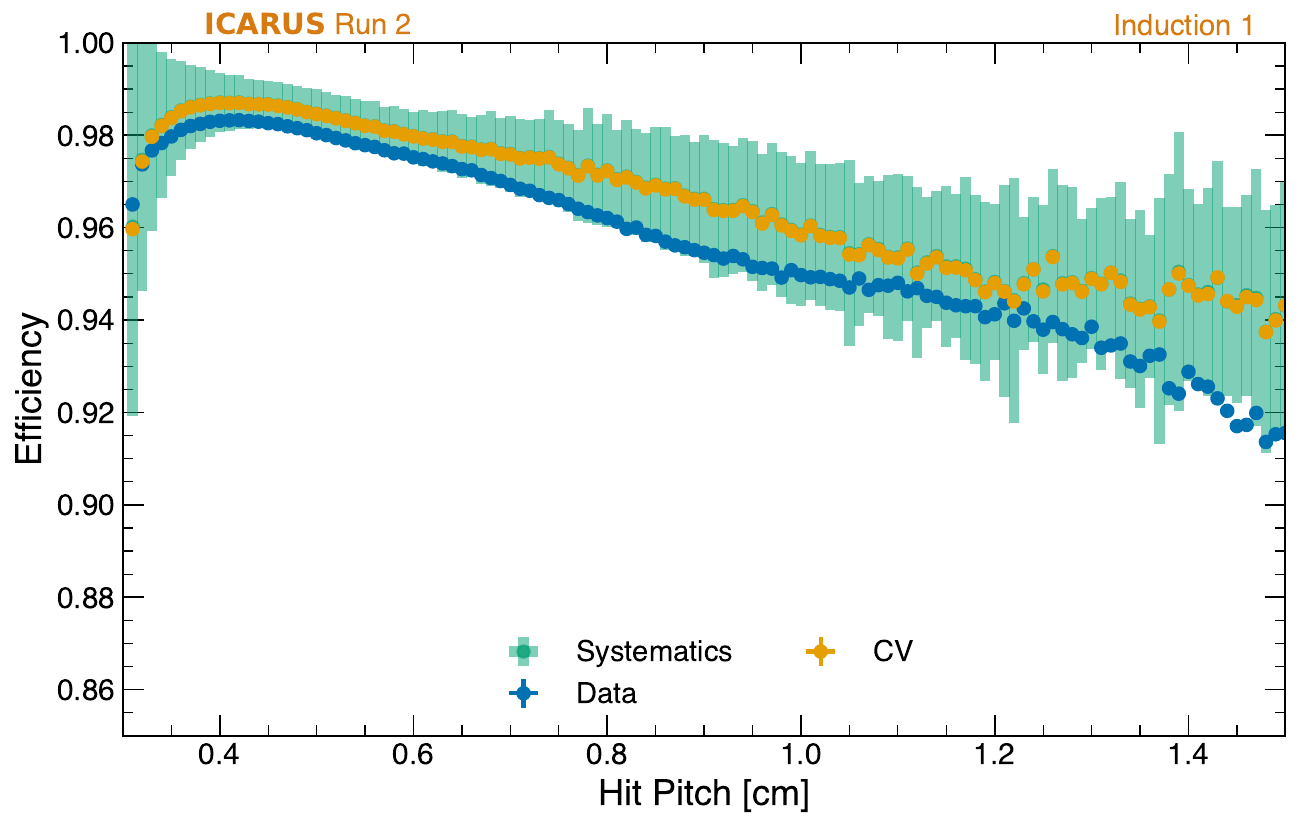}
    \includegraphics[width=0.8\columnwidth{}]{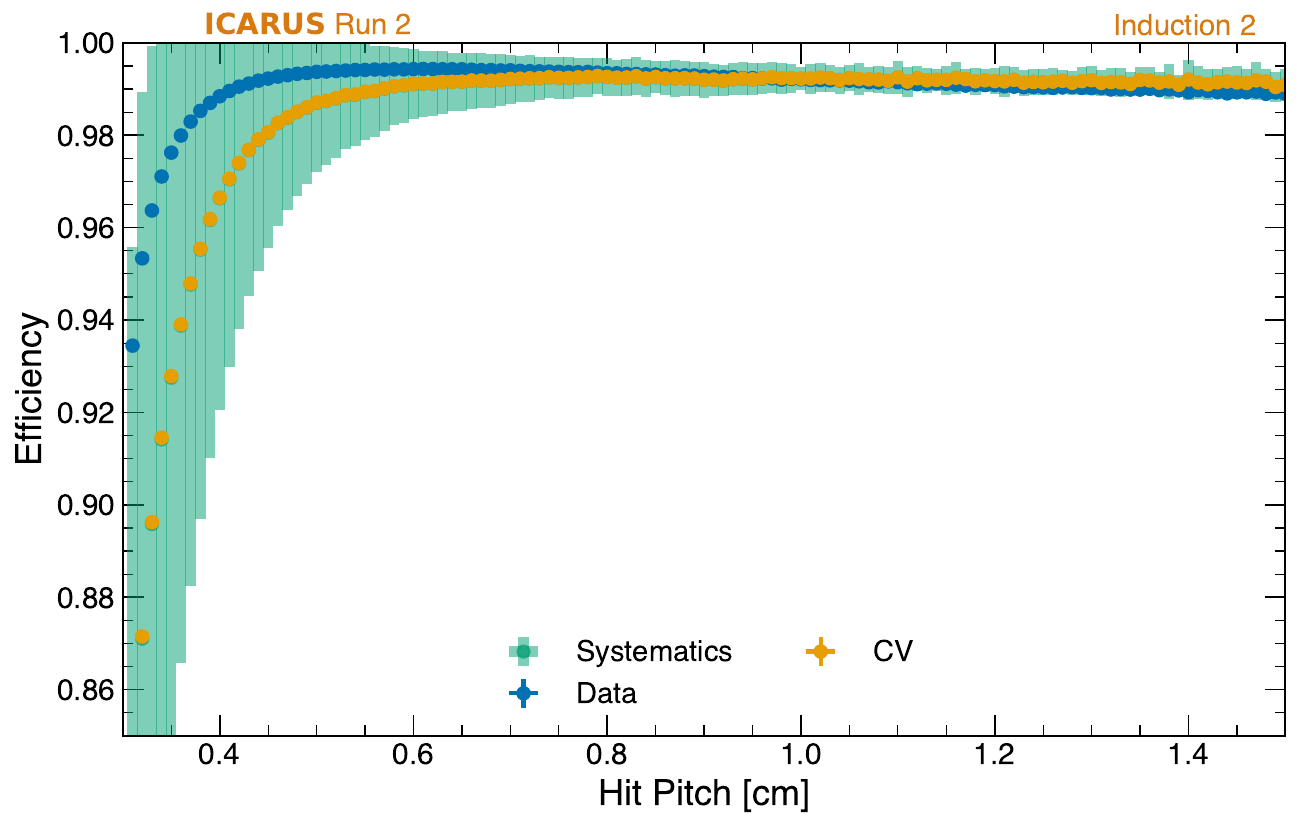}  
    \caption{\justifying Global results from hit identification efficiency studies for the Induction 1 (top) and Induction 2 (bottom) wire planes: the green band represents the hit efficiency coverage obtained by combining the detector systematic variations, to be compared with the measurements from data in blue. }
    \label{fig:hitseff}
\end{figure}

The efficacy of the detector variation systematics model is studied by exploiting two different low-level reconstruction variables: estimation of the wire signal (hit) identification efficiency and the charge/energy deposition ($dQ/dx$ and $dE/dx$) distributions. Cosmic muon tracks crossing the cathode are considered for both studies and the analysis is performed separating the three different wire planes, to take into account their specific characteristics. The hit efficiency is evaluated as a function of the average hit pitch, which is defined as the length of the 3D track segment seen by a single wire. The green bands in Fig.~\ref{fig:hitseff} indicate the overall MC hit efficiency for both induction planes, and are obtained by combining all the detector systematics variations included in the analysis. Data hit efficiency is overlaid, demonstrating that the detector systematics model provides the desired coverage.
The desired coverage is also observed in the collection plane, not pictured, where the hit efficiency exceeds 99$\%$ for both data and MC for almost all the studied hit pitch values.

\begin{figure}[!htb]
    \centering
    \includegraphics[width=0.9\columnwidth{}]{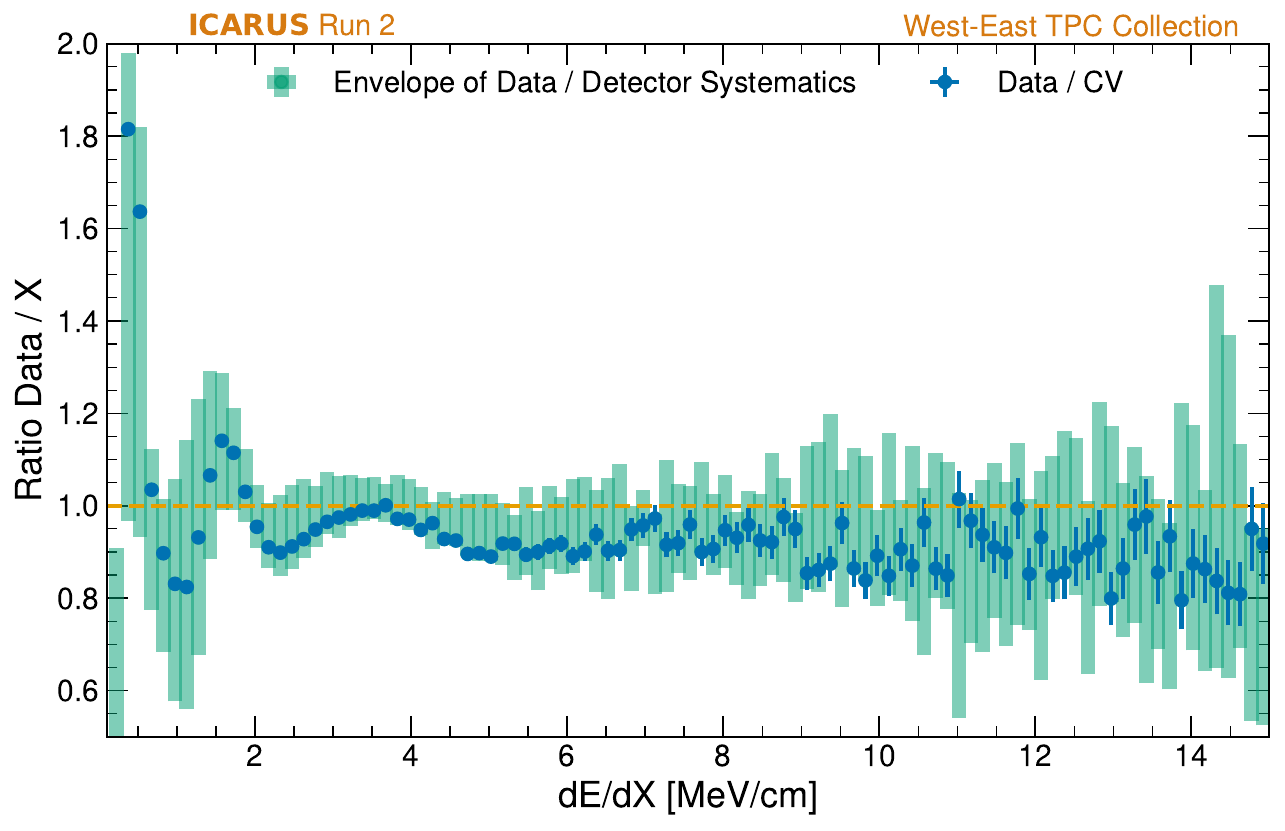}
    \caption{\justifying The envelope of the PDF ratio of data dE/dx distributions over CV and variations,considering the final sub-set of detector systematics, is compared with the measurements performed in the WEST-EAST TPC collection plane.}
    \label{fig:dEdx}
\end{figure}

In a similar way, to evaluate whether the shapes of $dE/dx$ (for the collection wire plane only) and $dQ/dx$ (for all the wire planes) are properly accounted for by the detector systematics, the probability density function (PDF) of the dQ/dx and dE/dx distributions are obtained. A comparison is made between the central value MC and the different systematics detector variations; in particular, an envelope is constructed by comparing all the PDF ratios of data/MC across the different TPCs and planes, in a bin-by-bin basis. An example is shown in Fig.~\ref{fig:dEdx} where the desired data/MC agreement (dashed orange line) is well contained within the green envelope across the entire dE/dx range for the WE TPC collection.

Uncertainty in the simulation of particle interactions within the detector volume could lead to additional uncertainty in the predicted spectra and are not explicitly included in the systematic variations described here. To verify that the impact of such variations is covered by our systematics model, a mock data study is performed in which the contribution of the impacted samples is artificially varied. Specifically, this analysis targets final states with zero pions, but truth-level studies reveal 5-10\% of selected MC events have true pions with initial energies above threshold in the final state, most of which are likely moved below threshold by reinteractions in the detector material. To study the impact of uncertainty in the reinteraction modeling, mock data is generated with the CCN$\pi$ component increased(decreased) by 100\% and the NCN$\pi$ component decreased(increased) by 50\%, as the two are expected to be anti-correlated. These represent extreme variations that are much larger than the actual uncertainty. Fits to this mock data do not result in any dramatic shifts of systematic parameters or any false oscillation signals, so we conclude that variations in particle interaction modeling are negligible for this analysis. 

Additional mock data studies are performed to confirm that the systematic variations included in the analysis provide sufficient coverage of potential alternatives to the models used to generate the prediction. One method of generating mock data is to use alternative interaction models that are not used in the central value MC; for example, alternative models of final-state interactions (FSI), quasielastic (QE) interactions, and meson exchange current (MEC) interactions are used to generate mock data, which is then fit using the central value MC for the prediction. The alternative QE model is based on a fit of the $z$-expansion CCQE axial form factor to lattice QCD and the alternative MEC model adjusts the angular distribution of outgoing nucleons in the MEC interaction. An alternative method of generating mock data is to reweight the prediction from GENIE, in leading proton kinematic variables, to match the prediction from alternative generators NEUT~\cite{Hayato:2021heg}, NuWro~\cite{Golan:2012wx}, and GiBUU~\cite{Buss:2011mx}.
In all of these mock data studies, no false oscillation signal is observed and the systematic pulls are not dramatic. Frequently the CC MEC normalization and QE RPA parameters are pulled to compensate for a normalization effect in the mock data. As described in Section~\ref{sec:results}, this effect is also seen in the full data fits.
Finally,  combinations of systematics parameters can be purposefully constructed to generate mock data that is qualitatively similar to an oscillation signal. Even in these  artificial worst-case scenarios, the $\Delta\chisq$ between fits with oscillation parameters floating and fixed to zero do not indicate the presence of a statistically significant oscillation.

\subsection{Systematics Summary}

\begin{figure}[!htb]
    \centering
    \includegraphics[width=\columnwidth{}]{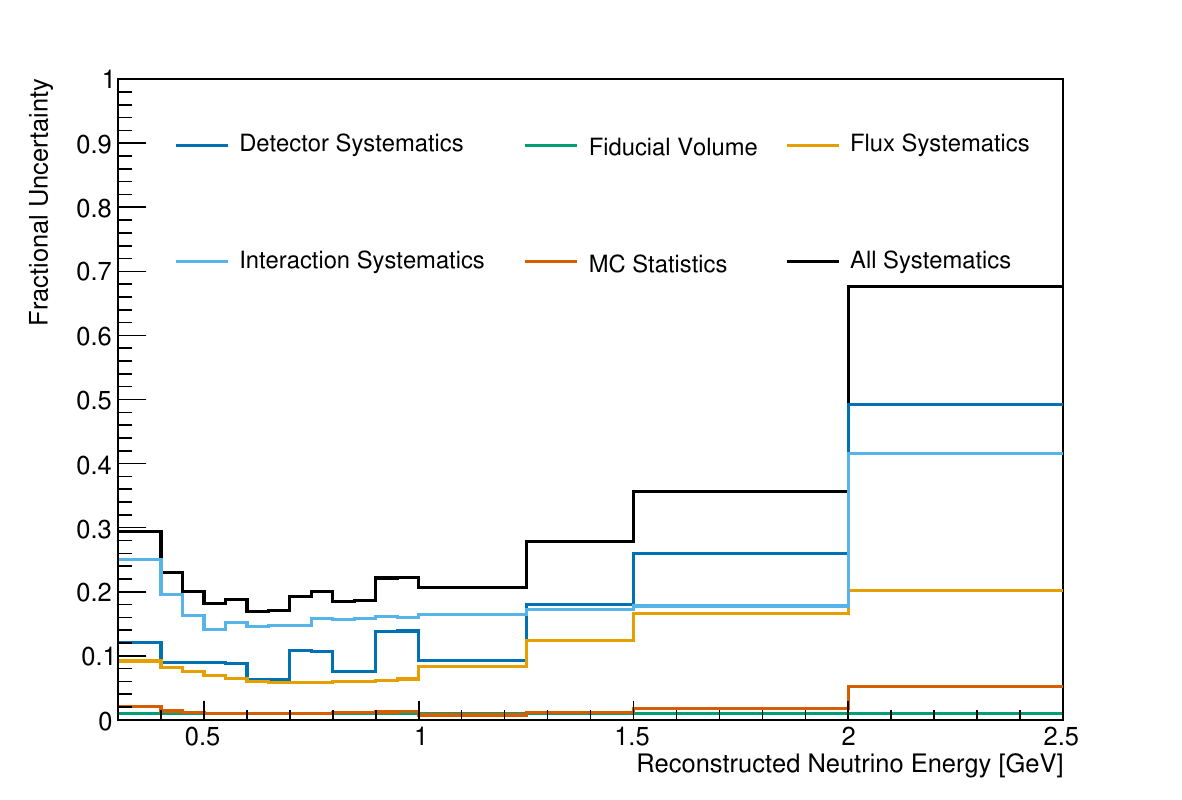}
    \includegraphics[width=\columnwidth{}]{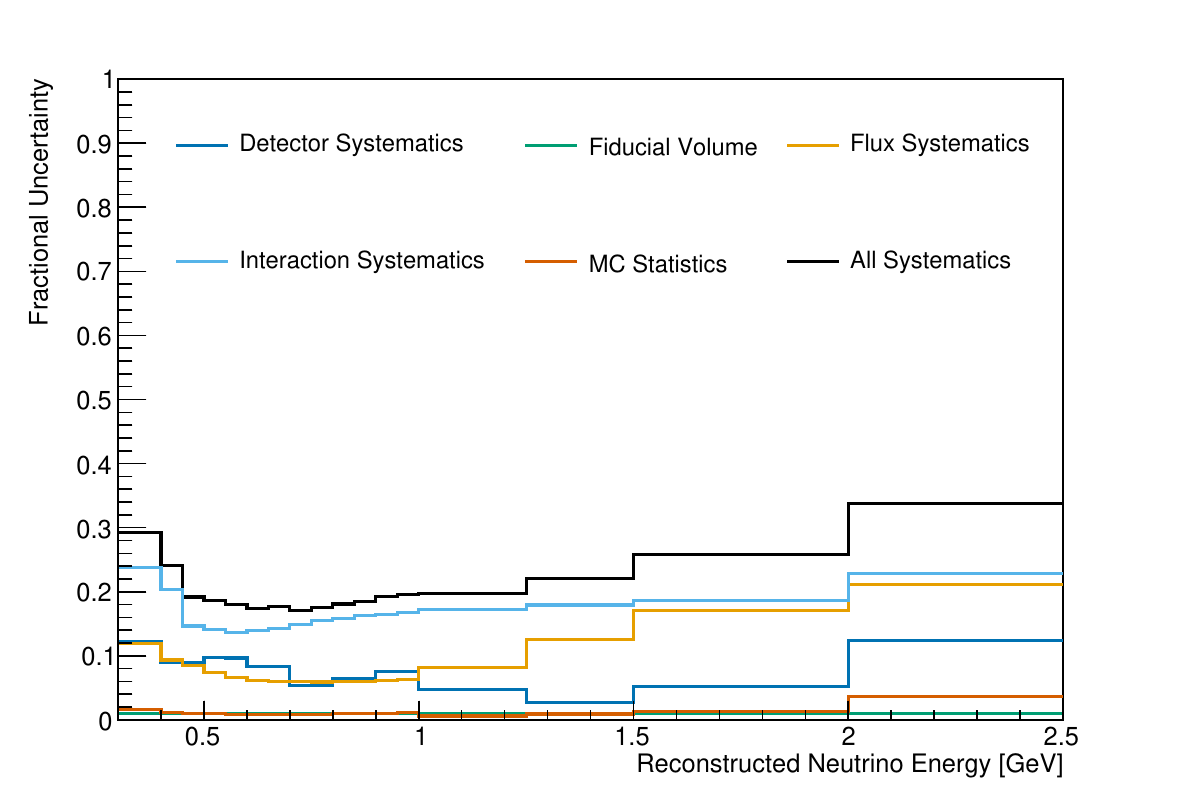}  
    \caption{\justifying Fractional size of systematic uncertainty as a function of reconstructed neutrino energy for Pandora (top) and SPINE (bottom).}
    \label{fig:systfrac_all}
\end{figure}

\begin{figure}[!htb]
    \centering
    \includegraphics[width=\columnwidth{}]{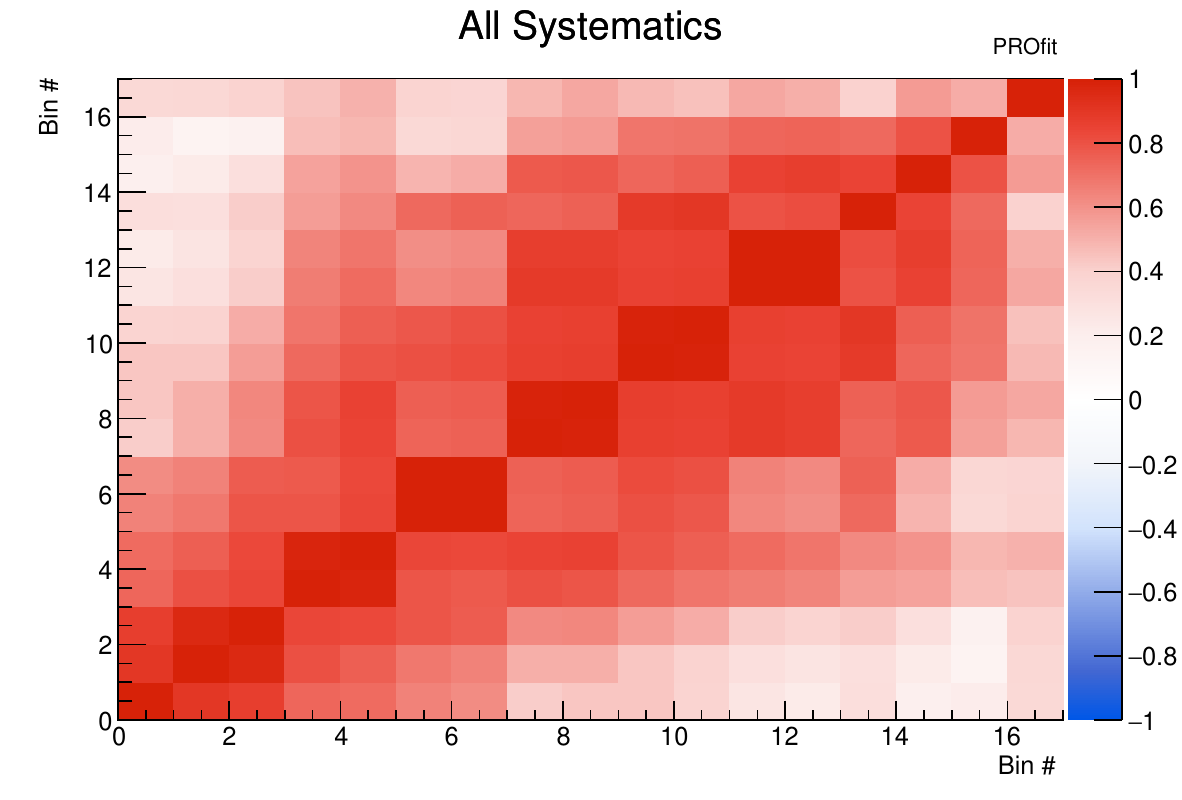}
    \includegraphics[width=\columnwidth{}]{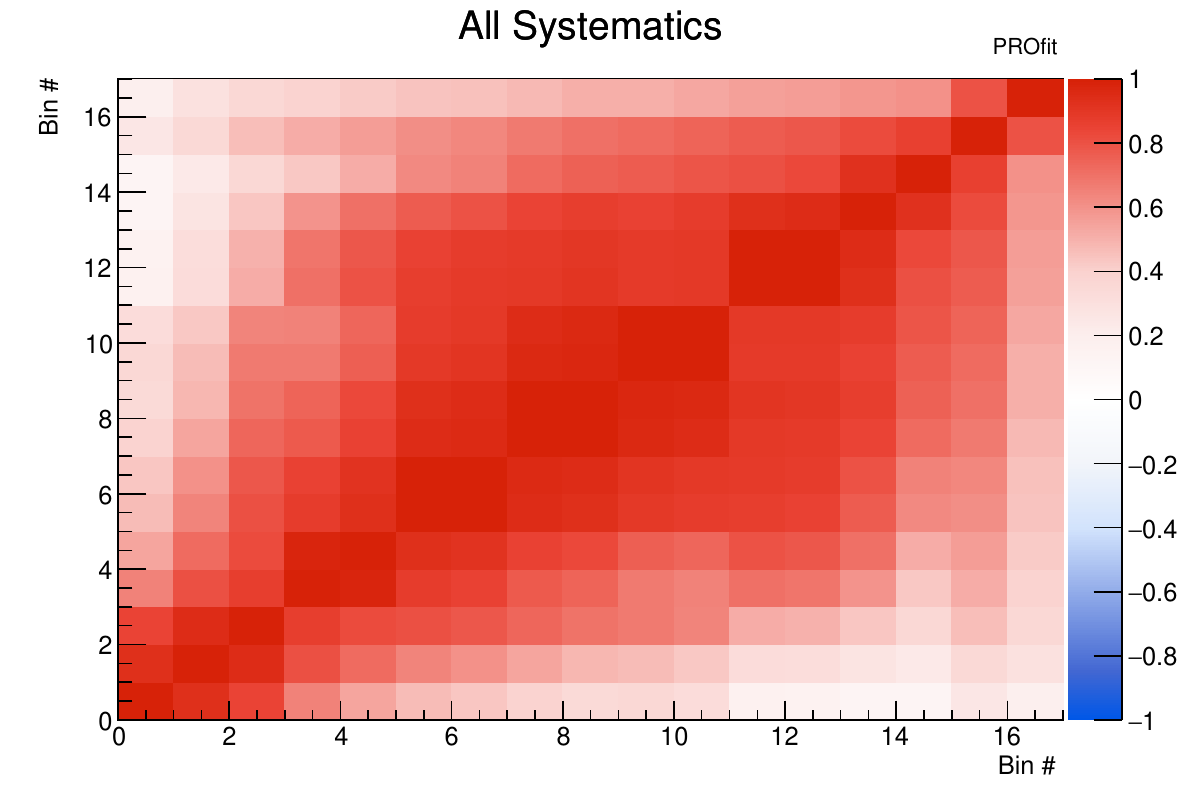}  
    \caption{\justifying Correlation matrix including all systematic uncertainties, for Pandora (top) and SPINE (bottom). Axes are labeled with bin numbers using the fit binning in reconstructed neutrino energy.}
    \label{fig:syst_corr_full}
\end{figure}

The total uncertainty in the predicted event rate resulting from systematic variations is around 20\% at the peak of the selected neutrino distribution, with the fractional contributions as a function of reconstructed neutrino energy shown in Fig.~\ref{fig:systfrac_all}. The systematics coming from uncertainty in the neutrino interaction model are the largest contribution, with uncertainty from the flux and detector models having approximately equal contributions for most energies. The full correlation matrix, including all systematics variations, is shown in Fig.~\ref{fig:syst_corr_full}.

\section{\label{sec:fit}Fitting and Statistical Analysis}
The suite of necessary statistical analyses are performed using PROfit~\cite{profit_github}, a new general-purpose oscillation and fitting framework that has been developed for SBN analyses. PROfit finds the set of oscillation and systematic parameters best describing the data, given prior uncertainties on the models used in the simulation.  The analysis is performed in the context of a two-flavor approximation within the 3+1 sterile neutrino model and results are presented in terms of $\dmsq$ and $\sinsqtwothmumu$, with the assumption $\theta_{ee}=\theta_{\mu e}=0$. Systematic uncertainties are incorporated as nuisance (pull-term) parameters governing the strength of correlated variations or through a covariance matrix; the choice made for each individual source of systematic uncertainty is described in Section~\ref{sec:syst}. Minimization of the combined Neyman--Pearson (CNP) $\chi^2_{CNP}$~\cite{Ji:2019yca} metric is performed and confidence intervals on the oscillation parameters are constructed using the Feldman-Cousins procedure~\cite{FeldmanCousins:PhysRevD.57.3873}. Signal unblinding follows a predetermined staged procedure in PROfit in which fits to both the SPINE and Pandora selections are performed simultaneously, with validation checks evaluated at each stage before proceeding to ensure a robust result.
Details of the PROfit framework and the statistical analysis choices in this analysis are provided in Appendix~\ref{app:profit}.

\section{\label{sec:results}Results}

\begin{figure}[!htb]
\centering
\includegraphics[width=\columnwidth]{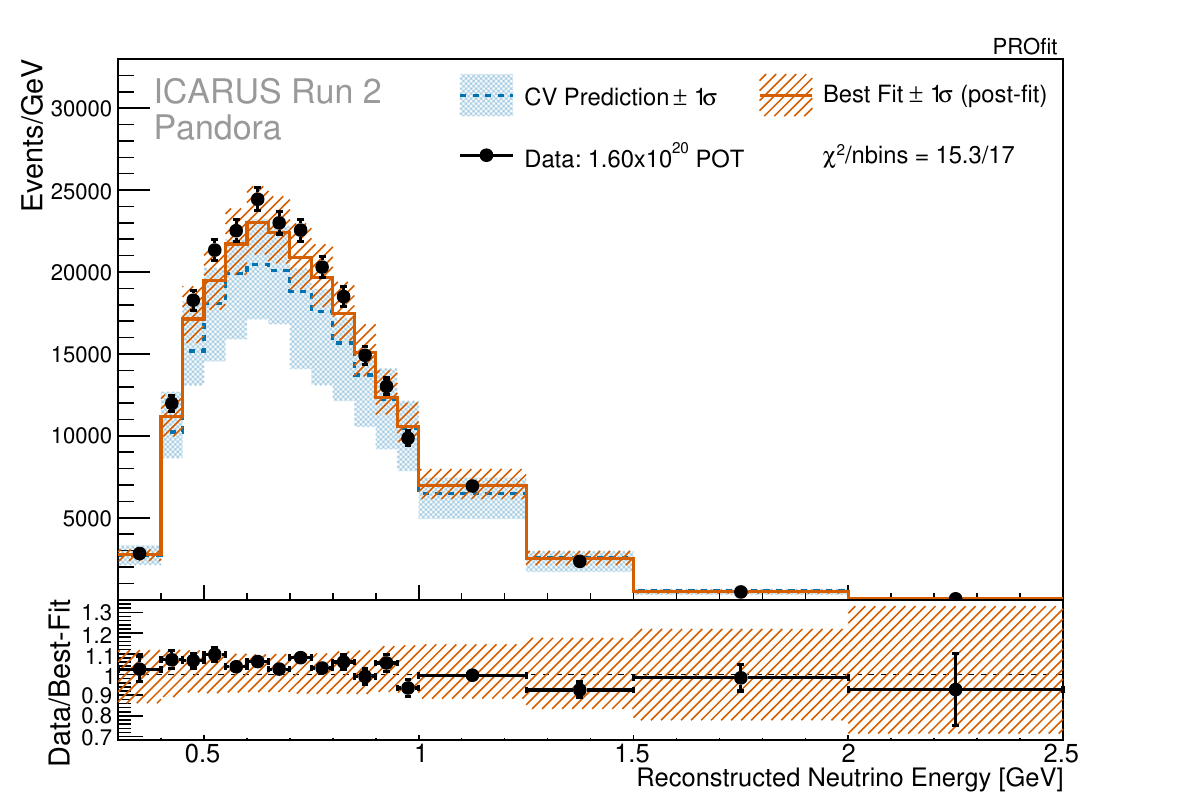}
\includegraphics[width=\columnwidth]{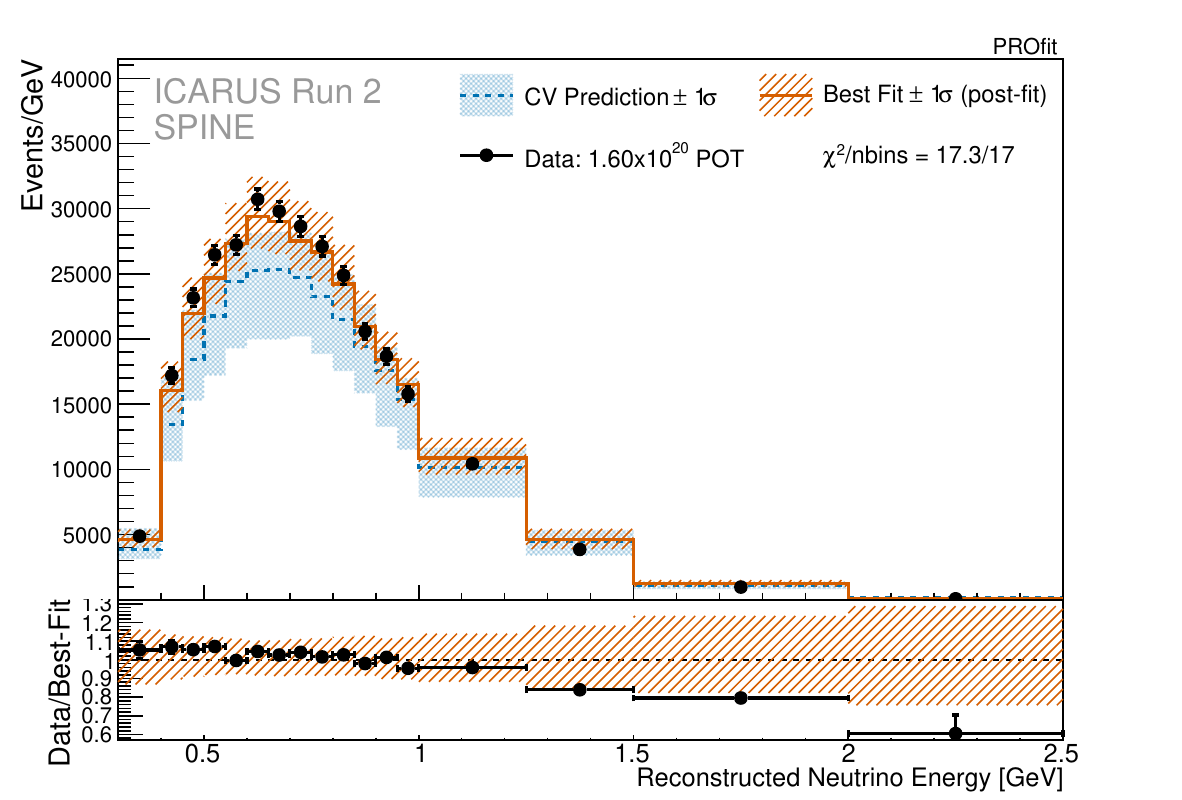}
\caption{\justifying Pandora (top) and SPINE (bottom) post-fit spectra and data/best-fit ratios for the full fit in which oscillation parameters are allowed to float. }
\label{fig:result_spec}
\end{figure}

\begin{figure}[!htb]
\centering
\includegraphics[width=\columnwidth]{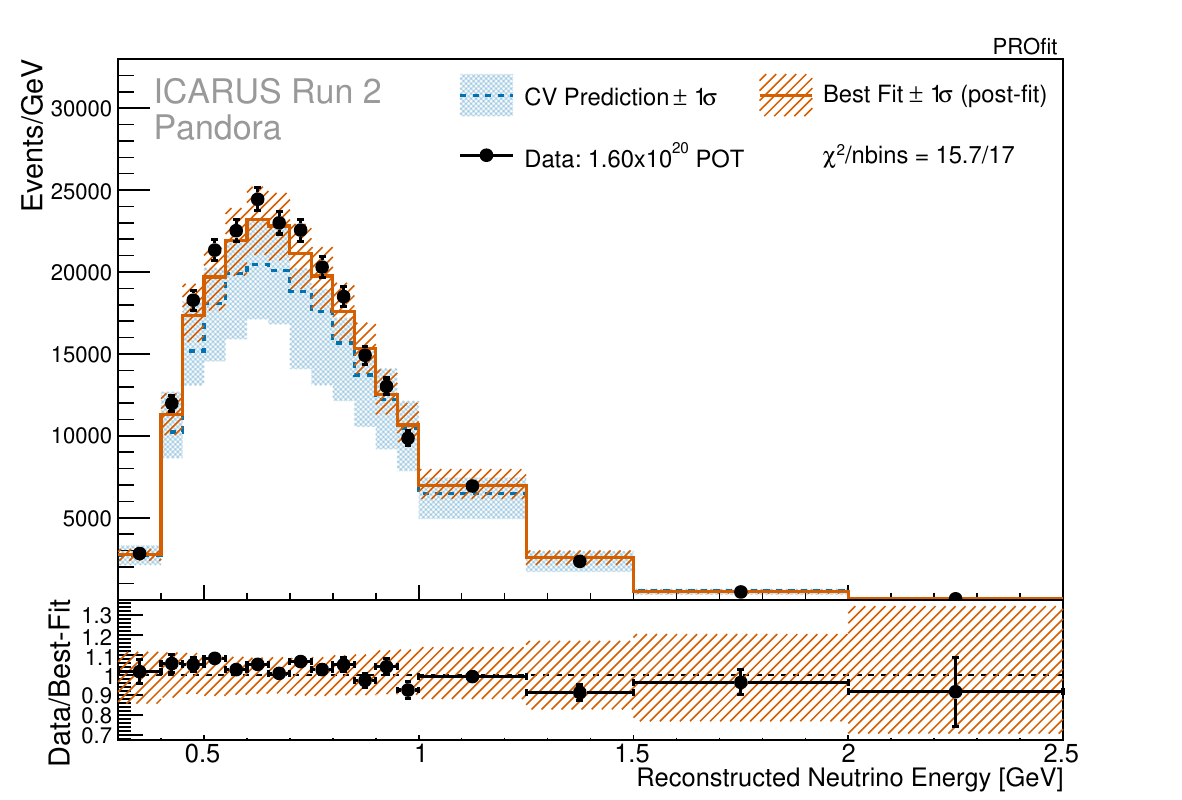}
\includegraphics[width=\columnwidth]{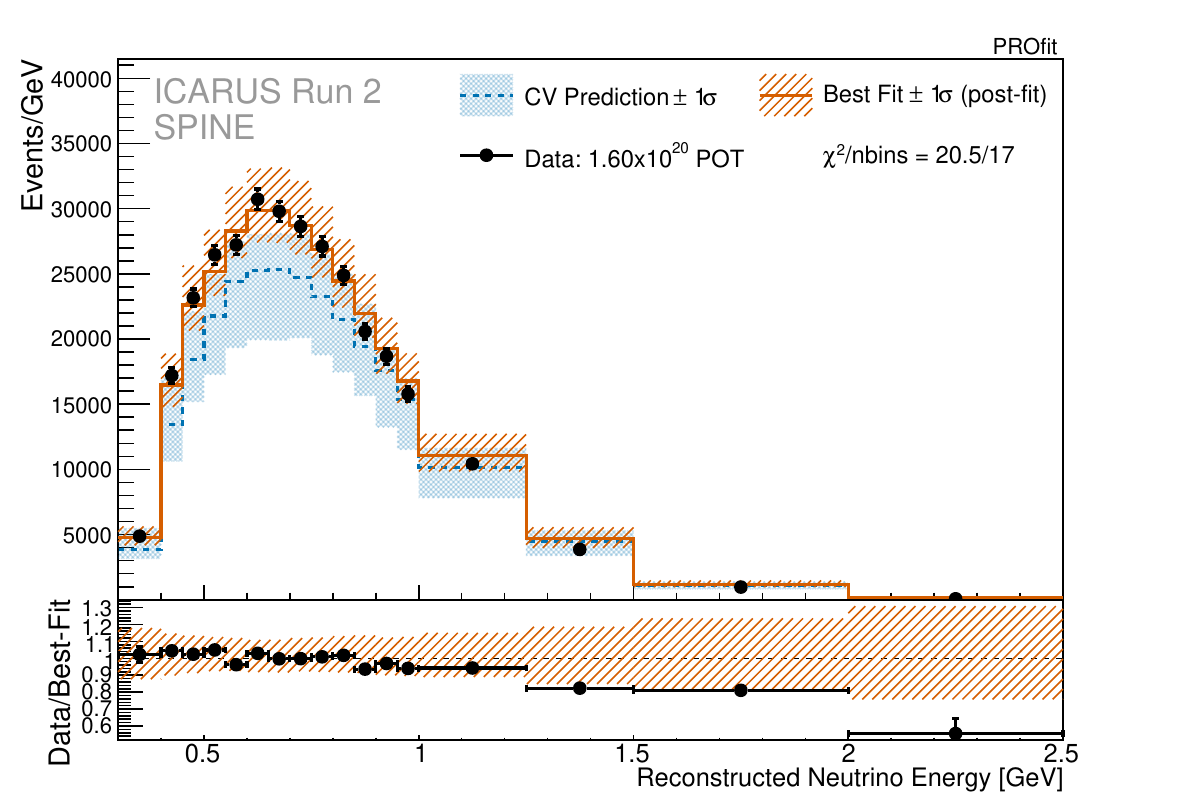}
\caption{\justifying Pandora (top) and SPINE (bottom) post-fit spectra and data/MC ratios for a fit to the null hypothesis, where oscillation parameters are fixed to zero.}
\label{fig:result_spec_null}
\end{figure}

No evidence for oscillation is observed in either the SPINE or Pandora fits. The best fit $\chi^2$/ndof is 15.3/17 for Pandora and 17.3/17 for SPINE. While the significance of oscillation is not statistically significant, the best-fit oscillation parameter values and the $\chisq$ values for both analyses are given in Table~\ref{tab:results_oscpars}.  
The post-fit spectra and uncertainties are shown in Fig.~\ref{fig:result_spec}. Post-fit spectra for fits to the null hypothesis, with oscillation parameters fixed at zero, are shown in Fig.~\ref{fig:result_spec_null}.

\begin{table}[b]
\caption{\justifying \label{tab:results_oscpars}
Fit results, including best fit oscillation parameters, the $\chisq$ for the nominal fit, and the $\Delta\chisq$ with respect to a fit to the no-oscillation null hypothesis. No statistically significant oscillation is observed.
}
\begin{ruledtabular}
\begin{tabular}{lcc}
      & Pandora & SPINE  \\ 
$\sin^22\theta_{\mu\mu}$ & 0.07 & 0.24  \\
$\Delta m^2_{41}$ [eV$^2$]& 10.2 & 13.5 \\
$\chisq$ (17 bins) & 15.3 & 17.3 \\
$\Delta\chisq$ & 0.4 & 3.2 \\
$p$-value & 0.910 & 0.421\\
\end{tabular}
\end{ruledtabular}
\end{table}

Feldman-Cousins $p$-values comparing the best-fit sterile neutrino hypothesis to the null (no oscillation) hypothesis are calculated as described in Appendix~\ref{app:profit}. The resulting $p$-values are 0.910 and 0.421 for the Pandora and SPINE selections, respectively, indicating no significant preference for sterile neutrino oscillations. Given the consistency with the null hypothesis, we report 90\% confidence level exclusion contours.

\begin{figure}[!htb]
    \centering
    \includegraphics[width=\columnwidth]{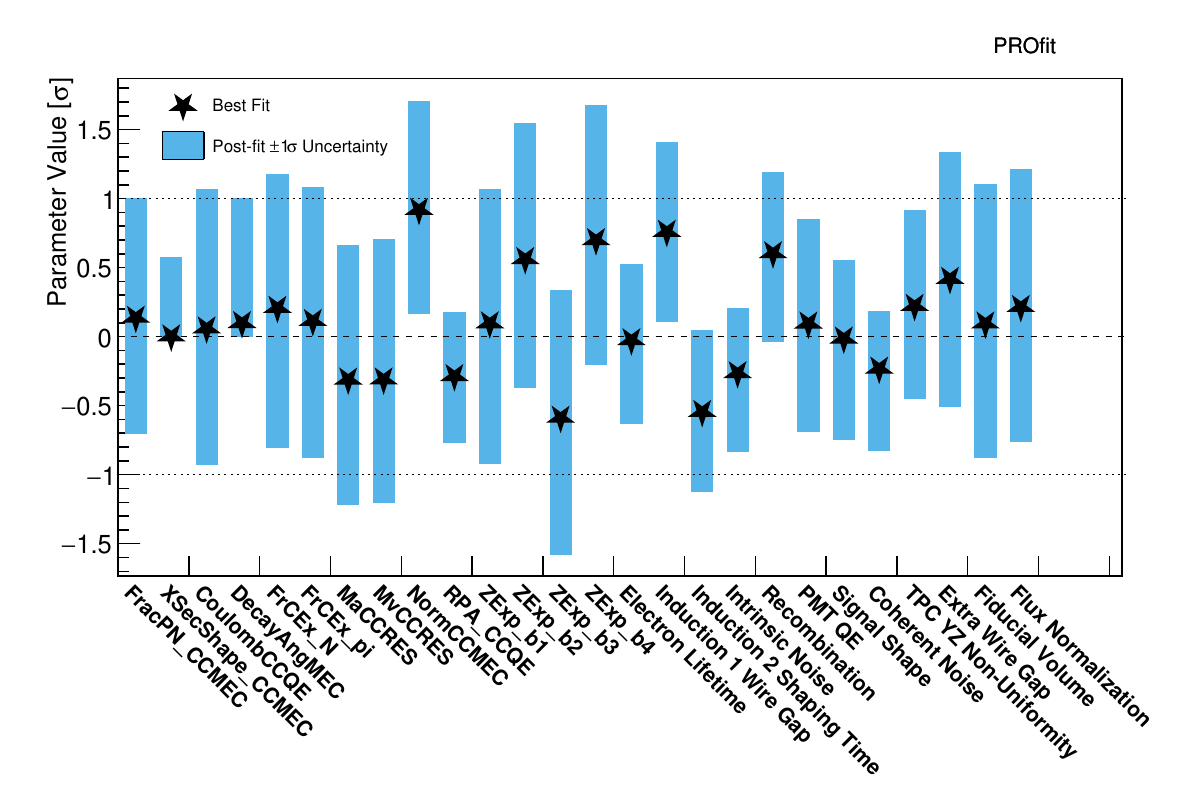}
    \includegraphics[width=\columnwidth]{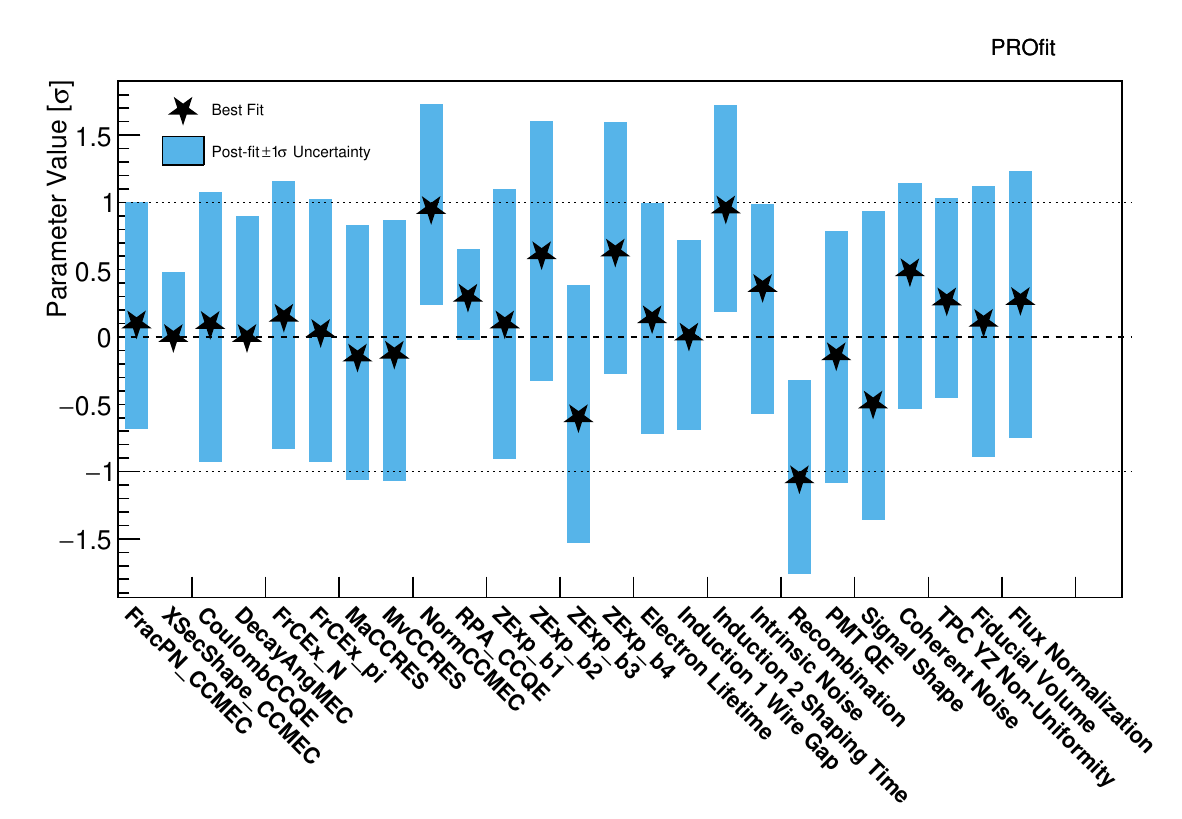}
    \caption{\justifying The post-fit nuisance parameter profiles for the Pandora (top) and SPINE (bottom) analyses. Pre-fit nuisance parameter values are by definition 0 (no pull), with pre-fit uncertainty defined as $\pm1\sigma$ (horizontal dotted lines).}
    \label{fig:result_pulls}
\end{figure}

PROfit diagnostic plots are inspected to understand the impact of systematic uncertainties in the analyses. All of the systematic parameter pulls are shown in Fig.~\ref{fig:result_pulls}.
The largest systematic parameter pulls in the full fit are $\sim$1$\sigma$, with the CC-MEC normalization being pulled to $+1\sigma$ in both analyses and the recombination and Induction 2 Shaping Time systematics being pulled to $-1\sigma$ and $+1\sigma$, respectively, in SPINE. CC-MEC normalization has a large prior uncertainty that impacts the event rate in the bulk of our $\numu$ distribution, so it is a natural way to accommodate the pre-fit normalization difference between data and CV MC, seen by comparing the data points to the CV prediction (dashed blue line) in Fig.~\ref{fig:result_spec}. This was observed in many mock data studies, where the CC-MEC normalization was frequently pulled to cover normalization differences. The negative pull on recombination in the SPINE analysis is not consistent with expectations from known detector parameters; the most likely explanation is that the impact of this parameter is degenerate with some other effect present in the data. The significant constraint on the RPA-CCQE parameter can be understood because this systematic has a large prior uncertainty in a specific energy range and thus can be constrained by the normalization of the data in that range.

\begin{figure}[!htb]
\centering
\includegraphics[width=\columnwidth]{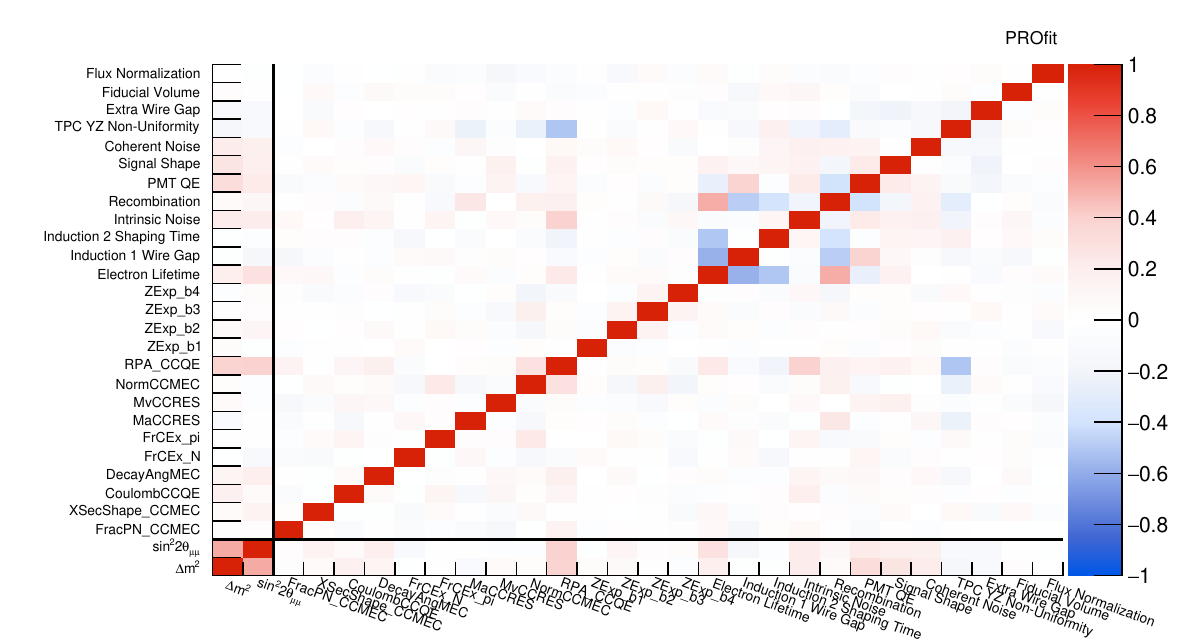}
\includegraphics[width=\columnwidth]{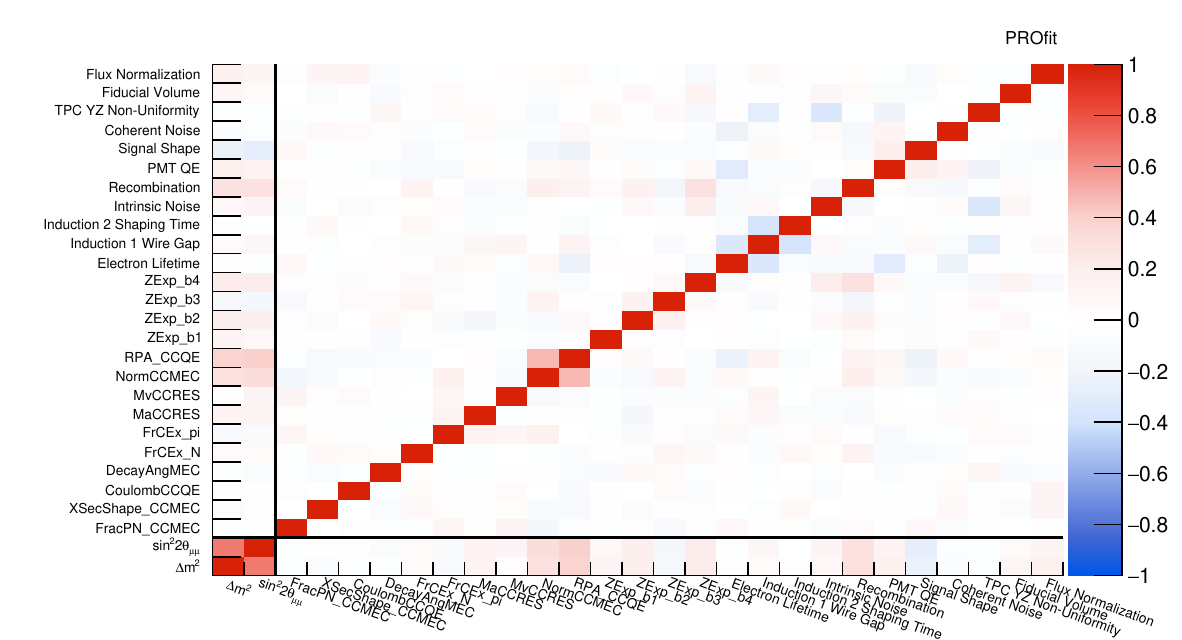}
\caption{\justifying The post-fit correlation matrices for all parameters from Pandora (top) and SPINE (bottom) when fitting with oscillations. Highlighted on the left is the two physics parameters.}
\label{fig:result_correlation}
\end{figure}

Figure~\ref{fig:result_correlation} shows the full correlation matrix for all parameters, including the oscillation parameters. Note that in the SPINE fit, two systematics parameters, the CC-MEC normalization and the RPA correction to the QE model, are correlated with each other and with the oscillation parameters. This is understood by noting that, as described above, these two parameters can combine to produce a normalization effect which acts oppositely to the normalization effect from a potential oscillation signal.

\begin{figure}[!htb]
\centering
\includegraphics[width=\columnwidth]{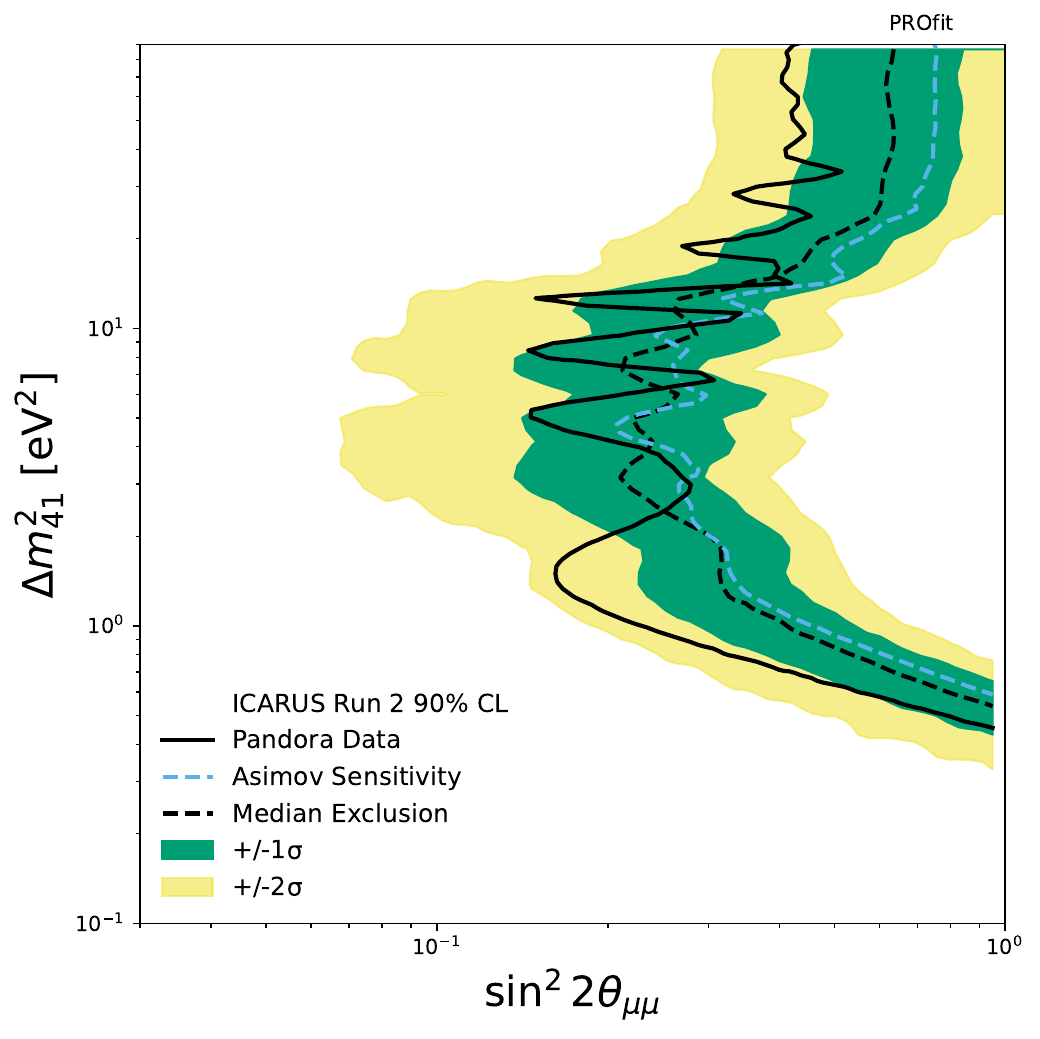}
\includegraphics[width=\columnwidth]{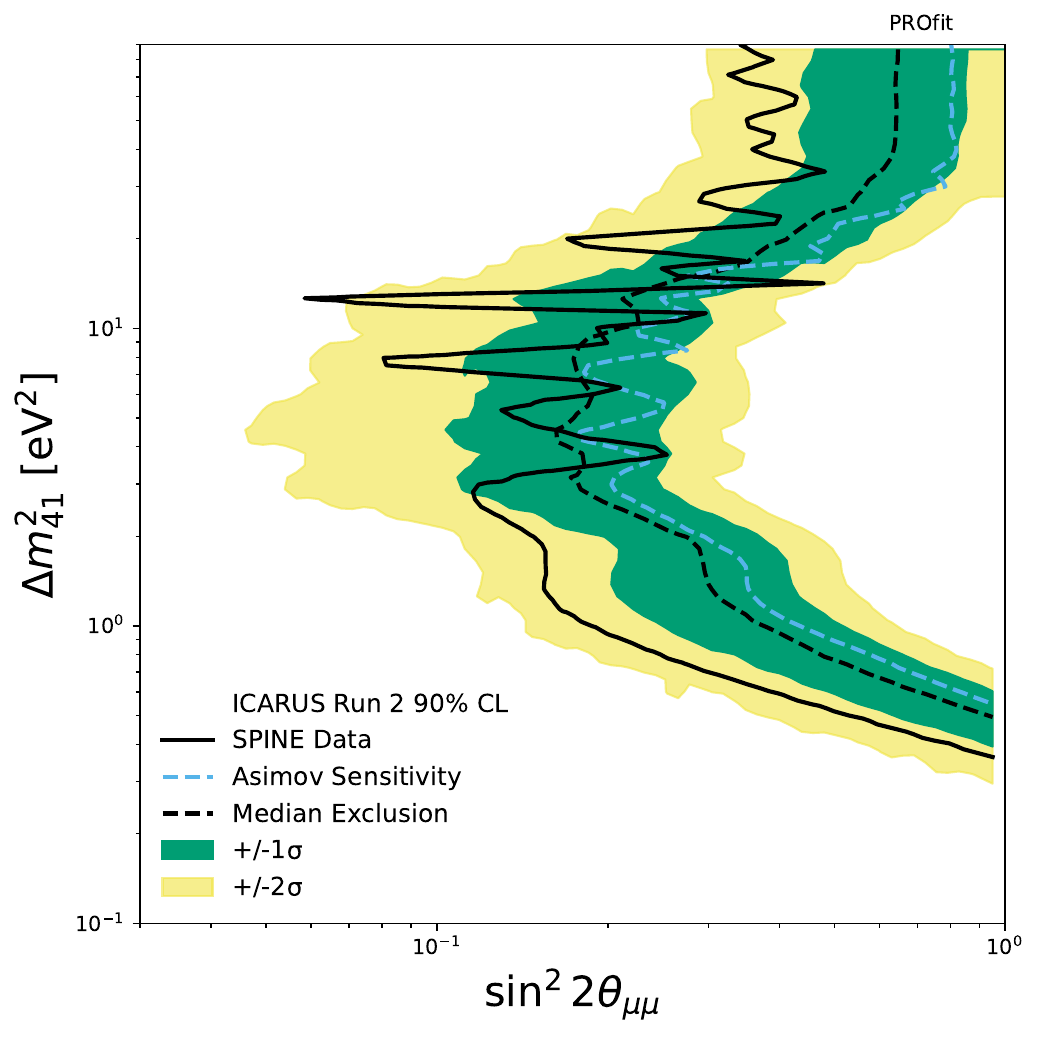}
\caption{\justifying Feldman-Cousins 90\% C.L. exclusion contours for Pandora (top) and SPINE (bottom), shown alongside the sensitivity contours, with $1\sigma$ and $2\sigma$ uncertainty bands for the median exclusion sensitivity. The uncertainty bands represent the range of exclusion contours within which 68.3\% and 95.4\% of universes excluding oscillation fall.}
\label{fig:results_exclusion}
\end{figure}

The final 90\% C.L. sensitivity contours and exclusion band shown in Fig.~\ref{fig:results_exclusion} are generated using Feldman-Cousins corrected critical $\chisq$ values as described in the last section.
For both analyses, the data contour falls within the 2$\sigma$ sensitivity band for nearly all points. 
The spiky behavior in the data contour as a function of $\Delta m^2$ is typical of short-baseline oscillation exclusion contours and results from from the sinusoidal variation in oscillation probability. The point with least exclusion power is closest to the data best fit point and the nearby point with strong exclusion is the opposite local extremum, which is less like the data and thus more strongly excluded. This also explains why the SPINE contour has somewhat more pronounced spikes than the Pandora contour, since the SPINE best fit has a slightly higher significance for oscillation and the preferred mixing angle for the SPINE fit is significantly larger.
At high values of $\Delta m^2$, the data contour falls near the $+1$ to $+2\sigma$ boundary of the sensitivity band. Oscillations in this region of parameter space would be observable at ICARUS primarily as a change in normalization; since the normalization of the data is slightly higher than that of the central value MC, while an oscillation result would result in a lower rate of data relative to prediction, it is clear why the data exclusion is stronger than the (MC-based) median sensitivity in this region. 

\section{\label{sec:discuss}Discussion and Future Prospects}
\begin{figure}[!htb]
\centering
\includegraphics[width=\columnwidth]{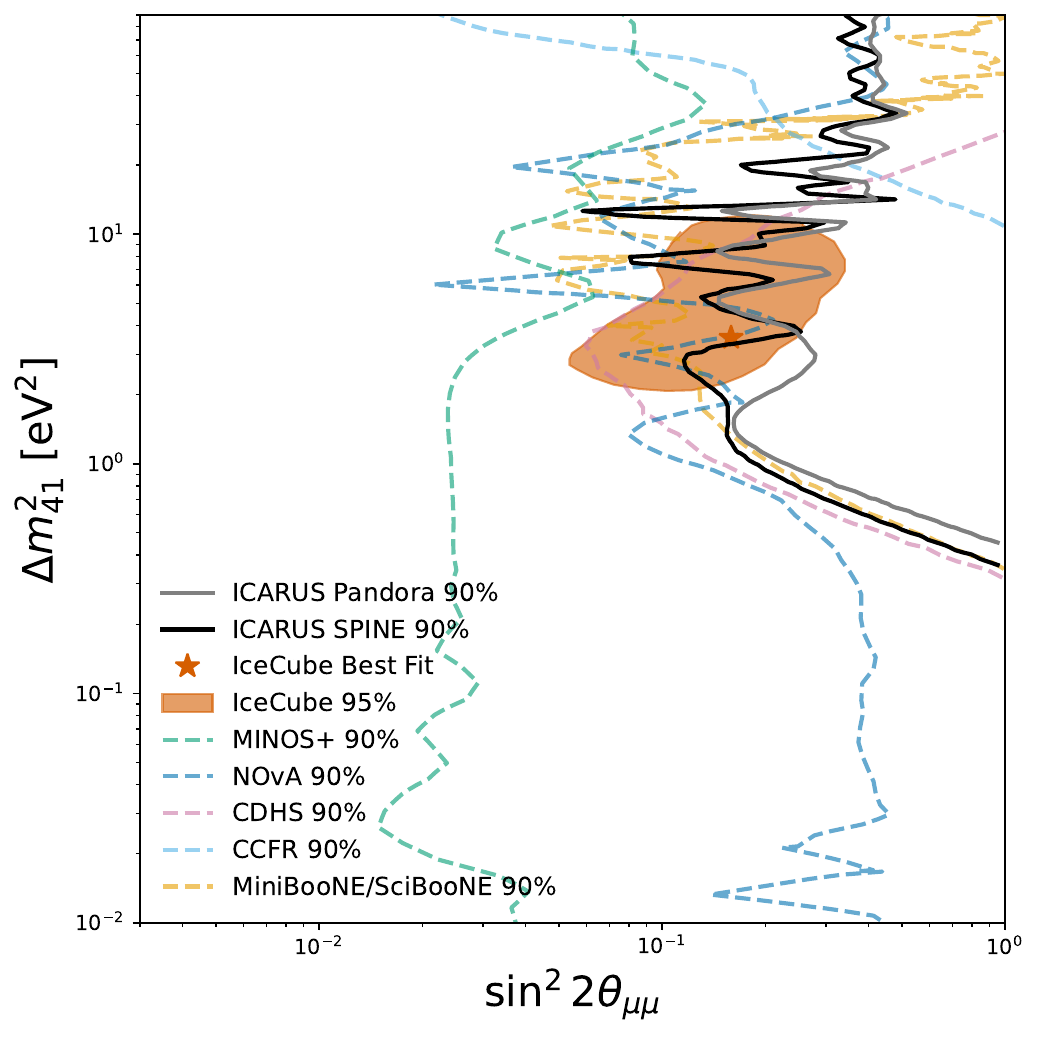}
\caption{\justifying 90\% C.L. (FC corrected) exclusion contours for Pandora (gray) and SPINE (black), shown in the context of existing limits and closed contours from other experiments~\cite{IceCubeCollaboration:2024dxk,MINOS:2017cae,NOvA:2024imi,MiniBooNE:2012meu,Dydak:1983zq,Stockdale:1984cg}.}
\label{fig:results_compare}
\end{figure}

The results presented here are consistent with and have comparable sensitivity to previous single-detector $\numu$ disappearance analyses, as seen in Fig.~\ref{fig:results_compare}, which compares the ICARUS 90\% C.L. exclusion contours to existing limits (and one closed contour) from other experiments. The global picture of short-baseline $\numu$ disappearance remains uncertain as there is mild tension between existing exclusion contours and the IceCube closed contour. Additionally, previous anomalous results suggesting potential $\nue$ appearance remain unexplained, with increasingly stringent limits on the 3+1 sterile neutrino model being set. While limits on 3+1 oscillation parameters, such as those reported in the analysis presented here, are still the benchmark by which sensitivity is evaluated, future results from ICARUS and SBN will move towards more comprehensive and sensitive exploration of the rich theory landscape. 

The ICARUS result presented here is systematics dominated, with the total systematic uncertainty of around 20\% in the predicted event rate being significantly larger than the statistical uncertainty.
As described in Section~\ref{sec:intro}, ICARUS serves as the far detector of the SBN program, the ultimate goal of which is to produce $5\sigma$ exclusion contours in a combined analysis of $\numu$ and $\nue$ disappearance and $\nue$ appearance in the BNB and NuMI beams, using data at different baselines from the near (SBND) and far (ICARUS) detectors. The SBN program is designed for constraint of systematic uncertainty, with the flux and interaction model uncertainties nearly 100\% correlated between the two LArTPC detectors and so significantly constrained in a two-detector oscillation analysis. Furthermore, improvements to the detector simulation, many of which were developed as systematic variations in the analysis reported here, will remove many residual data-MC discrepancies in ICARUS. The leading sources of detector model uncertainty in this analysis (spatial non-uniformity, recombination, electron lifetime, TPC noise, and TPC field response) are already better modeled or taken from data in the latest versions of ICARUS simulation. Combined with the high-quality data being collected in SBND, these improvements to the detector model will produce robust and world-leading two-detector oscillation analyses. 

\section{\label{sec:summary}Summary}

ICARUS observes no evidence for $\numu$ disappearance in the BNB in the Run 2 dataset, using both the Pandora and SPINE reconstruction frameworks to identify samples of $\sig$ interactions; 90\% exclusion contours are presented for both analyses in the context of the 3+1 sterile neutrino model. This result is consistent with previous single-detector $\numu$ disappearance searches. The result is systematics dominated, as expected in the absence of near detector data to provide constraint on the large \textit{a priori} uncertainties on the neutrino flux and interaction models. Future results will incorporate additional data and build on the tools and methods presented here to more fully constrain systematic uncertainties and more fully explore theories predicting baseline-dependent effects in SBN.

\begin{acknowledgments}
This document was prepared by the ICARUS Collaboration using the resources of the Fermi National Accelerator Laboratory (Fermilab), a U.S. Department of Energy, Office of Science, HEP User Facility. Fermilab is managed by FermiForward Discovery Group, LLC, acting under Contract No. 89243024CSC000002.

This work was also supported by Istituto Nazionale di Fisica Nucleare (INFN, Italy), EU Horizon 2020 Research and Innovation Program under the Marie Sklodowska-Curie Grant Agreement Nos. 822185, 858199, and 101003460 and Horizon Europe Program research and innovation programme under the Marie Sklodowska-Curie Grant Agreement No. 101081478, the research contract per Law 240/2010, Art. 24 (3)(a), and D.G.R. 693/2023 (REF. PA:2023-20090/RER—CUP:J19J23000730002) by FSE+ 2021–2027. Furthermore the support of CERN in the detector overhauling within the Neutrino Platform framework, in the detector installation and commissioning, is acknowledged.
This work was also supported by the Anusandhan National Research Foundation (ANRF, India) under the Ramanujan Fellowship (Grant No. RJF/2025/000203).

The ICARUS Collaboration would like to thank the MINOS Collaboration for having provided the Side CRT panels as well as Double Chooz (University of Chicago) for the Bottom CRT panels. 
We also acknowledge the contribution of many SBND colleagues, in particular for the development of a number of simulation, reconstruction and analysis tools which are shared within the SBN program.

\end{acknowledgments}
\appendix
\section{\label{app:pandora}Pandora}
This section provides a brief overview of the steps required to perform a  full event reconstruction using the Pandora software framework, from raw 2D hits to the hierarchy of interactions. A detailed overview of the software, including details on the reconstruction chain and the main algorithms employed therein is provided in~\cite{PandoraFirstPaper,PandoraMB2017,PandoraProtoDune2022}. Here, the software infrastructure is described in general, highlighting the specific algorithm optimizations performed for ICARUS analyses.

Pandora~\cite{PandoraFirstPaper} is a pattern recognition software widely used in LAr-based detectors, especially LArTPCs. 
The goal of this software is to provide a fully-automated and complete reconstruction of events, delivering a hierarchy of tracks and electromagnetic showers to represent final-state particles and any subsequent interactions or decays. 

A multi-algorithm approach is adopted to perform pattern recognition on the reconstructed images for the collected events. 
In the processing chain, each algorithm acts independently from the others and addresses a specific task in a particular topology, e.g., interaction vertex identification or shower 3D reconstruction. It is the ensemble of algorithms that gradually builds up a picture of the underlying events and collectively provides a robust reconstruction. 

The inputs to Pandora reconstruction are the collection of \textit{hits}, i.e. signals detected on single wires at a definite drift time, and information on the detector geometry. 
The starting point is a set of 3 x 2D images (where $x$ coordinate represents the drift time and $y$ the wire number), one per readout plane. 
Reconstruction begins with two-dimensional analysis, where separate clusters of hits are created for each structure in the input image from each readout plane. Ideally, each cluster defined at this step should contain all and only the hits belonging to a single reconstructed particle. Clustering is done in stages, of which the first one privileges purity over completeness and puts together only the hits that are likely to belong to the same reconstructed particle. A series of topological algorithms is later used to make cluster merging and splitting decisions on the overall event topology (as opposed to individual clusters in isolation) and refine the 2D clustering.
Next, clusters of 2D hits from the three readout planes that represent individual track-like particles are identified. 
Finally, for each input 2D hit in a reconstructed particle, a 3D hit is created, with different mechanisms depending on the cluster topology~\cite{PandoraMB2017}. An analytic $\chisq$ minimization is used to match hit coordinates in one direction (say $x$) with the remaining ones ($y$ and $z$). 

\begin{figure}[!htb]
    \centering
    \includegraphics[width=0.8\linewidth]{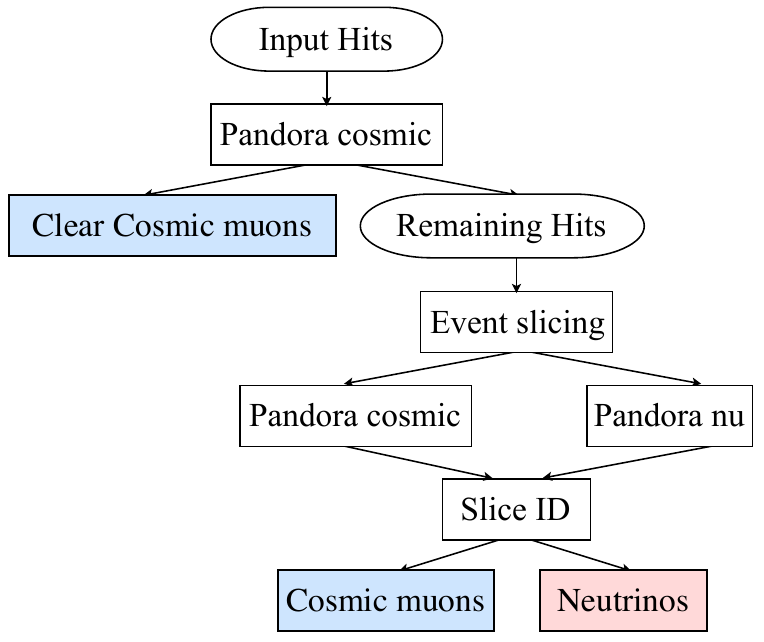}
    \caption{\justifying Schematic view of the processing chain for the Pandora event reconstruction. Inputs to Pandora 
    are indicated by elliptical
    boxes. The outputs of the reconstruction chain, that are saved and used in later steps of the processing chain, are indicated by colored boxes.}
    \label{fig:Pandora_scheme}
\end{figure}

Two multi-algorithm reconstruction paths 
are employed to provide hierarchy reconstruction: 
\begin{itemize}[topsep=3pt]
    \item \textit{Pandora Cosmic}, which is optimized for the reconstruction of Cosmic Ray (CR) muons and their delta rays. This path is strongly track-oriented, since the expected primary particles are muons with showers assumed to be their daughter delta rays. The interaction vertex is given by the high-y coordinate of the muon candidate track. A detailed description of the four main stages subsequently run as part of this reconstruction step is provided in~\cite{PandoraMB2017}.
    \item \textit{Pandora Nu}, is the alternative reconstruction path and is optimised for the reconstruction of neutrino interactions, starting from the interaction vertex and including all the particles emerging from that. The primary parent is the invisible neutrino and all the reconstructed particles are its 
    daughters. Further details on this path, including ICARUS dedicated optimization studies, are included in the following text.
\end{itemize}
The event processing chain, schematically represented in Fig.~\ref{fig:Pandora_scheme}, relies on both paths. 

First, an initial pass 
    of \textit{Pandora Cosmic} reconstruction is run 
    in order to extract a list of unambiguous candidate cosmic muons, tagged as \textit{Clear Cosmic} interactions, and discard the hits associated with them from further steps.
    Pandora identifies an interaction as a clear cosmic for being too far out-of-time with the beam, for example if it clearly extends well into the extra readout window pre- and post-beam (plus drift time), or if it appears to be fully through-going across the detector volume.

The 
standard rejection of cosmics from Pandora (Cut Mode = nominal) corresponds to i) $\cos \theta >- 0.6$, with $\theta$ being the angle with respect to the vertical, ii)  the containment along $y$, 
 and  iii) a time cut  requiring particles to be \textit{in-time} with respect to the beam.  With respect to the latter condition, particles are considered \textit{in-time} if  all their hits are within $\sim$1~ms drift--time interval as given by the trigger.  In addition, for a cathode crossing particle entering the detector, Pandora will stitch together the two pieces in the adjacent TPCs, extracting the crossing time that has to be compatible with the trigger. 

However, muons from $\nu$ interactions that are almost vertical can be tagged as CR muons due to the cut on $\cos \theta$.  
After dedicated studies comparing the efficiency of background rejection with the capability to retain neutrino candidates, it was decided to loosen the timing requirements for the cosmic ray tagging and remove the condition on the vertical angle $\theta$  (CUT Mode = cautious). With such a choice the tagging of cosmic rays is still effective, with virtually no loss of  neutrino events. 

The remaining cosmic-removed hit collection 
    is used for {\textit{Event Slicing}}. 
    This step is aimed at collecting and separating 3D hits belonging to single neutrino interactions from cosmic ray remnants, i.e. residual background contaminations. All the 3D clusters deemed to be associated with a given \textit{slice}, representing a single interaction in the detector, are grouped together. Ideally each slice should contain all and only the hits belonging to a single particle hierarchy. As part of this sequence and prior to the slicing step itself, some of the reconstruction steps included in the \textit{Pandora Nu} reconstruction pathway are performed as they are preparatory to an effective event slicing. Among them are 2D reconstruction, 3D track and shower reconstruction, 3D clusters recovery and 3D hit reconstruction.
    
    The cosmic-filtered 
    sets of 2D hits on the readout planes associated with a given slice are the input of the dedicated reconstruction performed in the next steps, in which each slice is processed 
    separately using both \textit{Pandora Nu} reconstruction and \textit{Pandora Cosmic} reconstruction. 
    This second pass of \textit{Pandora Cosmic} is identical to the former and is aimed at providing a proper reconstruction to leftover cosmic-ray contaminations. 
    
    A key step that distinguishes the \textit{Pandora Nu} reconstruction path from the \textit{Pandora Cosmic} path is \textit{3D interaction vertex reconstruction}. Here, the lists of 2D clusters from different readout planes are used to identify the neutrino interaction vertex, ensuring a correct identification and reconstruction of individual \textit{primary} particles as unique (reconstructed) \texttt{Particle Flow Particles (PFPs)}. This is done in three steps.
    First, all the cluster end-points are used to create up to four candidate vertices for each cluster pairing between different readout planes and 
    then, candidates not sitting on a hit or a registered detector gap are removed from the list.
    Finally, using a  BDT, the surviving candidates are assigned a score and the one with the best score is selected as the interaction vertex and saved for later steps. 
    
    The global score is evaluated combining three estimators: an energy, an asymmetry and a beam score. The energy score is extracted by comparing the energy of each cluster in each view with the vertex projection onto the same view and is aimed at assessing consistency between vertex candidates and reconstructed clusters on all the views. The larger the score, the more the vertex candidate is suppressed. This reflects the fact that primary particles, i.e. neutrino daughters, should point back to the true neutrino interaction vertex, while secondary particles may not, but should be less energetic. The asymmetry parameter is obtained by counting the number of hits upstream and downstream of the candidate vertex position and suppresses candidates incorrectly located along a single cluster as opposed to candidates close to cluster endpoints. The beam score uses the beam direction to favor vertex candidates with low values of $z$, i.e. the beam direction coordinate.
        
        The variables used to perform the classification and identify the most suitable vertex candidate include both properties of the reconstructed event and features of the candidate vertices. The former variables include the total number of hits, clusters and vertex candidates associated to the event, the fraction of hits likely originating from a shower based on cluster length and the associated reconstructed particle, an estimate of the energy and of the area of the event, extracted from the total clusters span along x and z directions and an estimate of how longitudinal the event is with respect to the beam direction. The vertex variables include all the quantities required to extract the score estimators mentioned above, i.e., a measurement of the transverse energy extracted comparing the displacement of all clusters in all views with respect to the candidate vertex projection and estimates of asymmetry parameters evaluated globally, locally and assuming the presence of a shower.
        In order to improve vertex selection for ICARUS, this BDT is trained on a MC sample of BNB $\nu_{\mu}+ \nu_e$ interactions in the ICARUS detector. 
    
    The subsequent steps are 3D track and shower reconstruction.  The former is described in detail above, since it is part of both neutrino and cosmic reconstruction paths, while \textit{3D shower reconstruction} algorithms are specific to \textit{Pandora Nu}. To select 2D shower candidate clusters, a set of geometrical and topological cuts are used to distinguish 2D track-like clusters from 2D shower-like clusters. Then, in each view, clusters representing candidate shower branches are merged with the appropriate candidate shower spines, for example, long shower-like clusters close to the interaction vertex, to provide high-completeness 2D shower clusters. Later, a rank-three tensor is diagonalized to properly match 2D information from different readout planes and build 3D unambiguous shower particles. The goal is to collect the clusters representing the same shower in each readout plane: predictions on the spatial extent of the shower from combinations of clusters on two planes are compared with the envelope of 
    matching clusters in the third plane. 
    Once 3D showers have been identified, 3D track reconstruction is repeated in order to recover potential track pieces dissolved to try to build showers. Finally, an attempt to recover groups of clusters that failed to satisfy the quality cuts for particle creation is done, lowering the thresholds used to match information coming from different planes.
    
    Before building up a hierarchy of reconstructed interactions, 3D track-like and shower-like particle reconstruction is refined. As usual, the first step is handled by a set of 2D algorithms and then 3D tools are employed. In ICARUS, three 2D and three 3D particle-refinement cluster mop up algorithms are subsequently run with the aim of merging unassociated clusters with assigned particle clusters, especially to increase the shower reconstruction completeness. A description of the algorithms is given in~\cite{PandoraMB2017}. Ideally, at this point only a negligible number of 2D hits should remain unassigned to reconstructed objects.
    
    The final step in the \textit{Pandora Nu} reconstruction is \textit{particle hierarchy reconstruction}. The procedure starts with the creation of a neutrino particle, then a 3D interaction vertex is assigned to the $\nu$ candidate and all the associated particles are added as primary daughters of the neutrino. Daughter particles of the primaries, i.e. secondaries, are later added to the existing $\nu$ daughters. Each particle in the final neutrino interaction is assigned a 3D vertex position which is equal to the neutrino interaction vertex for primary particles and the point of closest approach to their parents for secondaries.
    
    The last neutrino algorithm of the chain handles the classification of \textit{track-like and shower-like reconstructed particles}. Using a BDT, clusters are classified as track-like or shower-like based on a set of 13 reconstructed quantities, including \textit{geometric} quantities
    and \textit{calorimetric} features that rely on the measured charge as a proxy for energy.
    The BDT training is based on a sample of simulated $\nu_{\mu}+\nu_e$ BNB interactions containing equal numbers of track-like and shower-like particles. The BDT classification efficiency for MC is $\sim$80\% and consistent results are obtained with validation checks on NuMI MC events and data.
    The primary daughter particle with the highest number of hits dictates whether the neutrino candidate is identified as $\nu_{\mu}$ or $\nu_e$: the former if it is track-like, the latter if it is shower-like.

    Once both \textit{Pandora Nu} and \textit{Pandora Cosmic} reconstruction paths are completed, a choice of the most appropriate reconstruction for each slice in the event, called \textit{Slice classification} stage, takes place. This decision is again made by means of a BDT. The variables used to 
    define the \textit{neutrino score} rely on both reconstruction outcomes. 
    The list of BDT features extracted under the neutrino hypothesis includes the number of reconstructed neutrino daughter particles and the number of 3D hits in the slice, the $y$ coordinate of the $\nu$ vertex, the average direction of primary particles weighted by the number of space points associated to them, 
    and the output of a PCA on the space points reconstructed within a certain distance from the vertex. The variables computed under the cosmic ray assumption are in reference to the longest reconstructed track in the slice and include its direction projected along $y$, its vertical deflection, and both number and fraction of hits of the slice in this track. This is motivated by the fact that cosmic ray muons usually enter from the top of the detector and generate simpler, typically track-like, topologies relative to neutrinos from the beam. 
    The \textit{neutrino score} could be used to classify the slice as a \textit{Neutrino candidate} or a \textit{Cosmic Muon}, as is done e.g., in ProtoDUNE~\cite{PandoraProtoDune2022}. In SBN, ICARUS and SBND have adopted a conservative approach in which the neutrino-like reconstruction is always saved and, with the exception of BDT failures, the $\nu$ hypothesis is always selected. 

While particle identification (PID) is not part of the Pandora framework, a few details on the method employed to perform it are provided here.
The particle identification method used in ICARUS condenses spatial and calorimetric information extracted from reconstructed particles into a ``$\chisq$ score." 
The underlying assumption is that, if the incident particle stops in the TPC active volume, the energy loss $dE/dx$ as a function of the residual range $rr$, defined as the distance of a given energy deposition within a certain track from the endpoint of the track itself, is a powerful tool to distinguish charged particles of different species. 
In the $\chisq$ score definition, the measured $dE/dx$ for a given reconstructed track is compared, on a hit-by-hit basis, with the average response predicted under different particle hypotheses from the Bethe formula, including protons, charged kaons, charged pions, and muons. The total $\chi^2$, for a given readout plane, is the sum of the contribution of all the hits, except for the first and last in the selected track. This sum is divided by the total number of hits reconstructed in the same wire plane to extract the PID score:
\begin{equation}
    \text{PID} = \chi^2/ndof = \sum_{hit}\left[\frac{dE/dx_{meas}-dE/dx_{theory}}{\sigma_{dE/dx}}\right]^2/{n_{hits}}
    \label{eq:chi2_formula}
\end{equation}
where $\sigma_{dE/dx}$ is an estimate of the resolution on the measured $dE/dx$ per hit ($\sim$3\%). This technique was originally developed by ArgoNeut~\cite{Anderson:2012vc, Acciarri2013}. 

While a PID score can be extracted for all the wire planes, for the analysis presented here, only the collection plane is used. In addition, based on dedicated studies, unphysical hits with $dE/dx$ signals below 0.5~MeV/cm or above 100~MeV/cm are excluded from the $\chisq$ calculation, thereby increasing the purity of the sample without impacting its efficiency.
Furthermore, only the last 25~cm of the track is used to compute the $\chisq$, unless the track is shorter. It is important to note that this strategy relies on the characteristic Bragg peak of each particle; thus if a particle scatters, instead of stopping, the identification power is lost. The discrimination power is also undermined when the reconstruction fails to identify the entire track, losing some hits, especially close to the end point.    

\begin{figure}[!htb]
 \centering
 \includegraphics[width=0.85\linewidth]{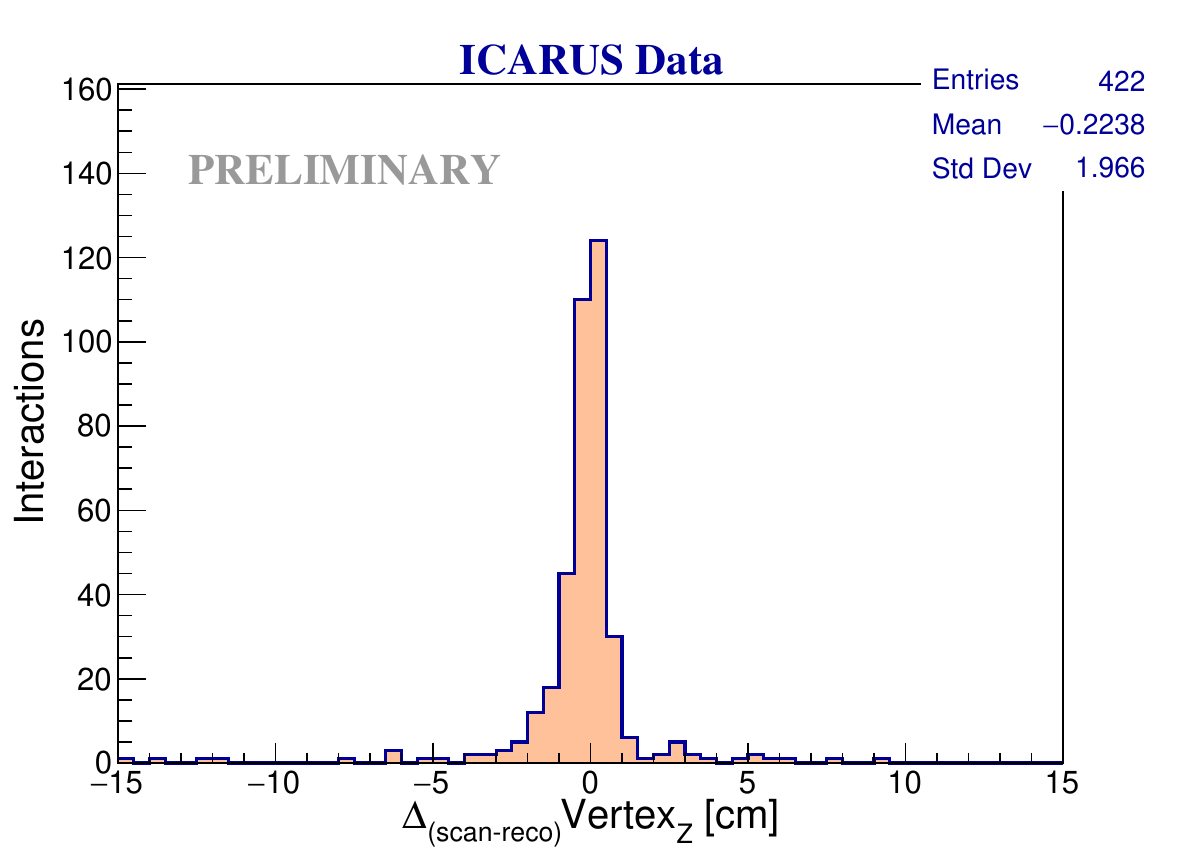}
 \includegraphics[clip,width=0.85\linewidth]{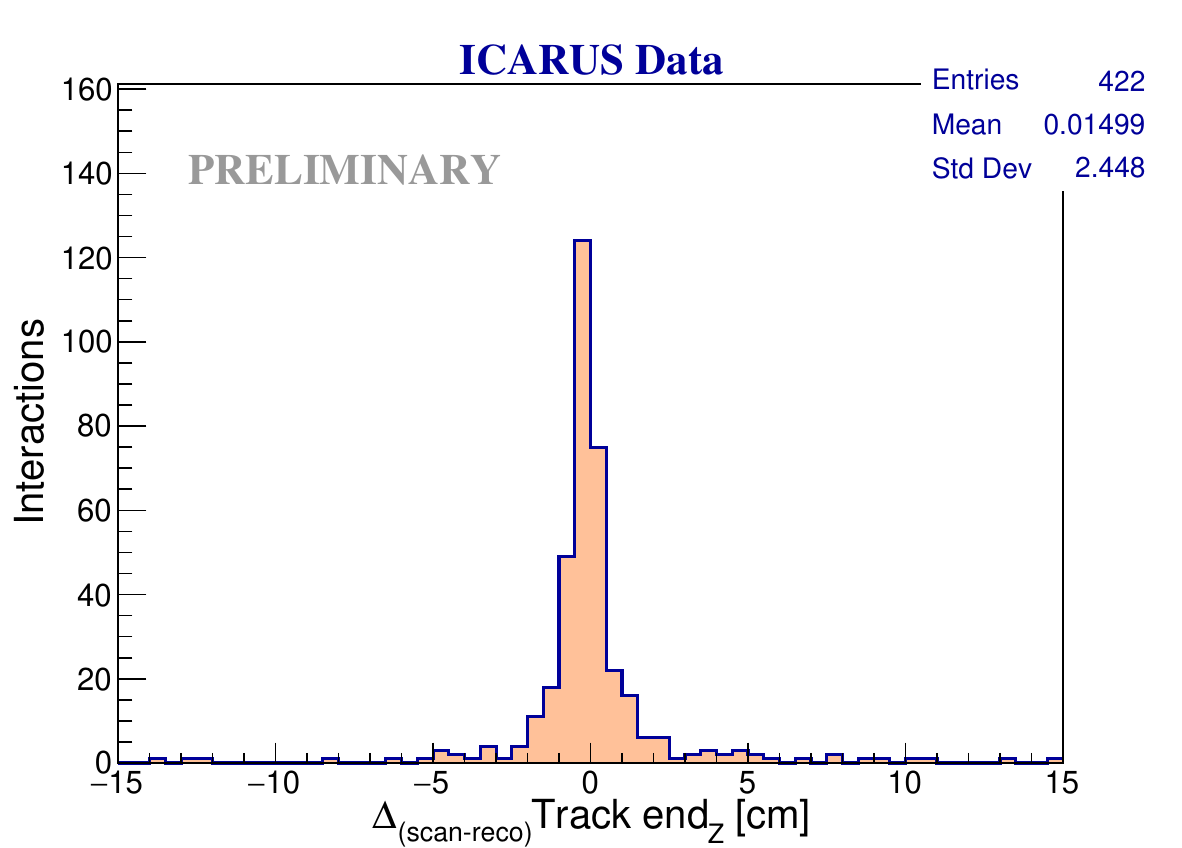}
\caption{\justifying The visual scan and Pandora-based automatic reconstructions are here compared for a sample of 422 candidate $\nu_{\mu}$ CC interactions from run 9435. In the top plot, a zoom in the range (-15,15)~cm in the difference between the vertex z coordinate obtained from the visual scan and automatic reconstruction is drawn. As a result 385/422 interactions are visible, meaning that the remaining 37 interactions lie on the tails of the distribution. In the bottom plot, a zoom in the range (-15,15)~cm in the difference between scan and reconstructed end of the muon track z coordinate is shown. As a result 371/422 interactions are visible.} 
\label{fig:scan_pandora_comparison}
\end{figure}

A dedicated visual study of events from a single run from Run~2, not included in the analysis presented here, was performed to select a sample of $\nu_{\mu}$ CC interactions from BNB. 
These events, combined with MC simulations, were used to validate Pandora-based event reconstruction, identify the most frequent pathologies -- particularly in vertex identification and track reconstruction for both muons and protons -- and to explore and test possible improvements.
This was also an essential step to tune our selection algorithms. 
Figure~\ref{fig:scan_pandora_comparison} shows the comparison between visual scan and Pandora-based reconstruction of the interaction vertex and the end point of the muon candidate along the beam direction, indicating that, for the bulk of the distribution, the automatic reconstruction and the manual measurement agree to within $\sim$7~mm, with a possible bias of a few millimeters. 

\section{\label{app:spine}SPINE}

\begin{figure*}
    \centering
    \includegraphics[width=0.95\textwidth]{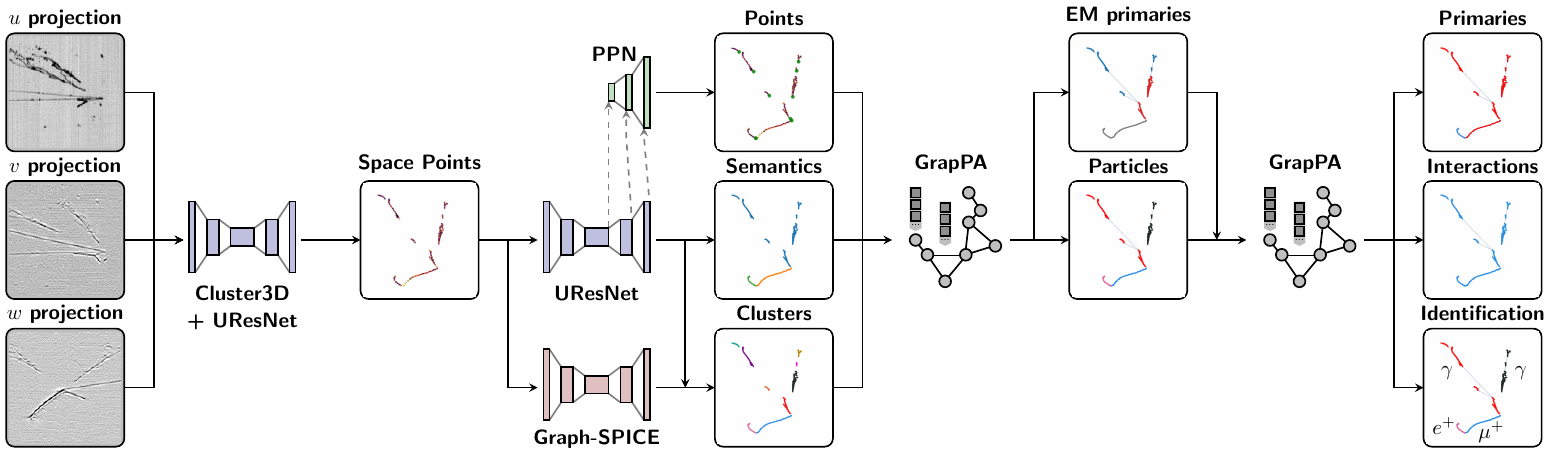}
    \caption{\justifying Schematic architecture of the end-to-end ML-based reconstruction chain (SPINE) used in the ICARUS experiment.}
    \label{fig:mlreco_schematic}
\end{figure*}

The end-to-end, ML-based reconstruction chain used in this analysis is referred to as SPINE (Scalable Particle Imaging with Neural Embeddings)~\cite{Spine}. SPINE consists of a hierarchical set of neural networks that are optimizable end-to-end. Two types of neural networks are employed: CNNs (Convolutional Neural Networks) and GNNs (Graph Neural Networks). The full chain of SPINE is shown schematically in Fig.~\ref{fig:mlreco_schematic} and each stage is summarized in Table \ref{fig:mlreco_nn}. This section reviews each stage of the SPINE reconstruction chain in order, starting from the voxel-level tasks and building up to the high-level particle- and interaction-level classifications.

\begin{table*}[!htbp]
    \centering
\begin{tabular}{lcc}    
\toprule
    Stage & Type & Description \\
    \midrule
    UResNet Deghost & CNN & Classification of space points as reconstruction artifacts or real charge depositions \\
    UResNet & CNN & Semantic segmentation (voxel-level classification of activity) \\
    PPN & CNN & Prediction of start/end points of showers/tracks \\
    Graph-SPICE & CNN & Coarse clustering of space points into particle fragments \\
    GrapPA-Shower & GNN & Clustering of shower fragments into complete showers \\
    GrapPA-Track & GNN & Clustering of track fragments into complete tracks \\
    GrapPA-Interaction & GNN & Clustering of particles into complete interactions with PID and primary designation \\
    \bottomrule
\end{tabular}
\caption{\justifying Summary of the neural networks used in the full ML (SPINE) event reconstruction chain in the ICARUS experiment.}
\label{fig:mlreco_nn}
\end{table*}

\subsection{Network Training}
\label{sec:training_datasets}

The training of neural networks requires a labeled dataset, where the true output is known. In the context of LArTPC data, the labeled dataset is provided by the simulation of the detector response to a set of known particles. In order to avoid potential biases due to assumptions of particular physics models governing cosmic ray and neutrino generation, the reconstruction chain is trained using a set of simulated events produced by two physics-agnostic generators:

\begin{itemize}
    \item \textbf{Multi-Particle Vertex (MPV)}: this generator produces a set of particles (between one and some specified number) originating from a common vertex. These particles include electrons, positrons, photons, muons, anti-muons, both negatively-charged and positively-charged pions, and protons.
    \item \textbf{Multi-Particle Rain (MPR)}: this generator produces a set of particles (between one and some specified number) that are isolated from one another. These particles include electrons, positrons, muons, anti-muons, and protons.
\end{itemize}

\noindent
The MPV generator is meant to emulate a pseudo-neutrino interaction with multiple particles emitted from a common vertex, whereas the MPR generator produces interactions that are similar to those produced by cosmic rays. For each particle, the generators are configured with a range of kinetic energies that can be assigned as well as a maximum particle multiplicity. The kinetic energy is drawn from a uniform distribution within the configurable range. 

\subsection{Tomographic Reconstruction}
\label{sec:tomographic_reconstruction}

The SPINE reconstruction chain operates on a 3D set of points representing the charge depositions in the detector. Due to the detector's readout being intrinsically 2D, some additional processing is required to reconstruct the input 3D image. The process of reconstructing a higher-dimensional image from a set of lower-dimensional projections is generally known as tomographic reconstruction, and is a concept that is widely used in medical imaging. In the context of LArTPC data, the 3D space points are reconstructed from the hits found in each wire plane of the detector.

\begin{figure}
    \centering
    \includegraphics[width=0.45\textwidth]{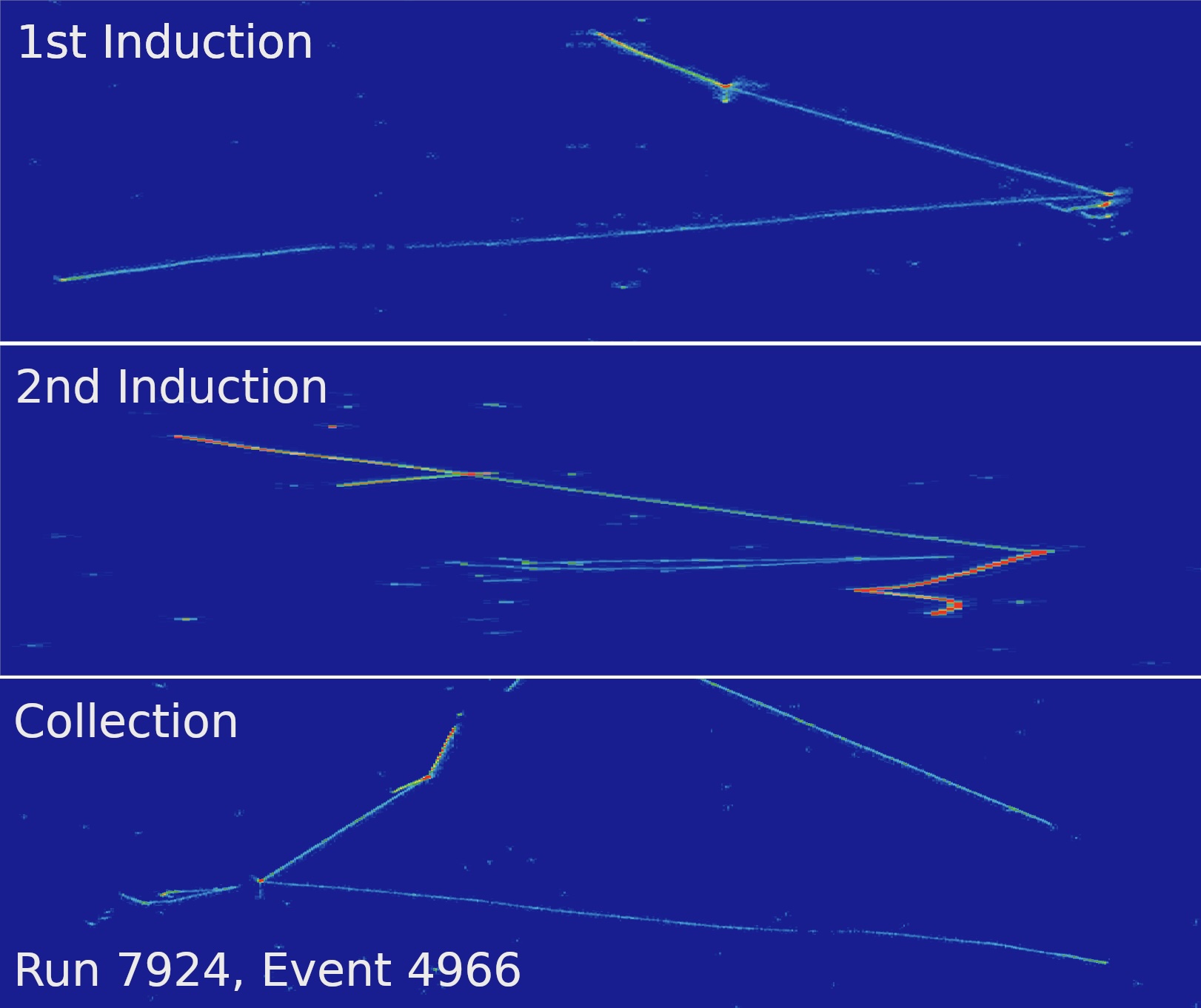}
    \includegraphics[width=0.45\textwidth]{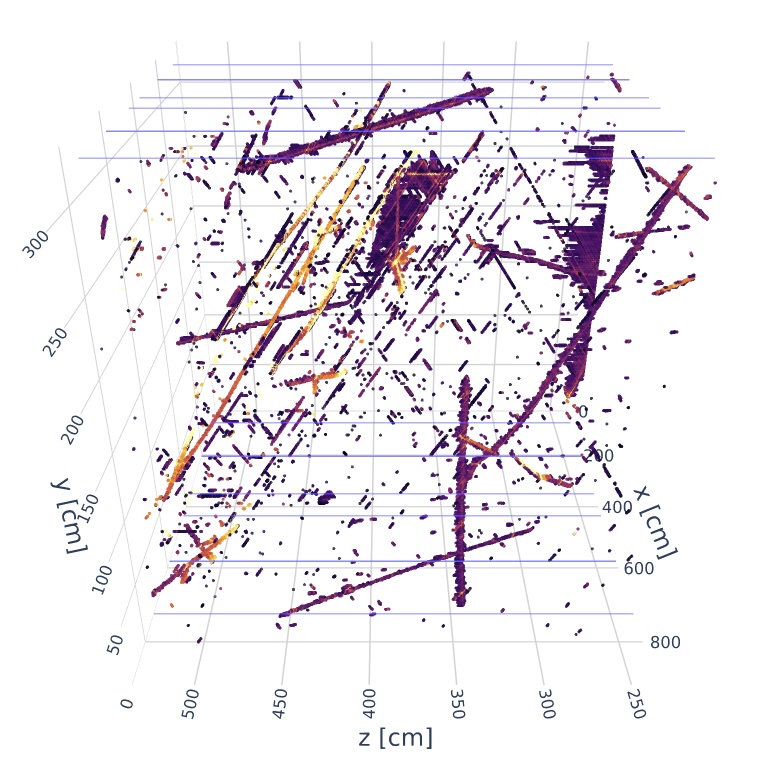}
    \caption{\justifying Example of a 2D event display of the same neutrino interaction in the ICARUS detector (left) and the corresponding 3D space points reconstructed by the Cluster3D algorithm (right). The image shown is a data event from Run1.}
    \label{fig:cluster3d}
\end{figure}

As described in section \ref{sec:TrackEventReco}, the set of hits selected by the Cluster3D algorithm are used in input to the tomographic reconstruction, tuned to be highly efficient at the cost of a higher number of incorrect combinations. These incorrect combinations are known as tomographic artifacts, or ``ghost points,'' and they are especially numerous for tracks parallel to the wire planes where the number of valid combinations of 2D hits is large. Figure~\ref{fig:cluster3d} shows an example of a 2D event display of a neutrino interaction in the ICARUS detector and the corresponding 3D space points reconstructed by the Cluster3D algorithm. Of particular note is the presence of the ghost points in the 3D space points, which pose a challenge to visually identifying the true activity in the image. The first stage of the point classification task is to identify and remove these ghost points.

\subsection{Point Classification}
\label{sec:point_classification}

\subsubsection{Ghost Point Removal}
\begin{figure}
    \centering
    \includegraphics[width=0.45\textwidth]{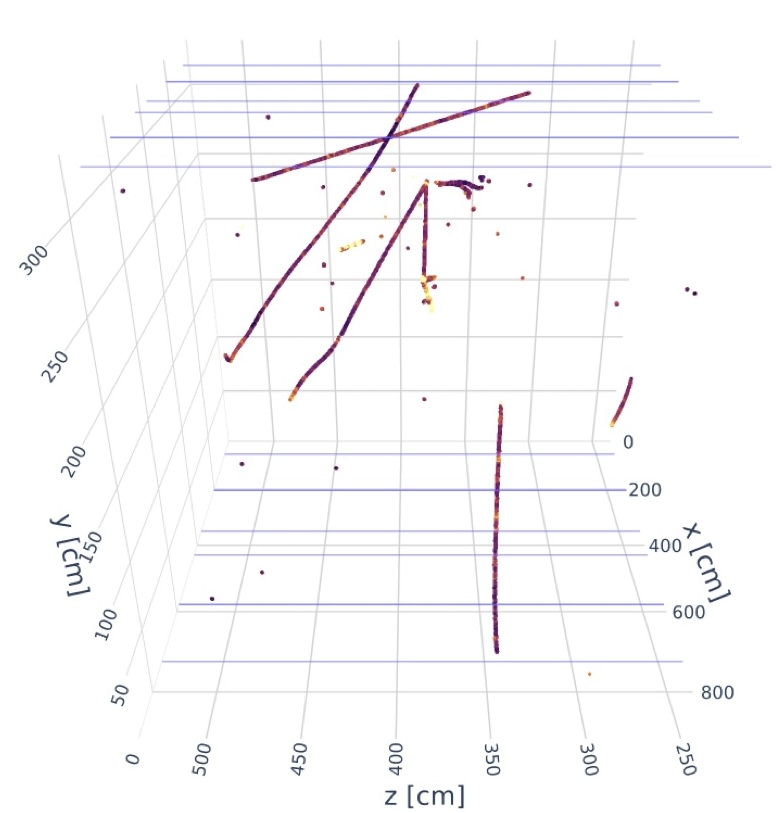}
    \caption{\justifying Example of the 3D space points after the deghosting step. The image shown is a data event from Run1 and the same as in Fig.~\ref{fig:cluster3d}.}
    \label{fig:deghosted}
\end{figure}

As a first step, SPINE reconstruction is tasked with the removal of the tomographic artifacts, or ghost points, from Cluster3D (``deghosting''). The neural network used for this implements the U-ResNet architecture, which was first popularized due to its success in biomedical imaging~\cite{Ronneberger2015}, to perform a binary classification of the space points, or semantic segmentation. Modifications were made to adapt U-ResNet to the sparse format of LArTPC data in contrast to regular images where each pixel contains information~\cite{Domine2020b}. After deghosting, the reconstructed charge from Cluster3D is redistributed to enforce conservation of the overall charge. This is done by counting the number of times a given 2D hit is associated with a 3D space point and distributing the charge equally. The charge assigned to each space point is the average of the charge measured on each plane that contributed to the hit. An example image showing the 3D space points after the deghosting step is shown in Fig.~\ref{fig:deghosted}.

\subsubsection{Semantic Segmentation and Point Proposal Network}
The next step in the point classification task is to classify the deghosted 3D space points into categories based on the type of parent activity that created them. The neural network used for this is a U-ResNet architecture, this time with five semantic classes of space points: Michel electron, delta ray, electromagnetic shower, track, and low-energy deposition. Each point is assigned a score for each class, and the class with the highest score is taken as the prediction. The semantic type of the space points is used in later stages of the reconstruction.

\begin{figure}
    \centering
    \includegraphics[width=0.45\textwidth]{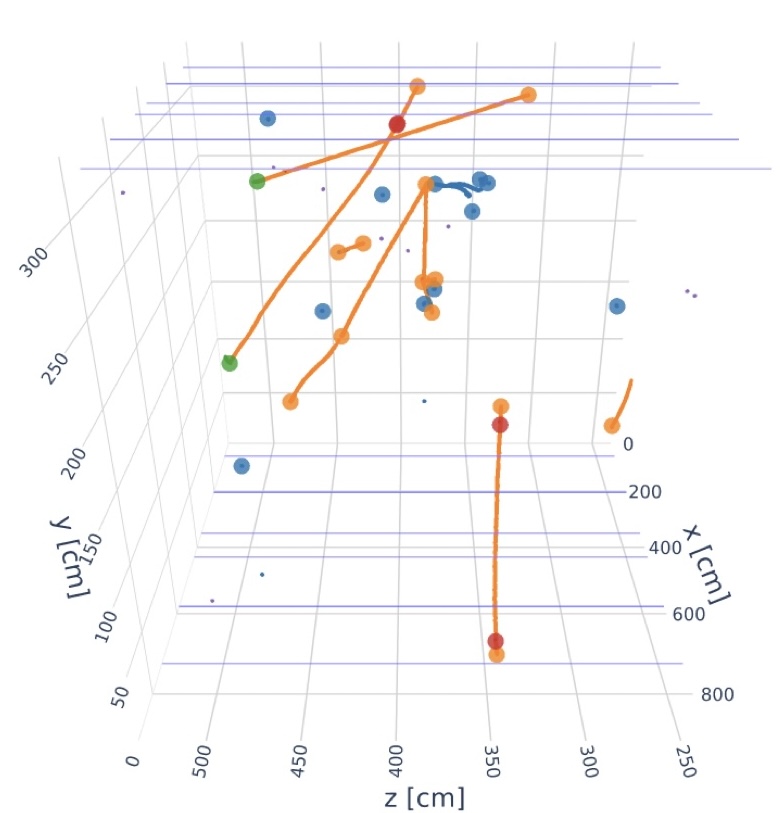}
    \caption{\justifying Example showing the 3D space points after the semantic segmentation and PPN stage. The space points are colored by the semantic type assigned by this stage (notably orange represents tracks, blue represents showers, green represents Michel electrons, and red represents delta rays). Also shown are the points of interest proposed by the PPN colored by the semantic class. The image shown is a data event from Run1 and the same as in Fig.~\ref{fig:cluster3d} and \ref{fig:deghosted}.}
    \label{fig:mlreco_points}
\end{figure}

The point proposal network (PPN) is a neural network that is tasked to propose the 3D space points that are the initial and terminal points of track-like objects and the initial points of shower-like objects~\cite{Domine2020a}. The PPN shares the same encoding backbone as the semantic segmentation network as shown schematically in Fig.~\ref{fig:mlreco_schematic}. The points of interest proposed by this network are valuable for high-level physics analyses (e.g.,\ particle start/end points and interaction vertex-finding) and are used in the clustering tasks in the later stages of the reconstruction. Figure~\ref{fig:mlreco_points} shows the 3D space points colored by semantic class and the points proposed by the PPN.

\subsubsection{Point Clustering}
The final step in the point classification task is to cluster the 3D space points in a set of points belonging to the same activity, i.e., track and shower fragments. This network, named Graph-SPICE, also uses a U-ResNet architecture and is tasked with learning an embedding of the space points in a higher-dimensional space where the points belonging to the same activity are close together. In this embedded space, the points belonging to the same semantic class are used to build a $k$-nearest neighbors graph, which connects the points that are close together. Each edge is given a score based on the feature vectors of the two points it connects and the edge is activated if the score is above a threshold. Space points are clustered into fragments by finding the connected components of the graph. The output of this stage is a set of point clusters, each representing a track or shower fragment.

\subsection{Formation of Particles and Interactions}
\label{sec:particle_interaction_reconstruction}

\subsubsection{Particle Instance Clustering}
At this point in the SPINE reconstruction chain, groups of space points corresponding to particle fragments have been identified. The next step is to cluster these groups into particles. The neural network used for this task is a GNN that is tasked with a binary (on/off) edge classification, where an activated edge indicates that the two instances belong to the same particle. Two distinct GNNs are used for particle instance clustering: GrapPA-Track and GrapPA-Shower for tracks and showers, respectively. It is worth noting that these two GNNs have identical architecture and only differ in the tasks for which they have been trained. The details of the architecture of these GNNs are described in~\cite{Drielsma2021}.

The node features in each GNN are mostly geometric quantities such as the particle fragment direction vector and the most-likely PPN point. Edge features are also geometry-based and include the distance between the two fragments. The initial state of the graph has activated edges for all node pairs up to a maximal distance. This limitation is motivated by both the computational cost associated with fully connecting an image in a detector as large as ICARUS and by physics concerns: beyond a certain length scale there cannot be legitimate connections. The result of these two GNNs is a set of complete particles of both the track and shower categories. Small gaps in tracks, for example due to track-breaking at the cathode, are expected to be resolved in this stage. The shower-clustering step also identifies the primary shower fragment that is used to identify the shower start point, a critical feature for the interaction-clustering step.

\subsubsection{Interaction Clustering}
The final stage of the event reconstruction is to cluster the particles into interactions, defined as a group of particles originating from the same vertex. Particles that are directly connected to the primary vertex are considered primary particles, meaning they are immediate products of the interaction, whereas other particles are labeled as secondaries. The clustering of particles into interactions is easily cast as an edge-classification task, where an activated edge indicates two particles belong to the same interaction. In addition to the interaction clustering, the network is also tasked with predicting for each node the particle type (photon, electron, muon, pion, or proton) and the primary/secondary label. The neural network used for this task is a GNN and is described extensively in~\cite{Drielsma2021}. The architecture of this GNN differs from GrapPA-Track and GrapPA-Shower only in that it has more hidden features, which are useful for the extra tasks of particle identification (PID) and primary ID classification. The output of this stage is a full list of interactions in the event, each with a list of primary and secondary particles and their types.

\subsection{Optical Flash Association}
\label{sec:optical_flash_association}

The tool used for associating optical flashes to neutrino interaction candidates reconstructed with SPINE is a package originally developed for MicroBooNE called OpT0Finder.cThe OpT0Finder algorithm is not available at the moment for the Pandora reconstruction. The event analysis based on Pandora exploits a simpler TPC-PMT matching as described in section \ref{sec:selection}. OpT0Finder employs a likelihood-based approach to match the optical flash to charge activity associated with an interaction in the event. For each interaction, as defined by the upstream ML reconstruction, the algorithm calculates an allowed range of $x$ offsets that are allowed by the geometry. This reflects our inability to know beforehand the exact time of the interaction. For each of these allowable positions\footnote{This is not actually a brute-force algorithm, but rather an optimization algorithm that profiles across the drift direction.}, the algorithm calculates a hypothesis flash based on the expected light yield of the interaction, the distribution of the ionization charge in the interaction, and the position of the PMTs in the detector. The hypothesis flash and the observed flash each have a likelihood computed using a joint probability distribution of the photoelectron count in each PMT given the $x$ position of the flash. Maximizing the likelihood between the two allows the assignment of a score to each potential interaction-flash pair. 

A matching algorithm then selects the best pairings based on the scores while simultaneously skipping pairs with an already matched flash or interaction. The result of this association process is the assignment of a time stamp to each interaction which was successfully matched to a flash. This time stamp can be used to precisely tag an interaction as being in-time or out-of-time with respect to the beam spill, and therefore can be leveraged as a tool for cosmic rejection.

\section{\label{app:profit}PROfit}
The PROfit framework supports a broad class of oscillation models. For the 
$\nu_\mu$ disappearance analysis presented here, we adopt the  short-baseline effective two-flavor approximation within the full 3+1 sterile neutrino model. Thus no matter effects are included and the oscillation probabilities are parametrized in terms of a single mass-squared splitting \dmsq\ and three effective mixing angles defined by the elements of the extended PMNS matrix:
\begin{equation}
\begin{aligned}
\sinsqtwothmumu &= 4|U_{\mu 4}|^2(1-|U_{\mu 4}|^2), \\
\sinsqtwothee &= 4|U_{e 4}|^2(1-|U_{e 4}|^2), \\
\sinsqtwothmue &= 4|U_{e 4}|^2|U_{\mu 4}|^2.
\end{aligned}
\label{eq:mixing_angles}
\end{equation}
Only two of the three effective mixing angles are independent. The corresponding survival and transition probabilities take the form:
\begin{equation}
\begin{aligned}
P(\nu_\mu\rightarrow\nu_\mu) &= 1-&\sinsqtwothmumu \sin^2\left(\frac{\dmsq L}{4E}\right), \\
P(\nu_e\rightarrow\nu_e) &= 1-&\sinsqtwothee \sin^2\left(\frac{\dmsq L}{4E}\right), \\
P(\nu_\mu\rightarrow\nu_e) &= &\sinsqtwothmue \sin^2\left(\frac{\dmsq L}{4E}\right).
\end{aligned}
\label{eq:osc_prob}
\end{equation}

For this $\nu_\mu$ disappearance-only analysis, we set $|U_{e4}|^2 = 0$ (equivalent to $\theta_{ee}=\theta_{\mu e}=0)$, such that only $P(\nu_\mu\rightarrow\nu_\mu)$ is non-trivial. As such there are two distinct degrees of freedom corresponding to the physics parameters of our model, \dmsq and $\sin^2\theta_{\mu\mu}$. 

Charged-current \numu\ interactions are weighted by $P(\nu_\mu\rightarrow\nu_\mu)$, charged-current \nue\ interactions remain unoscillated, and neutral-current interactions are oscillated using $P(\nu_\mu\rightarrow\nu_\mu)$. The latter choice introduces a small approximation given the 0.5\% intrinsic \nue\ component of the beam and the subdominant NC background in the selected sample (see Fig.~\ref{fig:selected_datamc_ereco}). Future analyses incorporating full 3+1, \nue\ appearance or explicit NC oscillation channels will expand to the full suite of oscillation probabilities including NC oscillation terms dependent on $|U_{\tau 4}|^2$.

PROfit reads analysis ntuples containing one entry per reconstructed neutrino interaction candidate for both data and simulation. While the framework accommodates an arbitrary number of detectors and oscillation channels, this analysis uses only the ICARUS detector and a single $\numu$ disappearance channel. This channel is decomposed into five subchannels based on true interaction type: $\nu_\mu$ CC, $\nu_e$ CC, NC, in-time cosmics (cosmic rays coincident with a neutrino interaction), and out-of-time cosmics (estimated from off-beam data). The subchannels are kept separate when applying oscillation weights and systematic shifts, then collapsed into a single spectrum for the $\chi^2$ calculation.

While oscillation probabilities can be applied event-by-event, this becomes computationally prohibitive for high-statistics simulation samples. Instead, a binned approach is adopted: oscillation weights are calculated in 150 equally spaced bins of true $L/E$ spanning $0-3$~km/GeV. This granularity was validated against full event-by-event reweighting and found to produce negligible differences for this single-detector ICARUS analysis.

The treatment of systematic uncertainties is central to any oscillation analysis. PROfit supports two distinct approaches: covariance matrix systematics and nuisance parameter (or pull term) systematics. 

In the nuisance parameter formalism, the effect of $N_\text{nuis}$ systematic variations on each bin is calculated explicitly. This approach offers greater interpretability, best-fit values and posterior uncertainties for each nuisance parameter are directly extractable, at the cost of increased computational complexity from the $N_\text{nuis}$ additional fit dimensions ($\lambda^i$), each with their own prior central value ($\lambda_\text{CV}^i$) and prior uncertainty ($\sigma_{\lambda}^i$). The bin-by-bin response to each nuisance parameter is stored as a spline, one for each systematic and each bin, interpolating between pre-calculated $-3\sigma$, $-2\sigma$, $-1\sigma$, $0$, $+1\sigma$, $+2\sigma$, and $+3\sigma$ weight variations.

In the covariance matrix formalism, uncertainties and correlations on a central-value prediction of $N_\text{bins}$ bins are encoded in a symmetric positive-semidefinite matrix $M_\text{syst}$. A single covariance matrix can capture the combined effect of many systematic sources without introducing additional fit parameters, appearing only once in the test statistic and enabling efficient evaluation. The drawback is that systematics are treated as symmetric and Gaussian, and their individual contributions are less transparent in post-fit results. The covariance matrix is constructed from $N_U$ random universe throws as $C_{ij} = \frac{1}{N_U}\sum_{u}(h_i^u - h_i^\text{CV})(h_j^u - h_j^\text{CV})$, where $h_i^u$ is the content of bin $i$ in universe $u$ and $h_i^\text{CV}$ is the central-value prediction. $N_U$ is 1,000 for flux systematics, and 100 for covariance interaction systematics.

PROfit allows analyses to operate in either a pure covariance or pure nuisance parameter mode, or in a hybrid configuration where selected systematics are incorporated via covariance matrices while others are treated as explicit nuisance parameters, leveraging the strengths of both approaches. In all cases, systematic effects enter through the test statistic minimized in the fit. This analysis employs the combined Neyman--Pearson (CNP) $\chi^2_{CNP}$~\cite{Ji:2019yca} as the metric minimized. This metric encodes the effects of both the covariance and nuisance parameters:

\begin{equation}
    \chi^2(\dmsq, \sinsqtwothmumu, \vec{\lambda}) = \Delta^T M_\text{CNP}^{-1} \Delta + \sum_{k=1}^{N_\text{nuis}}\left(\frac{\lambda^k-\lambda_\text{CV}^k}{\sigma_{\lambda}^k}\right)^2,
\end{equation}
where $\Delta \equiv O - P(\dmsq, \sinsqtwothmumu, \vec{\lambda})$ is the residual between observed ($O$) and predicted spectra ($P$), and $M_\text{CNP}$ is the combined statistical and systematic covariance matrix $M^\text{CNP}_\text{stat}+M_\text{syst}$. We note $M^\text{CNP}_\text{stat}$ is a function of both physics and nuisance parameters itself through the prediction, $P(\dmsq, \sinsqtwothmumu, \vec{\lambda})$, and defined as 
\begin{equation}
    M^\text{CNP}_\text{stat}(\dmsq, \sinsqtwothmumu, \vec{\lambda})(i,j) \equiv \frac{3\delta_{ij}}{1/O^i + 2/P^i}.
\end{equation}

All 13 flux uncertainties, 23 interaction model uncertainties (as laid out in Tab.~\ref{tab:xsecvariations}), and the effects of finite statistics of the simulation samples are treated via the covariance matrix approach. Since oscillation fits are performed in bins of reconstructed neutrino energy (Fig.~\ref{fig:result_spec}), all covariance matrices are constructed in this same binning. Fractional covariance matrices are stored and converted to absolute covariances at the time of $\chi^2$ evaluation, since oscillation weights and nuisance parameter shifts modify the predicted spectrum.

The remaining interaction model, detector uncertainties and normalization uncertainties are treated as nuisance parameters. For each systematic, spectra corresponding to $\pm1\sigma$, $\pm2\sigma$, and $\pm3\sigma$ variations are used to construct splines describing the bin-by-bin response as a continuous function of the nuisance parameter. Some variations represent discrete alternative model choices rather than continuous parameters; in these cases the nuisance parameter is bounded to prevent extrapolation beyond $\pm1\sigma$. Normalization uncertainties are also implemented as nuisance parameters, allowing the fit to report preferred values and uncertainties for each.

PROfit employs a four-stage minimization procedure. First, Latin hypercube sampling of 5,000 points across the full parameter space (physics and nuisance parameters) identifies regions of low $\chi^2$. A particle swarm optimization~\cite{488968} then refines the most promising candidates, with 25 particle and 250 iterations converging rapidly on local minima. The best 15 candidates from both the particle swarm and latin hypercube are subsequently passed to the L-BFGS-B gradient-based minimizer~\cite{LBFGSB/doi:10.1137/0916069} to locate precise minima. Finally, to guard against convergence to a local rather than global minimum, PROfit exploits the periodic structure of the oscillation probability: for a given $L/E$, the transformation $\dmsq \to \dmsq + 4n\pi E/L$ leaves $\sin^2(\dmsq L/4E)$ invariant, for fixed $L/E$. A dedicated scan around harmonics of the current best-fit $\dmsq$ is therefore performed, which is particularly important for identifying fast oscillations at large \dmsq.

\begin{figure}[!htb]
    \centering
    \includegraphics[width=\columnwidth]{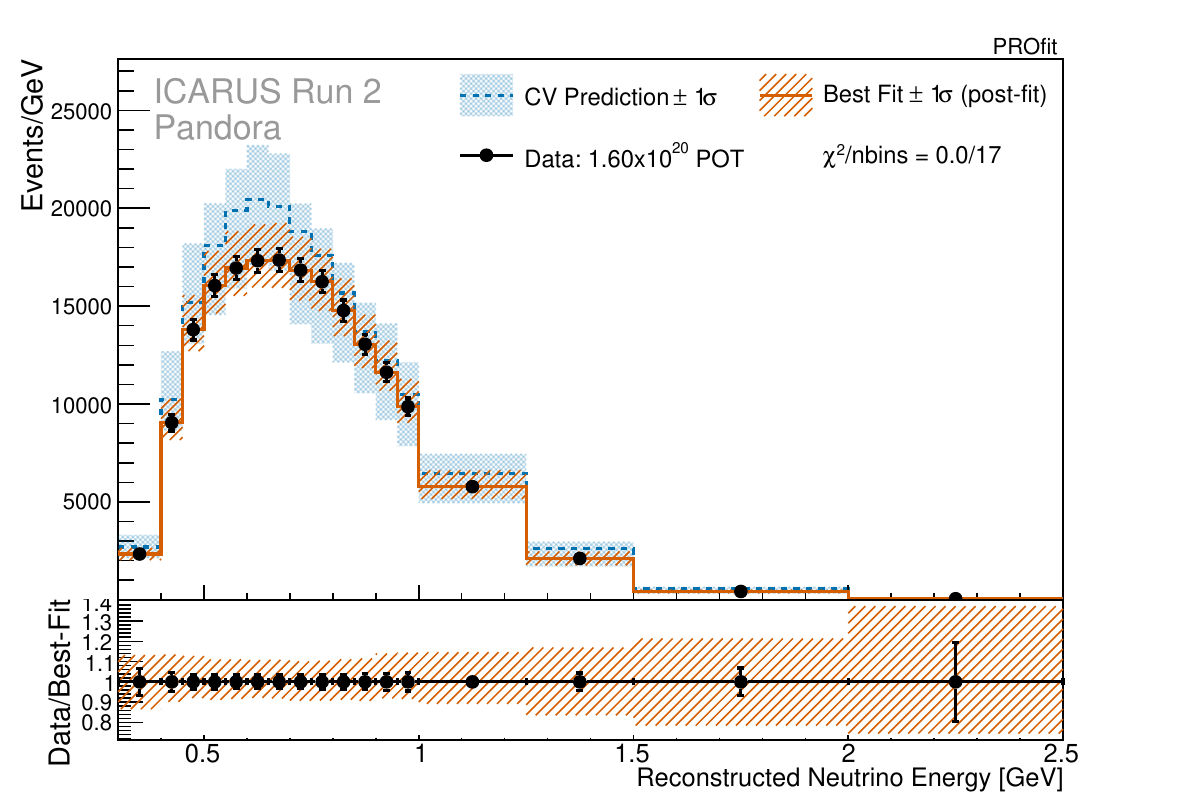}
    \includegraphics[width=\columnwidth]{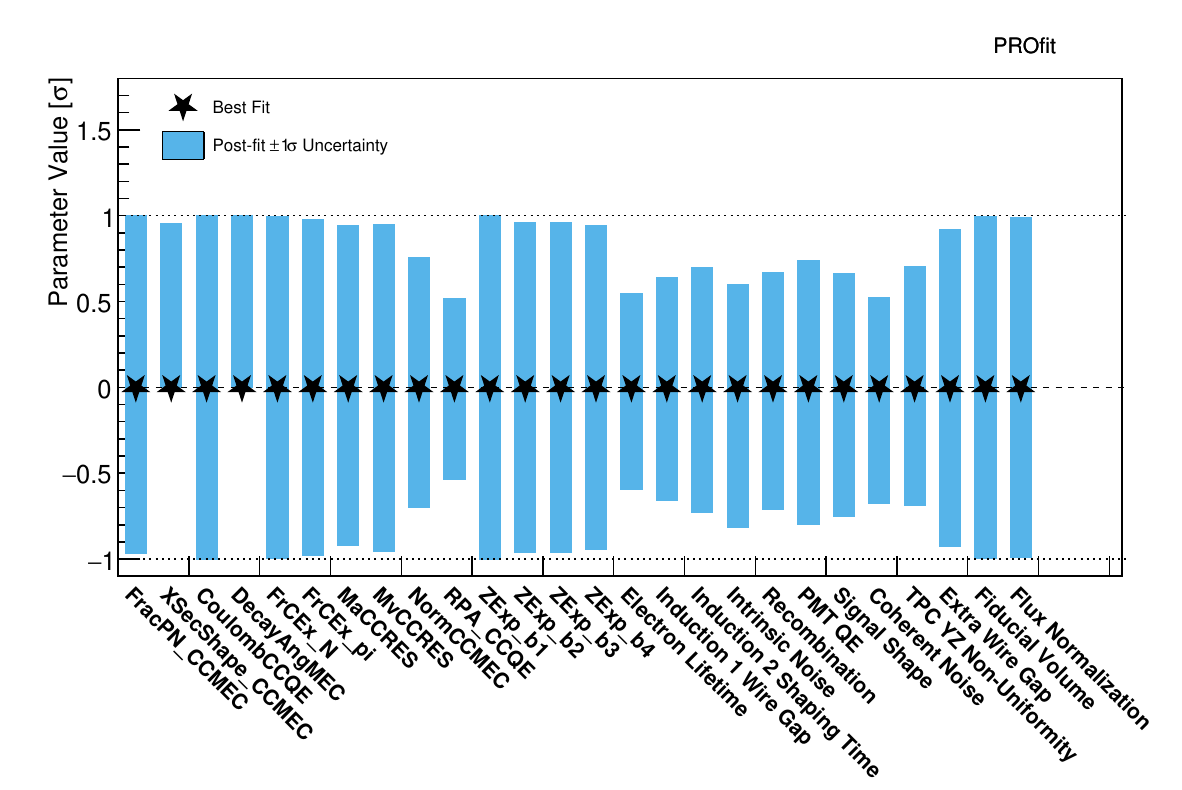}
\caption{\justifying Example PROfit output for an Asimov dataset with injected signal parameters $\dmsq = 4~\text{eV}^2$ and $\sinsqtwothmumu = 0.25$. (Top) Post-fit spectrum compared to the pre-fit central value prediction and data. (Bottom) Nuisance parameter best-fit values and $1\sigma$ uncertainties obtained via profiling. Pre-fit nuisance parameter values are by definition 0 (no pull), with pre-fit uncertainty defined as $\pm1\sigma$ (horizontal dotted lines).}    \label{fig:profit_example}
\end{figure}

An example fit to Asimov MC (the MC prediction with no statistical fluctuations) with an injected oscillation signal is shown in Fig.~\ref{fig:profit_example}, including pre- and post-fit spectra, uncertainties, and nuisance parameter values. The fit successfully recovers the injected signal.

Pre-fit uncertainty bands are obtained by generating an ensemble of pseudo-experiments in which nuisance parameters are sampled from their prior distributions and correlated fluctuations are drawn from the systematic covariance matrix; 5000 pseudo-experiments are used. Post-fit uncertainty bands must account for potentially correlated shifts in the nuisance parameters at the best-fit point. To capture these correlations, an adaptive Markov Chain Monte Carlo variant of the Metropolis Hastings method~\cite{bj/1080222083} is used to sample the nuisance parameter space in the vicinity of the best fit, with 25,000 burn-in steps followed by 20,000 sampling steps. In both cases, the displayed error bars represent the central 68\% interval of the resulting pseudo-experiment distribution in each bin.

Post-fit uncertainties on individual nuisance parameters are obtained separately by profiling over all other parameters while scanning each parameter of interest, mapping the $\chi^2$ surface in the vicinity of the best fit.

After the best fit parameters values are identified, confidence intervals on the oscillation parameters are constructed using the Feldman-Cousins procedure~\cite{FeldmanCousins:PhysRevD.57.3873}, the standard frequentist approach in the neutrino community. This method employs a likelihood-ratio ordering to construct confidence belts via a Neyman construction at each point in parameter space. The approach guarantees correct frequentist coverage by construction and naturally handles physical boundaries on parameters, providing a unified treatment that transitions smoothly between upper limits (when data are consistent with the null hypothesis) and two-sided allowed regions (when a signal is observed). At each grid point in the $(\dmsq, \sinsqtwothmumu)$ plane, an ensemble of pseudo-experiments is generated by sampling nuisance parameters from their prior distributions, drawing correlated fluctuations from the systematic covariance matrix, and applying Poisson fluctuations to produce simulated datasets. The nuisance parameters are sampled using their prior uncertainties according to the Highland-Cousins approach~\cite{Cousins:1991qz}. 

At each point in the $(\dmsq, \sinsqtwothmumu)$ grid, the test statistic is the likelihood ratio expressed as
\begin{equation}
\Delta\chi^2 = \chi^2(\dmsq, \sinsqtwothmumu, \hat{\vec{\lambda}}) - \chi^2(\widehat{\dmsq}, \widehat{\sinsqtwothmumu}, \hat{\vec{\lambda}}'),
\end{equation}
where the first term is the $\chi^2$ minimized over nuisance parameters $\vec{\lambda}$ with the oscillation parameters fixed at the grid point, and the second term is the global best fit with all parameters, including physics parameters, free. For each pseudo-experiment generated at a given grid point, this $\Delta\chi^2$ is computed, building an expected distribution of the test statistic under the hypothesis that the true parameters lie at that grid point.

When analyzing data, the observed $\Delta\chi^2_\text{data}$ is compared to this ensemble: the $p$-value is the fraction of pseudo-experiments with $\Delta\chi^2$ exceeding the observed value. A grid point is excluded at confidence level $1-\alpha$ if fewer than a fraction $\alpha$ of pseudo-experiments yield a larger $\Delta\chi^2$ than the data. Repeating this procedure across the full parameter grid produces Feldman-Cousins-corrected exclusion contours or, in the presence of a signal, allowed regions with correct frequentist coverage. For this analysis, between 1,500 and 6,500 pseudo-universes were samples across grids for both Pandora and SPINE analyses, with the higher numbers being sampled closer to the resulting confidence interval boundaries to improve the statistical precision closer to final result.

In addition to the Feldman-Cousins confidence intervals constructed from data, the expected sensitivity of the analysis is quantified in two complementary ways. First, the null hypothesis Asimov dataset is analyzed as a representative pseudo-experiment, yielding the Asimov sensitivity contour. Second, an ensemble of 5,000 pseudo-experiments are generated under the null hypothesis (no sterile neutrino oscillations) and Feldman-Cousins confidence intervals are constructed for each. From the ensemble of pseudo-experiments resulting in exclusion contours, we extract the median expected exclusion sensitivity as well as $\pm1\sigma$ and $\pm2\sigma$ bands, representing the expected spread of exclusion limits if the null hypothesis is indeed true.

A complete presentation of results therefore includes three elements: the observed exclusion contour from data, the Asimov sensitivity, and the distribution of expected exclusions under the null hypothesis. The Asimov contour provides a single representative sensitivity, while the ensemble bands indicate the range of statistical fluctuations expected in the measurement. Comparison of the observed contour to both of these expectations is essential for assessing whether any deviation from the null hypothesis is statistically significant or consistent with expected fluctuations.

Signal unblinding follows a predetermined staged procedure in PROfit in which fits to both the SPINE and Pandora selections are performed simultaneously. Validation checks are evaluated at each stage before proceeding to ensure a robust result. The unblinding stages are:
\begin{enumerate}
    \item \textbf{Fitter validation:} Verify that the fit executed without errors, the number of empty bins in data and simulation is acceptable, the covariance matrix is positive semi-definite, and the best-fit $\chi^2$ and parameter values are physical.
    \item \textbf{Goodness of fit:} Evaluate the best-fit $\chi^2$ and corresponding $p$-value, requiring $p > 0.05$ to proceed. No best-fit values themselves are known at this point.
    \item \textbf{Blinded pull terms:} Inspect the distribution of nuisance parameter pulls without identifying individual parameters. Pulls exceeding $3\sigma$ would halt the procedure; pulls exceeding $2\sigma$ prompt further investigation before proceeding. In mock data studies, pulls rarely exceed $1\sigma$.
    \item \textbf{Unblinded pull terms:} Inspect the best-fit values, $1\sigma$ uncertainties, and profiles for each nuisance parameter. Check and study any unexpected post-fit correlations.
    \item \textbf{Oscillation parameters:} Examine the best-fit values of $\dmsq$ and $\sinsqtwothmumu$, and evaluate the Feldman--Cousins $p$-value testing the no-oscillation hypothesis.
    \item \textbf{Confidence intervals:} Construct the Feldman-Cousins confidence region. If the $p$-value from the previous step exceeds 0.05, a 90\% exclusion limit is drawn; otherwise, an allowed region is reported.
    \item \textbf{Comparison to expectation:} Compare the observed result to the Asimov sensitivity and the ensemble-derived $\pm1\sigma$ and $\pm2\sigma$ bands.
\end{enumerate}
Stages 1--3 were first executed on a small 10\%  subset of the available data as an initial validation. Following successful completion, the full procedure was repeated on the complete Run~2 dataset.

\bibliography{icarusprd}
\end{document}